\documentclass[trackchanges,twocolumn,twocolappendix]{aastex701}
\usepackage{natbib}
\makeatother
\usepackage{gensymb}

\usepackage{amssymb}
\usepackage{amsmath}
\usepackage{enumitem}
\usepackage{multirow}
\newcommand{\citepcorp}[1]{(\kern-0.30em\citealp{#1})}

\begin{document}

\title{Supermassive black hole binaries in the multi-messenger context of ground and spaceborne VLBI \footnote{Contribution for the PASP special issue on BHEX \\ Email: b.hudson@tudelft.nl}}

\author[0000-0002-3368-1864]{Ben~Hudson}
%\altaffiliation{PASP Editor-in-Chief}
\affiliation{Faculty of Aerospace Engineering, Delft University of Technology, Kluyverweg 1, 2629 HS Delft, The Netherlands}
\email{b.hudson@tudelft.nl}

\author[0000-0002-0694-2459]{Leonid~I.~Gurvits}
\affiliation{Faculty of Aerospace Engineering, Delft University of Technology, Kluyverweg 1, 2629 HS Delft, The Netherlands}
\affiliation{Joint Institute for VLBI ERIC (JIVE), Oude Hoogeveensedijk 4, 7991 PD Dwingeloo, The Netherlands}
\email{leonid@gurvits.org}

\author[0000-0002-1271-6247]{Daniel~J.~D'Orazio}
\affiliation{Niels Bohr International Academy, Niels Bohr Institute, Blegdamsvej 17, 2100 Copenhagen, Denmark}
\affiliation{Space Telescope Science Institute, 3700 San Martin Drive, Baltimore, MD 21218, USA}
\affiliation{Department of Physics and Astronomy, Johns Hopkins University, 3400 North Charles Street, Baltimore, MD 21218, USA}
\email{dorazio@stsci.edu}

\author[0000-0002-3820-2404]{Christopher~Tiede}
\affiliation{Niels Bohr International Academy, Niels Bohr Institute, Blegdamsvej 17, 2100 Copenhagen, Denmark}
\email{christopher.tiede@nbi.ku.dk}

\author[0000-0001-9543-0414]{T. Marshall Eubanks}
\affiliation{Space Initiatives Inc, Princeton, WV 24740, USA}
\email{tme@space-initiatives.com}

\author[]{Erwin~Mooij}
\affiliation{Faculty of Aerospace Engineering, Delft University of Technology, Kluyverweg 1, 2629 HS Delft, The Netherlands}
\email{E.Mooij@tudelft.nl}

\begin{abstract}
\noindent Evidence of a gravitational wave background suggests the existence of a population of sub-parsec supermassive black hole binaries (SMBHBs), with characteristic angular separations on the order of 1--10\,$\mu$as. Spaceborne extensions of Very Long Baseline Interferometry (VLBI) and the next-generation ground arrays introduce the possibility of directly imaging these SMBHBs. In this work, a binary spectral energy distribution model is used to predict the detectability of SMBHBs with ground and spaceborne VLBI. We consider the Black Hole Explorer (BHEX), a proposed spaceborne VLBI mission, as our primary case study. We explore the detectable SMBHB parameter space and identify distinguishable binary signatures in the visibility domain. We find that for a flux-density-limited sample, ground array observations are more effective at detecting a wider region of the binary parameter space, with $M_\text{tot}\gtrsim10^9 \, M_\odot$ systems detectable out to redshift $z=0.075$ and beyond, depending on mass ratio. Conversely, inclusion of a spaceborne element such as BHEX, offering finer angular resolution ($\sim6\,\mu$as) and sampling of the \emph{(u,v)} plane not limited by Earth rotation synthesis, will provide significant benefits in constraining binary properties. An illustrative Fisher analysis shows improvements in characterisation of the separation and position angle of SMBHBs by a factor of $\sim4$. Near-future ground and/or spaceborne VLBI may achieve the first direct observation of a SMBHB, contributing significantly to multi-messenger studies of such systems with pulsar timing arrays and observations across the electromagnetic spectrum.
\end{abstract}

%% PASP uses Unified Astronomy Thesaurus (UAT) concepts:

\section{Introduction} 
\label{s:intro}

\noindent Pulsar timing array (PTA) observations provide evidence for a nanohertz gravitational wave background (GWB), suggesting the existence of a population of supermassive black hole binaries (SMBHBs) with sub-parsec separations \citep{antoniadis_international_2022,agazie_nanograv_2023,epta_collaboration_and_inpta_collaboration_second_2024,xu_searching_2023}. \cite{begelman_massive_1980} first described the formation and evolution of SMBHBs, and suggested the possibility of observing such systems with Very Long Baseline Interferometry (VLBI).

At the time of writing, there has been no conclusive, direct electromagnetic observation of a sub-parsec SMBHB. Various forms of indirect evidence have been proposed, such as: optical, ultraviolet (UV) and radio oscillations \citep[e.g.,][]{Graham+2015, dorazio_relativistic_2015, Charisi:2016, Liu+2019, 
ChenXin:2020,
liao_discovery_2020, oneill_unanticipated_2022}; astrometric oscillations with VLBI \citep{gurvits_milliarcsecond_2025}; and binary signatures on relativistic jet behaviour, particularly for the most studied candidate OJ287 \citep[][and references therein]{Britzen+2018MNRAS, valtonen_identifying_2025,  Traianou+2025AA-OJ287}. See \cite{dorazio_observational_2023} for a more in-depth review of such observational methods. Recently, the NANOgrav collaboration have also presented a search for continuous GW sources in the 15 year PTA dataset \citep{agarwal_nanograv_2026,Veronsi+2026}. Although none of the sources are statistically significant yet (two identified with $1 < $ Bayes Factor (BF) $ < 4$), this method holds promise for future binary candidate identification.

The next generation Event Horizon Telescope (ngEHT) will push angular resolution close to the limit of what is possible from the ground \citep{ayzenberg_fundamental_2025}. With observations at \(\sim\)345~GHz, and inclusion of more ground antennae, its angular resolution will approach 15~$\mu$as, \(\sim\)25\% improvement over the achieved to date EHT results \citepcorp{the_event_horizon_telescope_collaboration_first_2019}.  \cite{dorazio_repeated_2018} predict that with these capabilities, a few SMBHB systems may be detectable out to $z\leq0.2$. Ground-based VLBI is fundamentally limited in angular resolution by the Earth's diameter and opacity of its atmosphere at THz frequencies. It has long been recognised that a spaceborne system could remove these limits, and also achieve rapid and dense coverage in the \emph{(u,v)} plane. Two dedicated spaceborne VLBI (SVLBI) missions have flown at the time of writing: VSOP-HALCA \citep{Hirabayashi+1998Sci} and RadioAstron \citep{kardashev_radioastron-telescope_2013}, in addition to two ad-hoc experiments, TDRSS-OVLBI \citep{Levy+1986Sci} and LOVEX \citep{Hong+2025SCPMA-LOVEX}.

The Black Hole Explorer (BHEX) is a proposed spaceborne extension to ground-based arrays, observing at submillimetre wavelengths and aiming to achieve the finest angular resolution to date, $\sim6\,\mu$as 
%, in astronomical history   
\citep{johnson_black_2024}. BHEX will detect and measure characteristics of the photon rings that are predicted to exist in images of black holes, with M87* and Sgr\,A* as the primary targets. BHEX will also aim to resolve black hole shadows in additional SMBH targets \citep{zhang_accessing_2025} and to perform studies of AGN and relativistic jets in unprecedented detail.

The enhanced angular resolution of BHEX, and improving sensitivity of ground-based arrays, warrants an investigation into the possibility of performing the first direct detection of a sub-parsec SMBHB with VLBI. In our previous work, we preliminarily constrained the binary detection capabilities of BHEX, and demonstrated its ability to measure the relative position of the secondary component with respect to the primary, with in-beam relative astrometry \citep{hudson_towards_2026}. A post-Newtonian (PN) orbit fitting approach was also developed and used to show that with only three annual observations, BHEX could constrain the semi-major axis and eccentricity of low inclination binaries with observed orbital periods $P_\text{obs}\leq10$~years, to within 0.06 dex of their true values.

Here, we extend this work to evaluate binary detection prospects with ground and space VLBI using a binary spectral energy distribution (SED) model. We constrain the detectable binary parameter space, in terms of total mass ($M_\text{tot}$), redshift ($z$), mass ratio ($q = m_2/m_1$), and orbital parameters. These results can also be used to perform initial down-selection of binary candidate lists, and understand the inherent observational bias that will exist in future SVLBI surveys of SMBHBs. We note a high level synergy between the prospective VLBI observations of SMBHBs and their gravitational-wave (GW) signal. Together, these two observational techniques create a new field of multi-messenger studies.

The binary SED and toy image model is presented in Section \ref{s:image_model}. In Section \ref{s:vis_sig}, we predict the likely visibility response of a VLBI array to a binary structure, and identify unique, detectable signatures in the signal. Section \ref{s:vlbi_detect} includes an evaluation of the detectable binary parameter space with ground and SVLBI arrays. In Section \ref{s:charac}, we develop a framework based on a Fisher analysis to evaluate how accurately a given array can characterise binary properties. Finally, in Section \ref{s:multi-mess}, we evaluate the multi-messenger context of VLBI observations of SMBHBs, with particular reference to PTAs.

\begin{figure}
\centering
\includegraphics[width=\columnwidth]{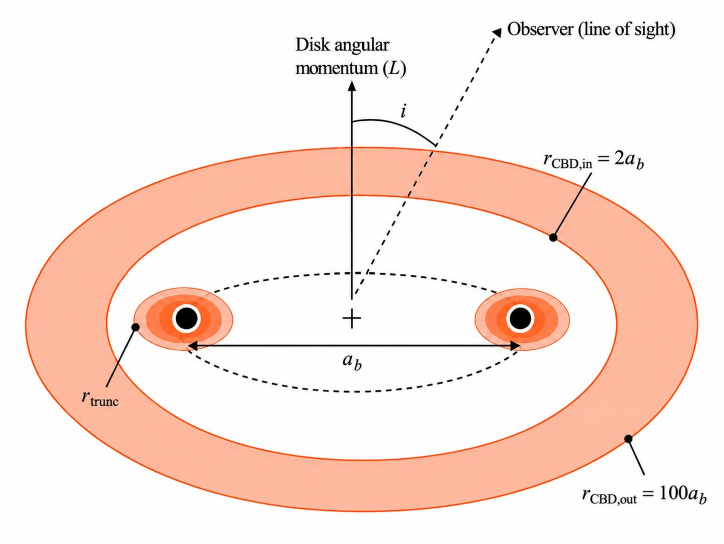}
\caption{Diagram illustrating key SMBHB model geometric properties (not to scale). Following \citet{tiede_hot_2025}, the outer CBD diameter is fixed to $100a_b$, where $a_b$ is the semi-major axis of the binary. Each minidisc extends out to the tidal trunctation radius, where $r_\mathrm{trunc} = 0.27\,q^{\pm0.3}\,a_b$ \citep{paczynski_model_1977,roedig_observational_2014}. We assume that the angular momentum of the minidiscs is aligned with the angular momentum of the CBD and that the binary orbit is circular.}
\label{f:diagram}
\end{figure}

For all calculations, we adopt a flat $\Lambda$CDM cosmology using the Planck18 realisation in
\texttt{astropy} \citepcorp{planck_collaboration_planck_2020}.

\section{Image Model} 
\label{s:image_model}

\noindent In recent years, the EHT has presented the first images of the SMBHs M87* and Sgr\,A* on an event horizon scale \citepcorp{the_event_horizon_telescope_collaboration_first_2019,event_horizon_telescope_collaboration_first_2022}. The shadow features observed are primarily formed by synchrotron emission fed by energy from the accretion disc and modified by the strong gravitational field close to the black hole. \cite{tiede_hot_2025} present a SED model, consisting of similar emission mechanisms, but in the context of SMBHBs. The binary advection-dominated accretion flow (BADAF) model is built of linear contributions from a circumbinary disc (CBD) and from individual minidiscs around each black hole. In the context of SMBHB observations with VLBI, we consider accretion regime IV \citep{tiede_hot_2025}, where all three discs are hot, generally optically thin advection-dominated accretion flows (becoming optically thick at low frequencies and/or small radii where synchrotron self-absorption begins, see Fig. \ref{f:ssa_region}). We choose this regime so that both minidiscs are bright at VLBI observing frequencies, allowing their relative positions to be tracked over time.
The BADAF model is built upon the \texttt{LLAGNSED}\footnote{\url{https://github.com/dpesce/LLAGNSED}} model of \cite{pesce_toward_2021}, used to calculate single black hole ADAF SEDs. This is a purely disc-based model and does not yet include any jet contribution. The model therefore describes emission from the inner accretion flow close to the black hole, rather than the partially self-absorbed core associated with the base of a relativistic jet, typically observed in VLBI.

\subsection{Model Parameters}
\label{ss:model_params}

\begin{figure*}
\centering
\includegraphics[width=\textwidth]{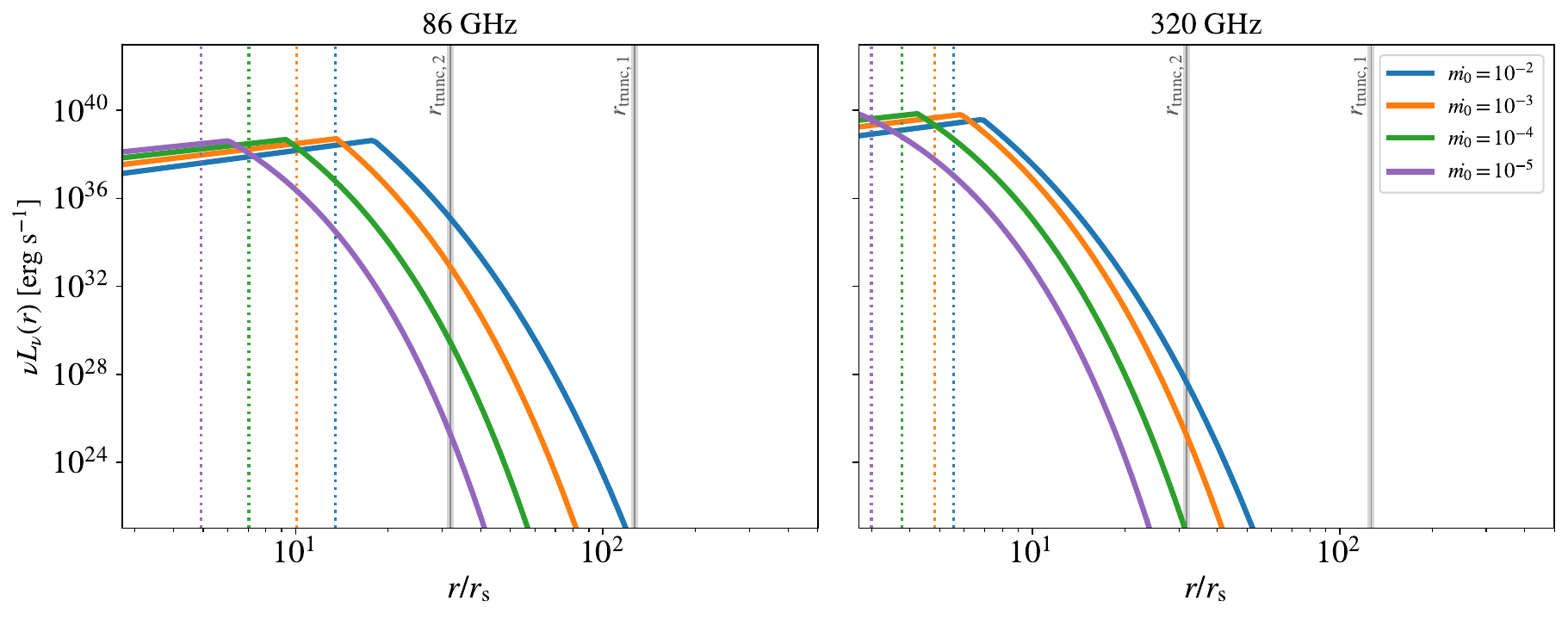}
\caption{Radial luminosity profile of either primary or secondary minidisc plotted as a function of Schwarzschild radii ($r_s$) from a single, $m=10^9 \, M_\odot$ black hole, at the specified rest-frame frequency and outer accretion rates. Rest-frame orbital period ($P_\mathrm{rest}=10$~years) and mass ratio ($q=0.1$) are only used to calculate the tidal truncation radii ($r_\mathrm{trunc}$) of each minidisc, shown in the vertical grey bands. The vertical dashed lines depict the approximate half-light radius for each accretion rate.}
\label{f:ssa_region}
\end{figure*}

\noindent The model sets an accretion rate ($\dot{m_0}$) onto the CBD at scales large compared to the binary semi-major axis, in units of Eddington rate. An accretion splitting function between the binary components is then assumed of the form, $\lambda(q)=\dot{m_2}/\dot{m_1}$, from \cite{kelley_massive_2019} using data originally generated by \cite{farris_binary_2014}. This function is used as a phenomenological model of binary accretion behaviour that emulates the preferential secondary accretion generally seen in hydrodynamic simulations \citep[c.f.,][]{ClyburnZrake:2026}. Mass loss is accounted for by a simplified correction factor of the form $(r_{s,1} / r_\mathrm{trunc,1})^{s}$, applied to $\dot{m_1}$ before calculating the SED, and equivalently for minidisc 2. Here, $r_s$ is the Schwarzschild radius and we take the fiducial value $s=0.3$. We adopt the same gas pressure to magnetic pressure ratio ($\beta$=10) as \citet{tiede_hot_2025}. Inclination ($i$) of the CBD is set to 0$\degree$ for the detectability analysis and given a separate treatment in Section \ref{ss:sensitivity}. We set $q=0.05$ as the minimum mass ratio for the parameter exploration as beyond this exists a transition to a regime where the secondary black hole and its smaller minidisc act as a perturbation within the disc of the primary, at which our approximation of a linear combination of three discs breaks down \citep{dorazio_transition_2016}. Figure \ref{f:diagram} illustrates the key geometry of the SMBHB model.

As stated above, the BADAF model does not yet include jet components, though \citet{tiede_hot_2025} demonstrate how the addition of a jet component emanating from the primary black hole produces a broadband spectrum consistent with that observed for the binary candidate PG1302$-$102. Modelling jet launching introduces a plethora of additional parameters to the binary SED, including those defining the emission characteristics, but also the viewing geometry of the system. There is also remaining uncertainty in the dynamics behind jet launching. In Section \ref{ss:jets} we briefly consider the implications of adding one or two jets to the model, but further evaluation of the impact to SMBHB detectability is left for future work.

Here, as in \citet{hudson_towards_2026}, for the binary image model we assume the superposition of two components, with circular Gaussian brightness distributions. The Gaussian assumption is adopted as a general depiction of an unresolved/partially resolved radio source. Fitting Gaussians to visibility data is a common approach in VLBI observations \citep[e.g., ][]{fomalont_sub-milliarcsecond_1999,gomez_probing_2022}. It also enables analytic evaluation of detectability. The position of these components is defined by their separation (calcuated from the system's observed orbital period and redshift), and an angle, measured east of north.

Figure~\ref{f:ssa_region} depicts an approximate radial luminosity profile generated from the BADAF model as a function of $\dot{m}_0$ and rest-frame frequency. The profile is calculated for a single accretion flow around a black hole with mass, $m=10^9\,\mathrm{M}_\odot$. The binary period ($P_\mathrm{rest}=10$~years) and mass ratio ($q=0.1$) are only used to calculate the tidal truncation radii ($r_\mathrm{trunc}$) and therefore do not impact the radial luminosity curves. At each radius, the model calculates the local synchrotron self-absorption (SSA) turnover frequency, $\nu_c(r)$, as the frequency at which the optically thin synchrotron luminosity equals the self-absorbed blackbody luminosity. For a specified rest-frame frequency, the SSA radius, $r_\mathrm{SSA}$, is then obtained by solving $\nu_c(r_\mathrm{SSA})=\nu_\mathrm{rest}$. Emission at this frequency is treated as optically thick at radii $r<r_\mathrm{SSA}$ and optically thin at radii $r>r_\mathrm{SSA}$. The dashed vertical lines show approximate half-light radii of the minidisc emission calculated from this radial luminosity model. The sharp turnover approximately marks the transition between optically thick and optically thin synchrotron emission and is associated with the radius $r_\mathrm{SSA}$ at which $\tau_\nu\simeq1$. For $\dot{m}_0=10^{-5}$, the estimated half-light radius is approximately $0.75r_\mathrm{SSA}$ at 86~GHz, with the two radii becoming similar at higher accretion rates. We therefore conservatively set the full width at half maximum (FWHM) of each Gaussian component in the image model to $2r_\mathrm{SSA}$. Because the estimated half-light radii are typically smaller than $r_\mathrm{SSA}$ across the parameter space considered here, this prescription produces broader Gaussian components than would be obtained by matching the FWHM directly to the estimated half-light radius. Broader components retain less correlated  flux on long baselines and therefore provide a conservative assumption for VLBI detectability. This choice also allows for uncertainty in the radial luminosity prescription and for departures of the true brightness distribution from an ideal Gaussian. For any systems with $r_\mathrm{trunc}<r_\mathrm{SSA}$, we set the FWHM to $r_\mathrm{trunc} = 0.27\,q^{\pm0.3}\,a_b$.

\subsection{(Sub)millimetre Emission}
\label{ss:emission}

\begin{figure*}
\centering
\includegraphics[width=\textwidth]{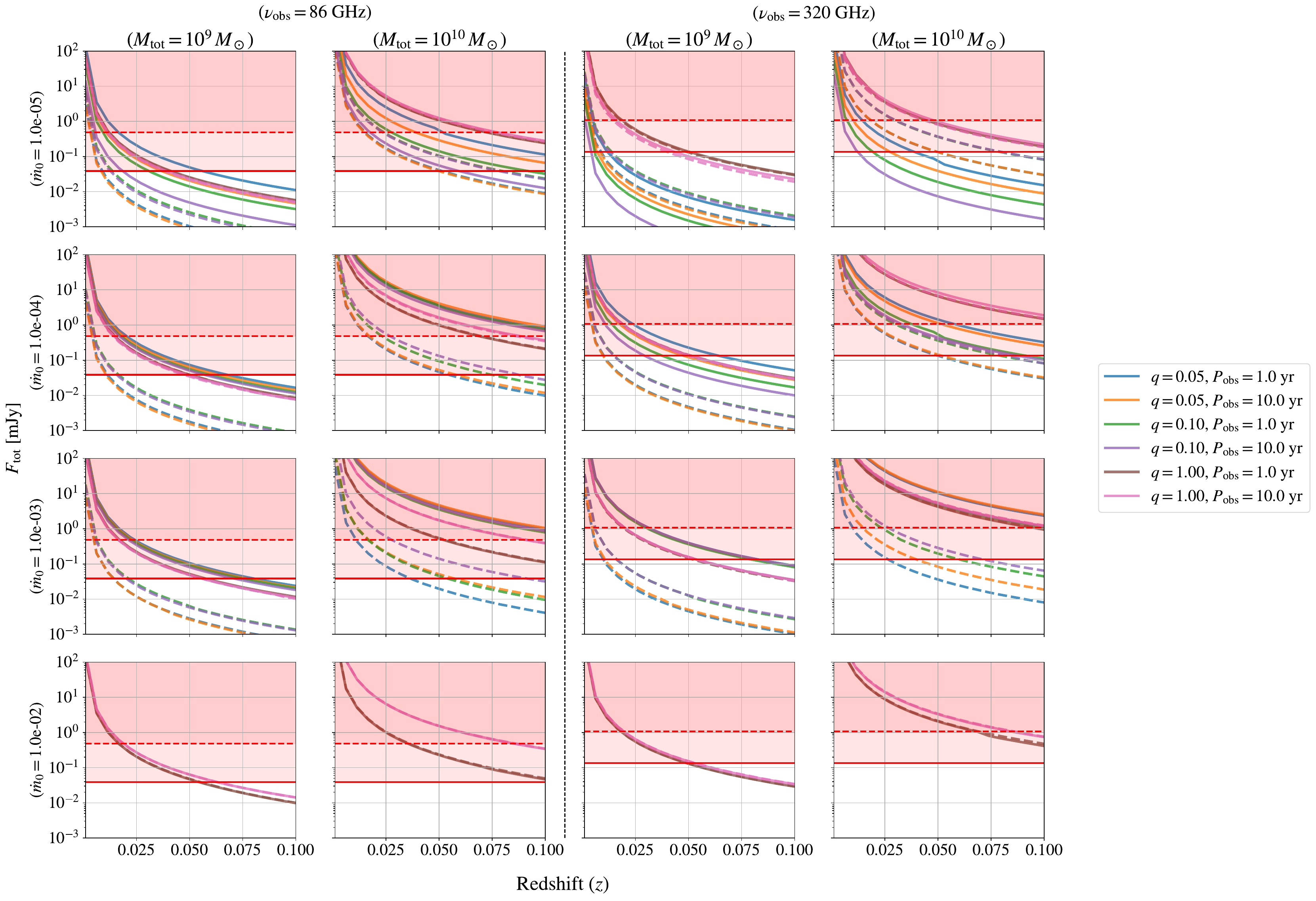}
\caption{Total flux density across redshift calculated using the BADAF model. The independent contributions of the primary (solid lines) and secondary (dashed) black holes are shown, consisting mainly of synchrotron emission. Indicative thermal noise for a ground (solid red) and space (dashed red) baseline are also illustrated.}
\label{f:total_flux}
\end{figure*}

\noindent Figure \ref{f:total_flux} depicts the total flux ($F_\mathrm{tot}$) from binary systems across redshift ($z$), total mass ($M_\textrm{tot}$), mass ratio ($q$), outer accretion rate ($\dot{m_0}$) in units of Eddington rate, and observed orbital period ($P_\text{obs}$) from Earth. For circular binaries, the dominant GW emission occurs at twice this orbital frequency. $F_\mathrm{tot}$ is the
zero-baseline flux density of the compact binary model evaluated at
the central observing frequency. Two observing frequencies ($\nu_\text{obs}$) are considered: 86 and 320~GHz, notional values within the bands supported by the BHEX receivers. 320~GHz is the highest frequency of BHEX and is therefore the frequency at which the finest angular resolution is achieved \citep{johnson_black_2024}. When comparing the binary detection performance to ground arrays such as the ngEHT, we use this value rather than the commonly considered 345~GHz. Indicative thermal noise-limited sensitivities are provided for ground and space baselines, noting that this plot shows total flux density, and not the correlated flux density on a given VLBI baseline. Thermal noise ($\sigma_{\mathrm{th}}$) is calculated as

\begin{equation}
\sigma_{\mathrm{th}} =
\frac{1}{\eta_{\mathrm{q}}}
\sqrt{
\frac{\mathrm{SEFD}_1 \, \mathrm{SEFD}_2}
{2 \, \Delta \nu \, \tau}
},
\end{equation}

\noindent where SEFD is the system equivalent flux density of the antenna, $\eta_{\mathrm{q}}$ is the quantisation efficiency, $\Delta \nu$ is the observing bandwidth and $\tau$ is integration time. Appendix \ref{A:arrays} provides the array characteristics used in these calculations. It is assumed that Frequency Phase Transfer (FPT) is used to maintain coherence across long integration times of $\sim$10~minutes at 320~GHz \citep{rioja_transformational_2023,issaoun_first_2025}. For the sensitivity on the baseline to the spaceborne radio telescope, a baseline between BHEX and ALMA is assumed (86~GHz: 0.48~mJy, 320~GHz: 1.09~mJy), and for the ground, ALMA and LMT (86~GHz: 0.04~mJy, 320~GHz: 0.14~mJy). %$\Longrightarrow 
$P_\text{obs}$ and $\nu_\text{obs}$ are converted to the rest-frame as a function of redshift, before calculation of the SED.

\begin{figure*}
\centering
\includegraphics[width=0.9\textwidth]{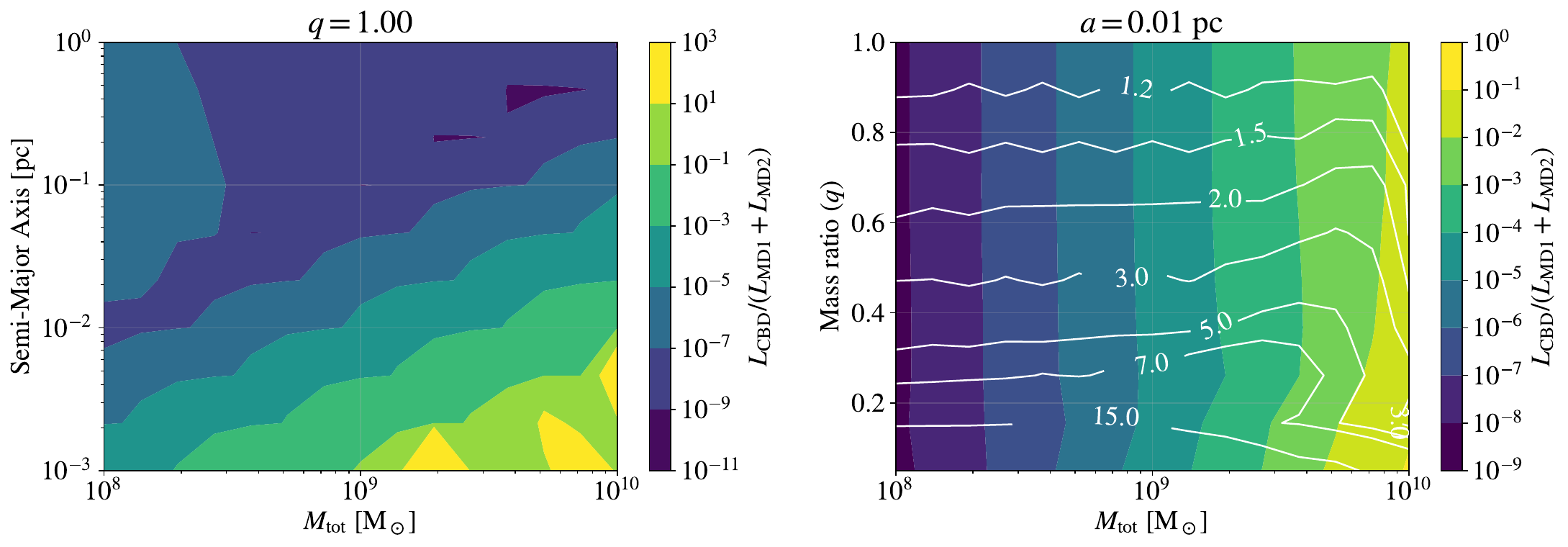}
\caption{Ratio of CBD luminosity to the total minidisc luminosity in the rest-frame, across mass ratio and separation. Rest-frame frequency ($\nu_\text{rest}$) set to 320~GHz, and outer accretion rate ($\dot{m_0}$) set to $10^{-4}$. White contours on the right panel depict the ratio $L_{MD1}/L_{MD2}$.}
\label{f:cbd_ratio}
\end{figure*}

The modelled SED is reasonably flat across the narrow (sub)millimetre band of interest for VLBI, illustrated by the small variations in total flux density between the 86 and 320~GHz columns. The primary benefit in observing at a lower frequency is the increased baseline sensitivity that can be achieved. This will be traded-off with the reduction in angular resolution in Section \ref{ss:sensitivity}. Increasing outer accretion rate has less of an impact than would perhaps be intuitively expected. For the primary and secondary components, the SED shifts upwards and right with increasing $\dot{m_0}$, as can be seen in Figures 3 and 4 in \citet{tiede_hot_2025}. In Figure \ref{f:total_flux}, for each binary system the regime of accretion is checked, and any that are not in region IV are excluded. The effect of this can be seen as $\dot{m_0}$ approaches $10^{-2}$ where lower mass ratio systems begin to enter the thin disc accretion regimes that produce little (sub)millimetre emission.

As expected, primary and secondary contributions are approximately equal for $q=1$. For lower mass ratios, a trade-off exists between preferential secondary accretion and mass. At $q=0.1$, at lower accretion rates, this trade-off balances such that the secondary's total flux density is similar to the primary's. At $q=0.05$, the lower mass of the secondary results in significantly lower luminosity. As this figure shows the total flux density, we will delay comments on detectability until Section \ref{ss:sensitivity}, where the correlated flux density on a specific baseline is calculated.

Near the critical accretion rates (where solutions start becoming thin), the CBD itself may generate potentially detectable (sub)millimetre emission. We do not plot the CBD contribution on Figure \ref{f:total_flux} for clarity, but Figure \ref{f:cbd_ratio} depicts the relationship between its luminosity and various model parameters, in the rest-frame. The CBD is also more sensitive to accretion rate and orbital period than the minidiscs, as varying $P_\text{obs}$ (and $M_\text{tot}$) will change the semi-major axis ($a_b$) and therefore the CBD geometry. For the remainder of this work, we neglect the CBD contribution, as more work is required to understand the sensitivity of this feature to the model parameters (e.g., the assumed 2$a_b$ inner and 100$a_b$ outer diameters of the disc). It is also only potentially detectable at the lowest redshifts and highest masses and accretion rates.

Over the small $P_\text{obs}$ range of our interest (see discussion in Section \ref{ss:ang_res}), the ratio of luminosities $L_\text{MD1} / L_\text{MD2}$ does not vary greatly. The inner edge of each minidisc is set at the innermost stable circular orbit (ISCO) and as such, the region of peak (sub)millimetre emission (close to the horizon) is relatively constant with the size of the minidisc. 

\section{Visibility Domain Signatures} 
\label{s:vis_sig}

\begin{table*}
\centering
\caption{Binary parameters encoded in visibility data, across baseline regimes.}
\begin{tabular}{llll}
\hline
\textbf{Visibility Feature} 
& \textbf{Expression} 
& \textbf{Binary Property} 
& \textbf{Baseline Regime} \\
\hline

Zero-baseline amplitude 
& $|V(0,0)| = F_{\mathrm{tot}}$ 
& $F_{\mathrm{tot}}$ (Total flux density)
& Short  \\

Null condition 
& $\mathbf{b}\cdot\boldsymbol{\rho} = n + \tfrac{1}{2}$ 
& $\rho$ and $\phi$ (Separation and position angle)
& All, unresolved components  \\

Modulation depth 
& $\displaystyle \frac{|V|_{\min}}{|V|_{\max}} \approx \frac{|F_1-F_2|}{F_1+F_2}$ 
& $f$ (Flux density ratio)
& All, unresolved components \\

High-baseline envelope 
& $\displaystyle \exp\!\left(-\frac{\pi^2\theta_i^2b^2}{4\ln 2}\right)$ 
& $\theta_i$ (FWHM of minidisc $i$)
& Long, partially resolved  \\

Angular anisotropy 
& $b^2\rho^2\cos^2(\gamma-\phi)$ 
& $\rho$ and $\phi$ 
& Short  \\

\hline
\label{t:nonim}
\end{tabular}
\end{table*}

\noindent A VLBI array samples the noise-corrupted visibility function of a target source by correlating the signals received at spatially separated antenna, which corresponds to the Fourier transform of the sky image. The separation between the antenna can be very large (including spaceborne elements), enabling the finest angular resolution of all astronomical techniques. With sufficient sampling in visibility vs. projected baseline space, properties of a source observed with VLBI can be estimated from the correlated visibilities. VLBI requires sufficiently compact structure in the source, such that the correlated flux density on a baseline is detectable.

\cite{fang_orbit_2022} model a binary as the superposition of two point sources. Our image model consisting of binary Gaussian components can be easily reduced to the point source model by setting $\sigma_i = 0$ in the following equation. The visibility response $V(u,v)$ of a binary Gaussian source is

\begin{equation}
\begin{aligned}
V(u,v) &= \frac{F_{\mathrm{tot}}}{1+f}\Big[
\exp\!\left(-2\pi^2 \sigma_1^2 b^2\right) \\
&+ f\,\exp\!\left(-2\pi^2 \sigma_2^2 b^2\right)
  \exp\!\left(-2\pi i\,\mathbf{b}\!\cdot\!\boldsymbol{\rho}\right)
\Big],
\end{aligned}
\label{e:vis_resp}
\end{equation}

\begin{equation}
    b^2 = u^2 + v^2,
\end{equation}

\noindent where $F_\mathrm{tot}=F_1 + F_2$, the sum of the two Gaussian components' total flux densities, $f = \frac{F_2}{F_1}$ is the flux density ratio, $\sigma$ is the standard deviation of each Gaussian component, $\mathbf{b}$ is the baseline vector in units of wavelength, $\boldsymbol{\rho} = \rho\,(\text{sin}(\phi),\, \text{cos}(\phi))$ is the angular separation vector of the two components, and $\phi$ is the position angle of $\boldsymbol{\rho}$, measured east of north. The FWHM is related to the Gaussian standard deviation by $\theta = 2 \sigma \, \sqrt{2 \text{ln}2}$. The indices 1 and 2 describe the primary and secondary black holes, respectively. This is the superposition of two Gaussian components with the primary fixed to the origin of the image reference frame \citep{thompson_interferometry_2017}. This is practically achieved during the model fitting process, such that the primary becomes the phase reference \citep{hudson_towards_2026}.

As performed in \citet{hudson_towards_2026}, modern model fitting techniques, such as those provided in \texttt{eht-imaging}\footnote{\url{https://github.com/achael/eht-imaging}} \citep{chael_interferometric_2018}, determine the combination of the binary parameters that most closely match the visibility data. However, it is useful to consider the theoretical basis of such model fitting in order to derive requirements for observations that provide strong constraints on binary features.

\begin{figure*}
\centering
\includegraphics[width=\textwidth]{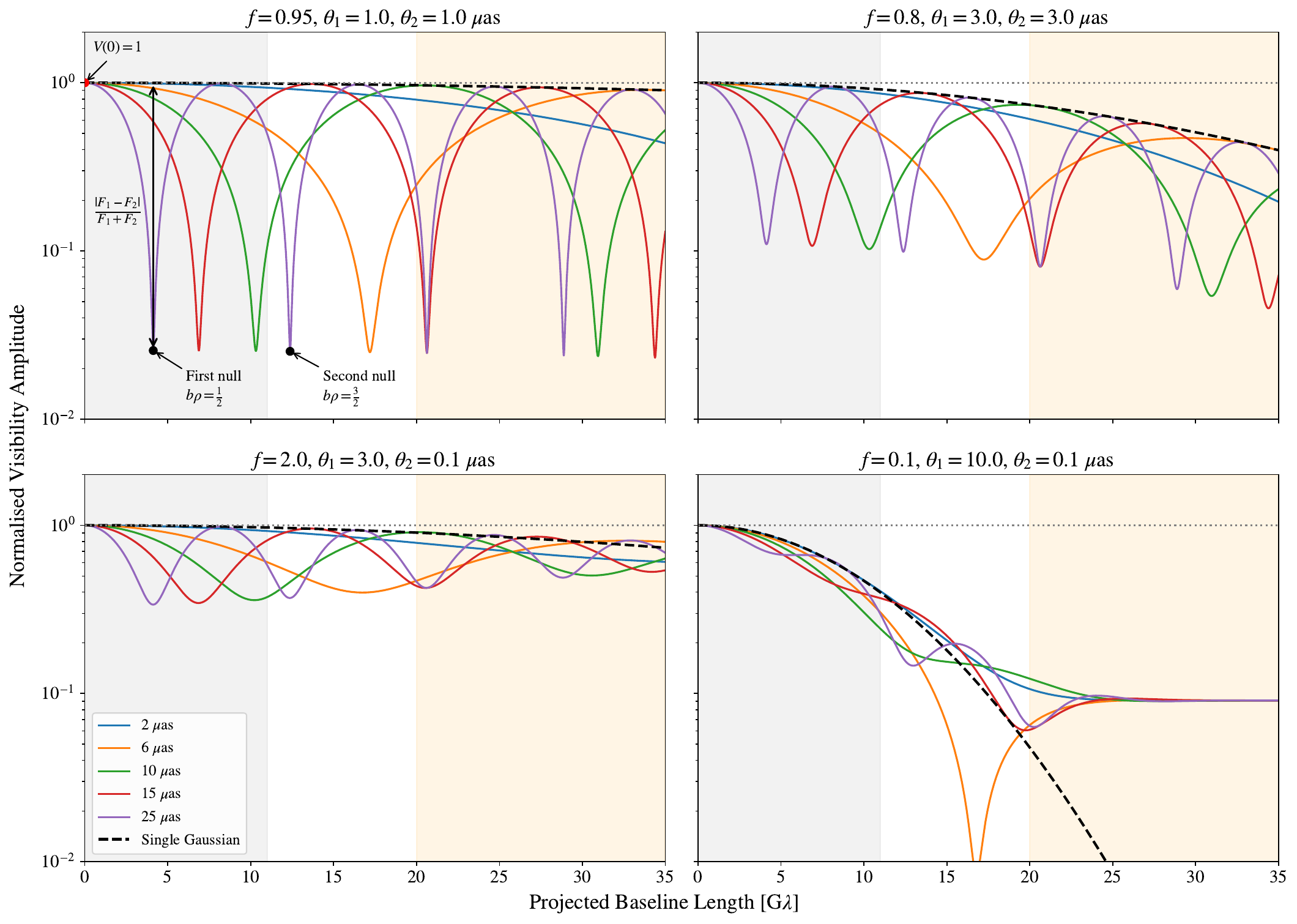}
\caption{Visibility amplitude response of binary systems with varying flux density ratio $f = \frac{F_2}{F_1}$, angular separation and FWHM ($\theta$) of each component. The grey shaded region depicts the region of projected baseline covered by the ground array. The orange shaded region shows the ground-space baseline regime for BHEX, when observing a source near-perpendicular to the orbital plane (such as M87*). The white region is the unsampled baseline lengths assuming this particular geometry.}
\label{f:bin_resp}
\end{figure*}

Figure \ref{f:bin_resp} depicts the normalised visibility amplitude response ($ |V| / F_\text{tot}$) of binary systems with varying characteristics, across the range of expected projected baseline lengths for BHEX, observing with a ground array comparable to the ngEHT. A projected space baseline length range of 20 - 35~G$\lambda$ is used, the maximum baseline occurring when the ground station is on the other side of the Earth to the spacecraft (i.e. $B = r_\text{orb} + 2R_E$, where $r_\text{orb}=20200$~km is the orbital altitude of BHEX), at 320~GHz. Ground baselines are calculated assuming a range of projected antenna separations from \(\sim\)250~km (NOEMA-PV baseline) up to the Earth's diameter. Plotted as a dashed line is the response of a single Gaussian source of the same total flux density and a FWHM equal to the average of the two components.

In general, increasing the FWHM of the Gaussian components increases the rate of drop off of the visibility amplitude with baseline length. Reducing the ratio $f$ decreases the amplitude modulation between minima and maxima. The top-left panel can be considered analogous to a near-equal mass binary. The limits of angular resolution can be seen in the lack of minima on BHEX baselines of the 2~\(\mu\)as separation system. The bottom-left panel illustrates a scenario where a smaller secondary over-accretes compared to the primary. The primary in the bottom-right panel would be in the region of event horizon-scale imaging for the ngEHT and BHEX. The inclusion of a secondary with a tenth of the angular size does however produce a noticeable impact on the response, but on baselines where the visibility  amplitude will have reduced to 10\% of $F_\text{tot}$.

For amplitude only measurements, restricted to a single baseline orientation ($\gamma$) or radially-averaged measurements, a degeneracy exists between the separation vector and angular sizes of the Gaussians, as both produce quadratic suppression of the visibility across baselines. For unresolved or partially resolved Gaussian sources, the separation can be estimated from the position of nulls through $\mathbf{b}\cdot\boldsymbol{\rho} = n + \tfrac{1}{2}$ (see Appendix \ref{A:vis_features} for derivation of this expression). However, as $f$ or $\sigma_i$ increases, the oscillations are suppressed such that clear minima may not be detectable. In this regime, without clear oscillatory behaviour, a single Gaussian, or more complicated structure, can mimic a binary. Observations across baseline orientations breaks this degeneracy as the suppression of a single Gaussian, even on short baselines, is not a function of $\gamma$ (see Appendix \ref{A:vis_features}). Table \ref{t:nonim} summarises the binary features that can be determined from visibility data, along with the baseline regions they apply to. 

SVLBI is particularly powerful at constraining such models as it has the ability to sample a wide range of spatial scales, as the baseline vector evolution is no longer constrained by the diameter of and rotational rate of the Earth. The characterisation described here can in theory be performed as a non-imaging study on a single baseline (although this wouldn't enable the use of closure phases, as discussed in the next section, or the use of FPT on baselines with stations that cannot observe in both an upper and lower band). Imaging studies require very dense \emph{(u,v)} coverage and as such, a dense ground array, which introduces significant scheduling and cost complications. For a mission such as BHEX, observations with the full ground array could be reserved for the primary science objectives (e.g., photon ring detection, imaging studies of jets), whilst preliminary, non-imaging binary investigations could be performed with a more limited array. The challenge for binary detection with BHEX is the need for there to be sufficient flux density on the space baselines to enable detection with the relatively small antenna of the spaceborne element.

\section{Detectability of SMBHB with VLBI} 
\label{s:vlbi_detect}

\noindent In \citet{hudson_towards_2026}, synthetic data simulations of BHEX observations were performed on a small number of toy binary systems to demonstrate the efficacy of the orbit fitting methodology. Using the VLBI synthetic data simulation and model fitting software packages \texttt{eht-imaging} and \texttt{ngehtsim} \footnote{\url{https://github.com/Smithsonian/ngehtsim}}, a binary Gaussian model was fit to visibility amplitudes and closure phases to determine the accuracy to with which the system could be characterised. Closure phases are measurements around triangles of three baselines that cancel out many station-based errors \citep{chael_interferometric_2018}. A chi-squared test was also used to determine whether the binary morphology was distinguishable from a single, best-fit Gaussian component. Here, we apply a similar process, but using an analytic model for visibility calculation in order to assess detectability across a wider range of binary parameter space.

\subsection{Requirements for Detection}
\label{ss:req}

\noindent The following requirements are set for detectability, inline with those used by \citet{dorazio_repeated_2018} and refined in our previous work \citep{hudson_towards_2026}:

\begin{enumerate}
    \item Both components are sufficiently bright in the (sub)millimetre to be independently detectable on ground or space baselines.
    \item The binary morphology can be confidently distinguished from that of a single, best-fit Gaussian. 
    \item $P_\text{obs}$ is short enough such that detection of orbital motion is possible over human lifetime.
\end{enumerate}

\noindent We note that morphologies other than a single Gaussian-like component could mimic a binary visibility response. For a real detection, many other scenarios will have to be rejected to build confidence that what has been observed is indeed a SMBHB. For a preliminary evaluation of binary detectability, the single Gaussian structure has been selected as the null hypothesis to be rejected because:

\begin{enumerate}
    \item For scenarios where one component is out-shone by the other (e.g., high accretion splitting fraction), a key question is whether the dimmer component can be distinguished from the primary;
    \item A single Gaussian is a simple representation of an unresolved source.
\end{enumerate}

\subsection{Angular Resolution}
\label{ss:ang_res}

\noindent The diffraction-limited, nominal angular resolution ($\theta_\text{res}$) of a VLBI array is given by $\lambda / B_\text{max}$, where $\lambda$ is the observing wavelength and $B_\text{max}$ is the maximum baseline length projected in the direction of the source. Requirement 2 defined above will set a minimum array angular resolution necessary to sufficiently resolve the binary system, such that it can be confidently distinguished from other source types.

\citet{dorazio_repeated_2018} set the requirement that the source must have an angular separation at least that of the array resolution, as a preliminary metric for detectability. In \citet{hudson_towards_2026}, we found that under specific noise assumptions, binary parameters could be  constrained on angular scales smaller than the nominal array angular resolution, a manifestation of the effect known in imaging interferometry as super resolution. Specifically, in \citet{hudson_towards_2026} we found that BHEX could confidently distinguish a binary structure (and estimate separation and position angle of the binary) from the best-fit, single Gaussian, down to separations of $\sim2\mu$as.

\begin{figure*}
\centering
\includegraphics[width=0.9\textwidth]{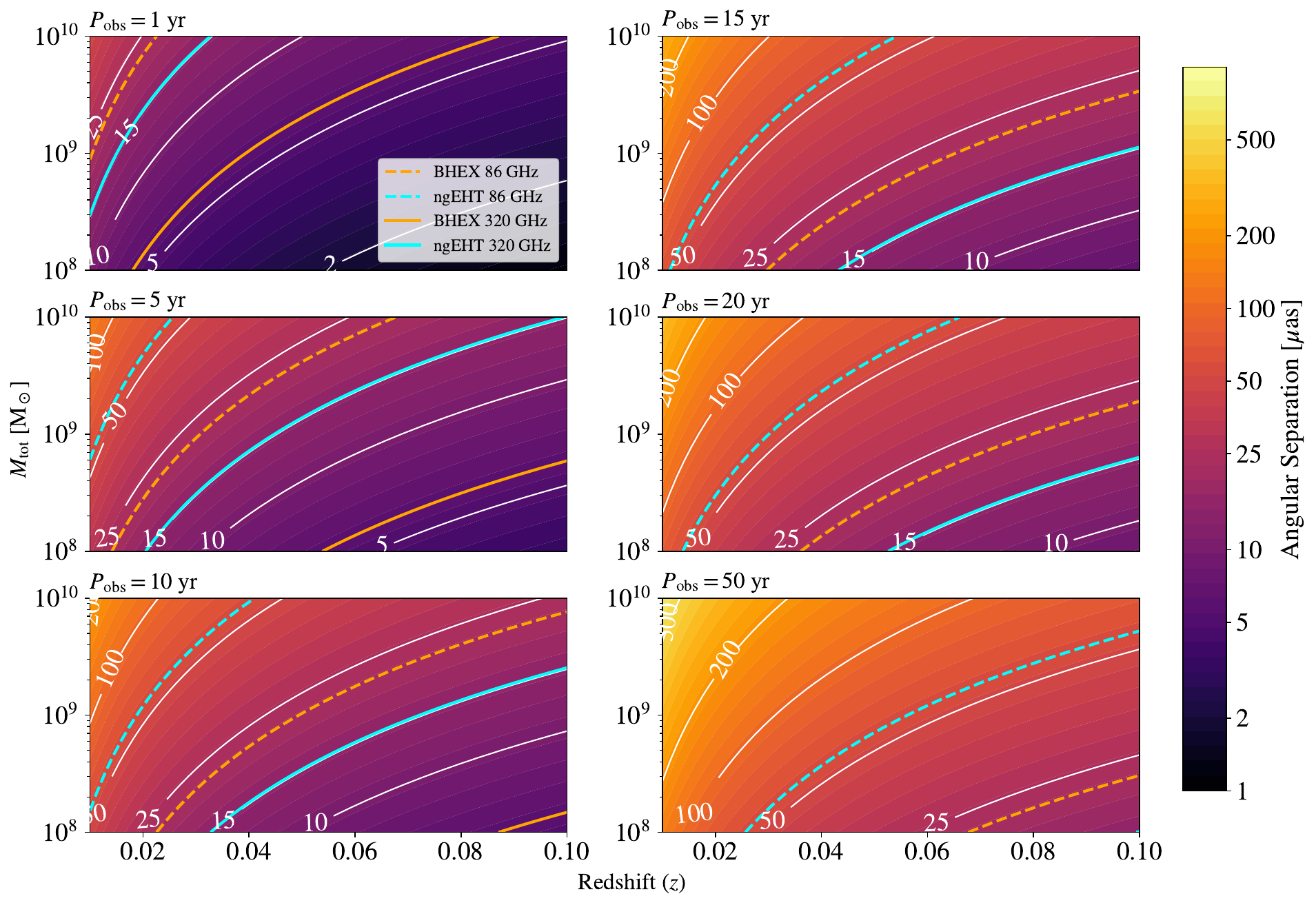}
\caption{Angular separation of binary systems as function of total mass, redshift and observed orbital period. Nominal angular resolution of BHEX and ngEHT also indicated.
\label{f:ang_sep}}
\end{figure*}

Figure \ref{f:ang_sep} depicts the angular separation of binary systems across redshift, for different $P_\text{obs}$. Also shown are the nominal array resolutions of BHEX and the ngEHT, at the observing frequencies of interest (recalling that super resolution may enable binaries with separations less than the angular resolution to still be detected). This figure can be used to associate the systems plotted on Figures \ref{f:total_flux} and \ref{f:visibility} with an angular separation, and evaluate whether the binary population is likely to be flux- rather than angular resolution-limited. Table \ref{t:example_seps} provides the angular separations for the example binary systems in Figure \ref{f:visibility} at two redshifts. As discussed in Section \ref{s:image_model}, the ratio of minidisc luminosities does not change greatly with $P_\mathrm{obs}$. We do, however, set a maximum $P_\mathrm{obs}$ of 10 years in Figures \ref{f:total_flux} and \ref{f:visibility} because, in \citet{hudson_towards_2026}, we demonstrated accurate orbit reconstruction over only a few years of BHEX and ngEHT observations up to this limiting orbital period (Requirement 3 above).

\begin{table*}
\centering
\caption{Angular separation for example systems depicted in Figures \ref{f:total_flux} and \ref{f:visibility}.}
\label{t:example_seps}
\begin{tabular}{llll}
\hline
\textbf{Total Mass}
& \textbf{Observed Period}
& \textbf{Redshift}
& \textbf{Angular Separation} \\
\hline

\multirow{4}{*}{$10^{9}\,M_{\odot}$}
& \multirow{2}{*}{$1\,{\mathrm yr}$}
& $0.025$
& $9.16\,\mu{\mathrm as}$ \\

&
& $0.100$
& $2.39\,\mu{\mathrm as}$ \\
\cline{2-4}

& \multirow{2}{*}{$10\,{\mathrm yr}$}
& $0.025$
& $42.50\,\mu{\mathrm as}$ \\

&
& $0.100$
& $11.08\,\mu{\mathrm as}$ \\
\hline

\multirow{4}{*}{$10^{10}\,M_{\odot}$}
& \multirow{2}{*}{$1\,{\mathrm yr}$}
& $0.025$
& $19.73\,\mu{\mathrm as}$ \\

&
& $0.100$
& $5.14\,\mu{\mathrm as}$ \\
\cline{2-4}

& \multirow{2}{*}{$10\,{\mathrm yr}$}
& $0.025$
& $91.56\,\mu{\mathrm as}$ \\

&
& $0.100$
& $23.86\,\mu{\mathrm as}$ \\

\hline
\end{tabular}
\end{table*}

Some of the parameter combinations shown in Figure \ref{f:ang_sep} are less likely to exist than others. For example, population estimates from PTA data suggest that it is unlikely that the most massive systems ($\sim10^{10} \, M_\odot$) at $P_\text{obs} \lesssim 6.3$~years are prevalent in the GWB \citep{agarwal_nanograv_2026}. This is discussed further in Section \ref{s:multi-mess}.

\subsection{Detectability}
\label{ss:sensitivity}

\noindent Using the BADAF SED and our image model, we can calculate the complex visibilities on a given VLBI baseline and assess the parameter space detectability against requirements 1 and 2, defined above. Visibilities consist of amplitudes and phases, the latter of which are described by the complex part (see Equation \ref{e:vis_resp}). Whilst phase and closure phase information contains additional constraints on source morphology, their practical use depends sensitively on array geometry, phase calibration, atmospheric stability, and the details of the model-fitting approach. Here we consider only the visibility amplitudes in the preliminary assessment of SMBHB detectability. This is a deliberately conservative choice as we cannot make use of the additional constraints provided by phase information.

Our metric for determining whether a given binary system is detectable is as follows:

\begin{itemize}
    \item The correlated flux density is detectable above the thermal noise with a minimum signal-noise ratio (SNR) requirement (notionally set to five).
    \item The visibility data enables a binary Gaussian to be confidently distinguished from a single Gaussian model.
\end{itemize}

\noindent The thermal SNR of each visibility measurement is defined as

\begin{equation}
    \mathrm{SNR}_{\mathrm{th},j}
    =
    \frac{|V(u,v)_j|}{\sigma_{\mathrm{th},j}},
\end{equation}

\noindent where $|V(u,v)_j|$ is the visibility amplitude and $\sigma_{{\mathrm th},j}$ is the thermal noise on baseline $j$. Only
measurements satisfying $\mathrm{SNR}_{\mathrm{th},j}>5$ are retained. We also set a requirement of having at least five usable visibility measurements before attempting to distinguish the binary source structure. Following \citet{hudson_towards_2026}, we include additional sources of uncertainty in the visibility amplitude error budget: a fractional uncertainty to account for systematic errors ($f_\mathrm{sys}$) and a fractional model-discrepancy term ($f_\mathrm{model}$). We adopt $f_\mathrm{sys}=0.1$, in line with the upper limit demonstrated by the EHT \citepcorp{event_horizon_telescope_collaboration_first_2022_II}. The model-discrepancy term accounts approximately for departures of the compact components from ideal Gaussian brightness distributions. We notionally set this to $f_\mathrm{model}=0.1$. Treating these contributions as independent uncertainties, the total uncertainty on visibility amplitude $A_j$ is

\begin{equation}
    \sigma_j^2
    =
    \sigma_{\mathrm{th},j}^2
    +
    \left(f_\mathrm{model}A_j\right)^2
    +
    \left(f_\mathrm{sys}A_j\right)^2.
\end{equation}

\noindent A single, circular Gaussian model is fitted to the visibility data (see Appendix \ref{A:bestfit} for detail on the model fitting). We calculate the chi-squared ($\chi_{\mathrm G}^2$) of the best-fitting single Gaussian model and the associated p-value ($p_\mathrm{single}$). We adopt the requirement $p_\mathrm{single}<0.01$ for detectability of binary structure, requiring rejection of the best-fit single Gaussian with greater than 99\% confidence. In general, detecting binary structure requires detecting the oscillations in the visibility response (see Figure \ref{f:bin_resp}). Therefore, increased sensitivity and sampling across the projected baseline range promotes more confident rejection of single Gaussian models.

\begin{figure*}
\centering
\includegraphics[width=\textwidth]{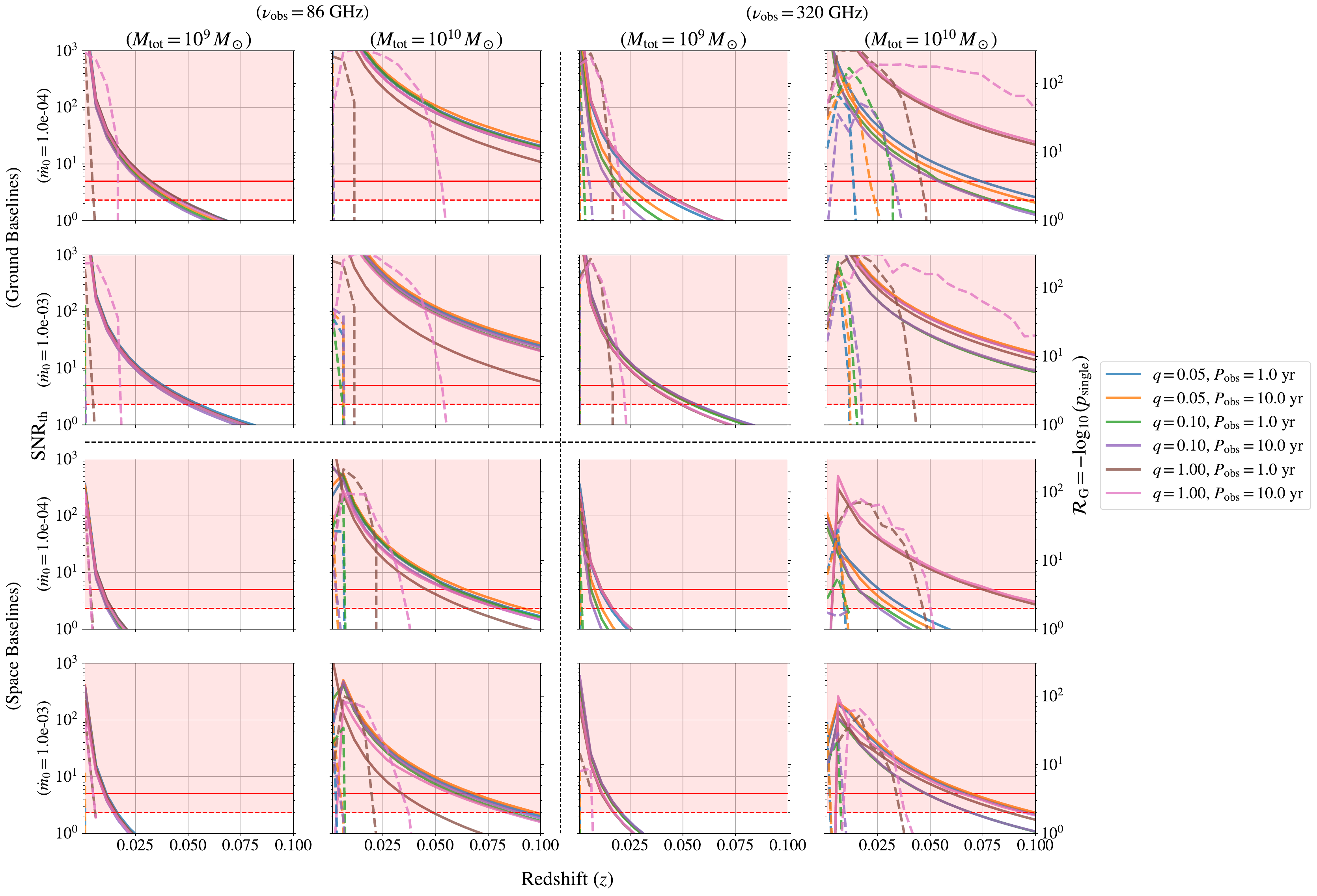}
\caption{SMBHB detectability on VLBI baselines. Solid lines/left axes: $\mathrm{SNR}_\mathrm{th}$  of the baseline with the fifth-highest correlated flux density. Dashed lines/right axes: 
$\mathcal{R}_\mathrm{G}$ representing the confidence of rejecting a best-fit, single Gaussian model. ALMA -- LMT is assumed for ground baselines, and ALMA -- BHEX for space baselines.
\label{f:visibility}}
\end{figure*}

Figure \ref{f:visibility} depicts the SMBHB detectability assessment on VLBI baselines. The shaded red region depicts the detectable parameter space according to our requirements above. Plotted in solid lines is the $\mathrm{SNR}_\mathrm{th}$ of the fifth-highest correlated flux density, corresponding to our need for five visibility measurements above $\mathrm{SNR} = 5$. The dashed lines show $p_\mathrm{single}$, plotted as $\mathcal{R}_{\mathrm G}=-\log_{10}\left(p_\mathrm{single}\right)$, with our requirement of at least 99\% confidence equivalent to $\mathcal{R}_{\mathrm G}\geq2$.

Observations on a single baseline are considered, with 12 observations-per-night (as is intended for BHEX) across four nights. Uniform coverage of the projected baseline length within the range depicted in Figure \ref{f:bin_resp} is assumed. Uniform baseline angle coverage across 180\(\degree\) is assumed for space baselines, and 45\(\degree\) for ground baselines, an approximate representation of the \emph{(u,v)} coverage limits imposed by ground synthesis. The same stations are assumed for the ground as in Section \ref{s:image_model}. The binary is fixed at a position angle of 45\(\degree\).

The correlated flux density of a binary Gaussian decreases with baseline length, as shown in Figure \ref{f:bin_resp}. The rate of this drop off depends strongly on the binary parameters, and is particularly evident for systems with high contrast between the primary and secondary components, in terms of flux density and angular size. As such, binaries of low mass ratio are perhaps unsurprisingly the hardest to distinguish from a single Gaussian model. The drop off in correlated flux density also means that in general, there is less flux density on the longest BHEX baselines and therefore the ground-space baseline has access to less of the binary parameter space than the ground baselines.

With this model, the most massive binaries ($M_\text{tot}=10^{10}$ \(M_\odot\)) are detectable and distinguishable as binary sources out to $z=0.1$ on ground baselines, for equal mass ratio systems ($q=1$) at $P_\mathrm{obs}=10$~years. The space baseline can detect the same system out to $z=0.05$. As total mass decreases to $M_\text{tot}=10^{9}$ \(M_\odot\), the detectable region reduces to $z<0.025$ for ground baselines, and $z<0.01$ for space baselines. Naturally, confirming binary structure is a more stringent requirement than detecting some correlated flux density above the thermal noise. Furthermore, Figure \ref{f:visibility} depicts the evolution of a baseline between only two antenna to show the difference between a space and ground baseline. A larger array of $N$ stations forming $N\,(N-1)\,/2$ baselines, and evolving over time, would more strongly constrain the binary nature.

A higher $\dot{m_0}$ generally makes detecting low $q$ systems easier. However, in some cases, increasing $\dot{m_0}$ reduces the detectable parameter space. This is because a higher outer accretion rate increases $r_\mathrm{SSA}$, resulting in less compact Gaussian components with less correlated flux density on long baselines. The interplay between this effect and the behaviour of the SED at different $\dot{m_0}$ explains the unintuitive response of the detectability curves to outer accretion rate.

As was found in Section \ref{s:image_model}, detecting any correlated flux density (solid lines in Figure \ref{f:visibility}) on the baseline is generally more achievable at 86~GHz due to the increased baseline sensitivity. However, 320~GHz is more favourable for distinguishing a binary source (dashed lines), due to the finer angular resolution providing access to smaller scale structure. The benefits of BHEX resolution can also be seen here. Although the space baselines have access to less of the parameter space, their ability to dinstinguish a $q=1.0$, $P_\mathrm{obs}=1$~year binary is as good as the ground baselines due to the improved angular resolution. Naturally, BHEX would become more beneficial for resolving binaries as the orbital period is decreased for the same total flux density.

At very low redshift in Figure \ref{f:visibility}, oscillation in the SNR lines can be observed where in some cases they quickly rise to a maximum and then drop off with redshift. This is an artefact of our requirement that detections above the noise floor occur on at least five baselines (and we plot the fifth highest SNR), and incomplete sampling of the projected baseline space. As the size of the Gaussian components increases as redshift decreases, there is significant drop-off in correlated flux density (see Figure \ref{f:bin_resp}), making it harder to detect the system on long baselines. This is particularly prevelant for BHEX. Furthermore, for the closest and most massive systems, the angular scale of the black hole event horizon approaches the array's nominal resolution. In this regime, particularly at higher frequencies where the emission becomes more optically thin, the assumed Gaussian brightness distribution becomes increasingly approximate. 

Correlating these results with Figure \ref{f:ang_sep}, we find the population to be flux- rather than angular resolution-limited, in agreement with \citet{dorazio_repeated_2018}. This is demonstrated by the larger detectable parameter space of the ground baselines compared to the space baselines in Figure \ref{f:visibility}. Because of this, increased sensitivity is more valuable for binary detection than finer angular resolution (within the limits of the binary parameter space explored here), making ground baselines more powerful for this activity than space baselines. Conversely, the opposite is true for characterisation of the source, as is discussed in Section \ref{s:charac}.

\subsection{Inclination Effects}
\label{ss:doppler}

\noindent For systems with small separations, the orbital velocities of the black hole components can approach relativistic speeds. \cite{dorazio_relativistic_2015} demonstrate that the light curve variability of PG1302$-$102 can be explained by relativistic Doppler boosting of the emission from an unequal-mass binary. In this section, we assess whether this phenomena will have a significant impact on the parameter space of binaries that could be detectable with VLBI.

\begin{figure*}
\centering
\includegraphics[width=0.9\textwidth]{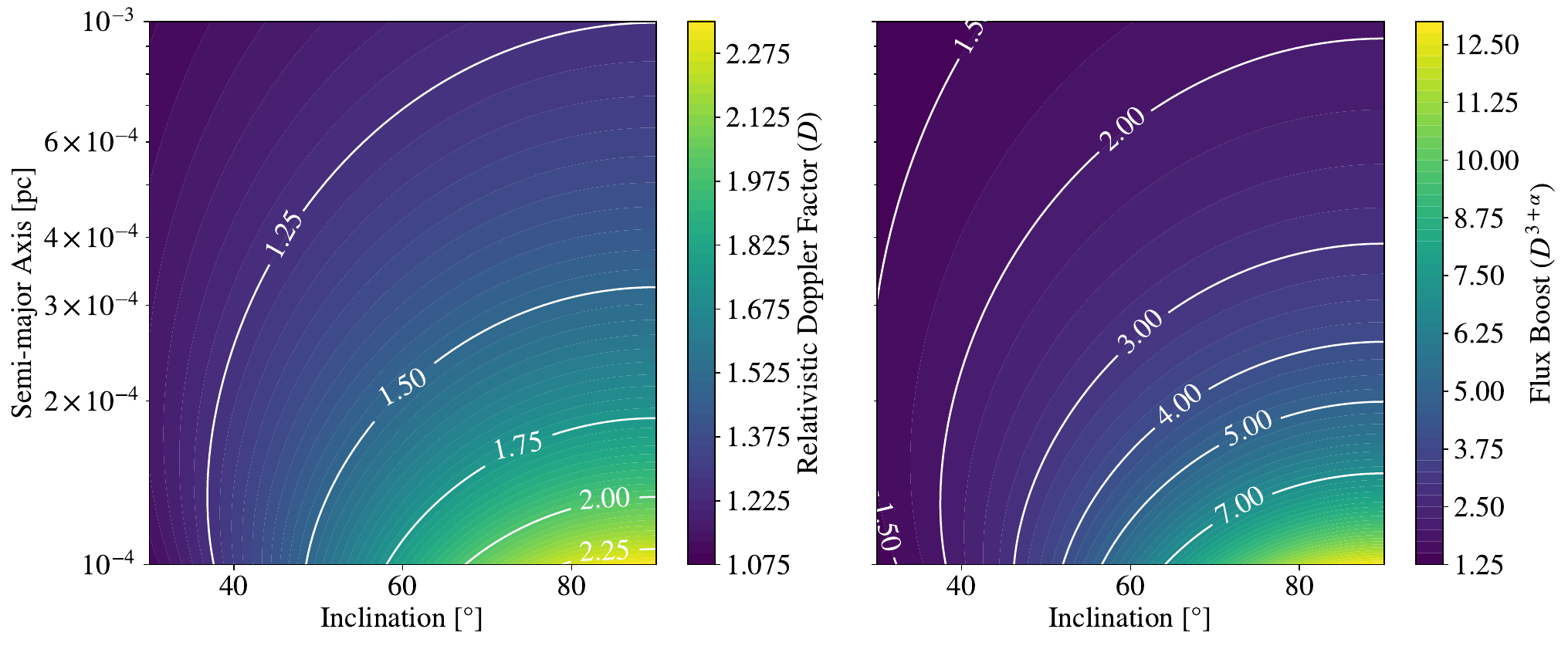}
\caption{Doppler shift factor (left) and effective flux density boost (right), assuming $\alpha = 0$, consistent with compact radio cores observed at (sub)millimetre wavelengths. Total binary mass $M_\text{tot}=10^9 \, M_\odot$.}
\label{f:doppler}
\end{figure*}

Figure \ref{f:doppler} shows the Doppler factor ($D$) and associated boost in observed flux density, for a system with $M_\text{tot}=10^9$ \(M_\odot\), for varying inclination and semi-major axis, assuming a flat radio core with a spectral index, $\alpha = 0$. For inclined orbits, at any given time one of the black holes will have a velocity component in the line-of-sight (LOS), boosting emission. As Figure \ref{f:doppler} shows, this effect approaches an order of magnitude increase in flux density for separations $\sim 10^{-4}$~pc and $i > 80\degree$. \citet{hudson_towards_2026} determined that BHEX could confidently detect binaries down to $\sim2\,\mu$as. At $z=0.001$, a $10^{-4}$~pc system has an angular separation of $4.8\,\mu$as. At $z=0.005$, this has dropped below $1.0\,\mu$as, and out of BHEX detectability. A $10^{-4}$~pc binary has an angular separation $> 2\,\mu$as out to $z=0.025$ ($\sim100$ Mpc). 

Therefore, an order of magnitude Doppler boost will only apply to BHEX-observable binaries in the very local universe, for the most massive and highly inclined systems. For binaries out to higher redshift, a factor of 2 increase in flux density (and therefore SNR) may be expected due to Doppler boosting. Although in theory this has the potential to increase detectability of one component, it of course also reduces the flux density of the other. However, it is conceivable that for a short $P_\text{obs}$, observation at one epoch may show the component moving towards the observer, whilst at a later epoch, when the true anomaly has progressed, the other component may instead be detectable. This would result in a time changing visibility structure that reflects not only the change in projected orbit on the sky, but an orbital phase-correlated change in relative brightness, as controlled by our parameter $f$ in Table \ref{t:nonim} and Fig.~\ref{f:visibility}.

The periodic modulation caused by the Doppler boost effect is a key signature in the identification of binary candidates from optical and ultraviolet surveys \citep{dorazio_relativistic_2015}. \citet{davis_reliable_2024} evaluate the prospects of the Rubin Observatory's Legacy Survey of Space and Time (LSST) for identification of SMBHBs. They determine that the orbital period of the binary can be reliably recovered roughly 70\% of the time, although a conservative estimate of the false-positive detection rate is $\sim$40\% of isolated, single quasars. Therefore, whilst such surveys hold promise for binary candidate identification and characterisation, it is likely that consideration alongside other multi-messenger signals will be required to isolate a strong and small enough selection of systems for follow-up VLBI observations (see Section \ref{s:multi-mess}).

Subtler variability caused by Doppler boost in VLBI observations could be predicted from variability in optical data of the same system, assuming the two bands trace emission close to the same black hole. Taking a standard $\alpha_\mathrm{opt}=-0.5$, the ratio of amplitude of brightness modulation is $(3-\alpha_\mathrm{opt}) / (3 - \alpha) \simeq 1.17$.  Interestingly, this method could also be applied to unresolved SMBHB systems as a means of providing additional binary candidate evidence. However, such variability is likely to be small (on the order of 10's \%) and therefore difficult to isolate from intrinsic source variability for all but the closest, most inclined systems \citep[see Figure \ref{f:doppler} and][]{kelley_gravitational_2021}.

As well as Doppler boosting, inclination will affect the projected separation of the SMBHB, as viewed from the Earth. As inclination increases, the angular separation compared to a face-on orbit will decrease when either of the black hole components is nearest the observer. For example, a system with a 10 $\mu$as face-on separation would have a minimum separation of $\sim7~\mu$as at $i=45\degree$, and $\sim1.7~\mu$as at $i=80\degree$, tending towards zero as $i\rightarrow90\degree$. This will of course affect the visibility signature of the system when observed at these positions and make it more difficult to confirm binary structure. Although Doppler boosting will increase the flux density of the approaching black hole, the reduced separation will make resolving two, distinct components more difficult if the separation is less than the array's nominal resolution. The interplay of these effects for VLBI detectability requires more exploration and we leave this for future work. However, the reader is referred to the synthetic data simulations performed by \citet{hudson_towards_2026} which included a highly-inclined test case, but without the Doppler boosting effect.

\subsection{Jet Launching}
\label{ss:jets}

\noindent The BADAF SED model does not include jet components for either the primary or secondary black holes. As such, the analysis performed to this point only considers emission close to the event horizon, as is the case for unjetted sources.

Adding jet emission could simply increase the correlated flux density of the source, largening the detectability regions shown in Figure \ref{f:visibility}. However, this depends on the angular scale of the observation, and the distance between the core and the jet base. \citet{kim_limb-brightened_2018} examined the base of the M87* jet at the scale of single Schwarzschild radii. Jets emanating from this close to the core would be visible in the image for some of the parameter space explored in this work, however much uncertainty remains around this phenomenon.

Jetted black holes may also bring into view binaries where the cores are not detectable on their own, as might be the case for OJ287 \citep{valtonen_identifying_2025}. \citet{gutierrez_non-thermal_2024} discuss the possibility of dual jet launching from sub-parsec binaries and potentially detectable signatures of a binary source.

More work is required to understand the relationship between jet launching and the black hole, particularly in the context of sub-parsec binaries, in order to explore a sensible jet parameter space in evaluation of VLBI detectability.

\section{Binary Characterisation} 
\label{s:charac}

\noindent As discussed in \citet{hudson_towards_2026} and highlighted in Requirement 3 above, a major piece of evidence in confirming a SMBHB detection is observation of orbital motion. This requires observation of the candidate source over sufficient time to first detect relative motion between the components. Once sufficient samples of the binary motion have been captured, an orbit can be reconstructed and its goodness-of-fit used to reject other possible relative motion models (e.g., linear relative motion). See \citet{hudson_towards_2026} for a more detailed treatment of this issue, and evaluation of the minimum number of observations required to discern curved orbital motion between two components.

The orbit fitting methodology developed in \citet{hudson_towards_2026} requires samples of the relative position of the secondary component with respect to the primary. The accuracy with which this can be estimated heavily impacts the accuracy and confidence with which an orbit can be fitted. In \citet{hudson_towards_2026} we performed a preliminary evaluation of the impact of measurement noise and model misspecification on the estimate of the secondary position, using synthetic data simulations with \texttt{ngehtsim} and \texttt{eht-imaging}. Here, we present a generalised, analytical approach that can be used for parameter space exploration.

The ability of VLBI observations to constrain specific binary parameters can be evaluated using a Fisher information analysis. A Fisher analysis is a statistical method used to measure the amount of information an observable random variable (i.e. visibility $V$) carries about an unknown parameter. Formally, the Fisher information matrix is the variance of the gradient of the log-likelihood function. We focus specifically on the visibility amplitudes (as discussed in Section \ref{ss:sensitivity}), but this could be extended to include phase information. We note again that using only visibility amplitudes is a conservative assumption, as the phase will also carry a binary signature that in some cases may be stronger than that in the amplitude data. In Appendix \ref{A:fisher}, we derive the following form of the discrete Fisher matrix for binary visibility amplitudes,

\begin{equation}
F_{ij}
=
\frac{1}{\sigma_{|V|}^2}
\sum_{k=1}^{N}
\frac{\partial |V_k|}{\partial p_i}
\frac{\partial |V_k|}{\partial p_j}.
\end{equation}

\noindent Where $V_k$ are measured visibilities on baseline $k$, assuming constant variance of the measurement $\sigma_{|V|}$. $p$ is the parameter set describing the binary appearance (see Appendix \ref{A:fisher}). The full Fisher matrix for separation ($\rho$) and position angle ($\phi$) is assembled in Appendix \ref{A:fisher} and the expressions for calculating the uncertainty in these parameters are also provided.

\begin{figure*}
\centering
\includegraphics[width=0.9\textwidth]{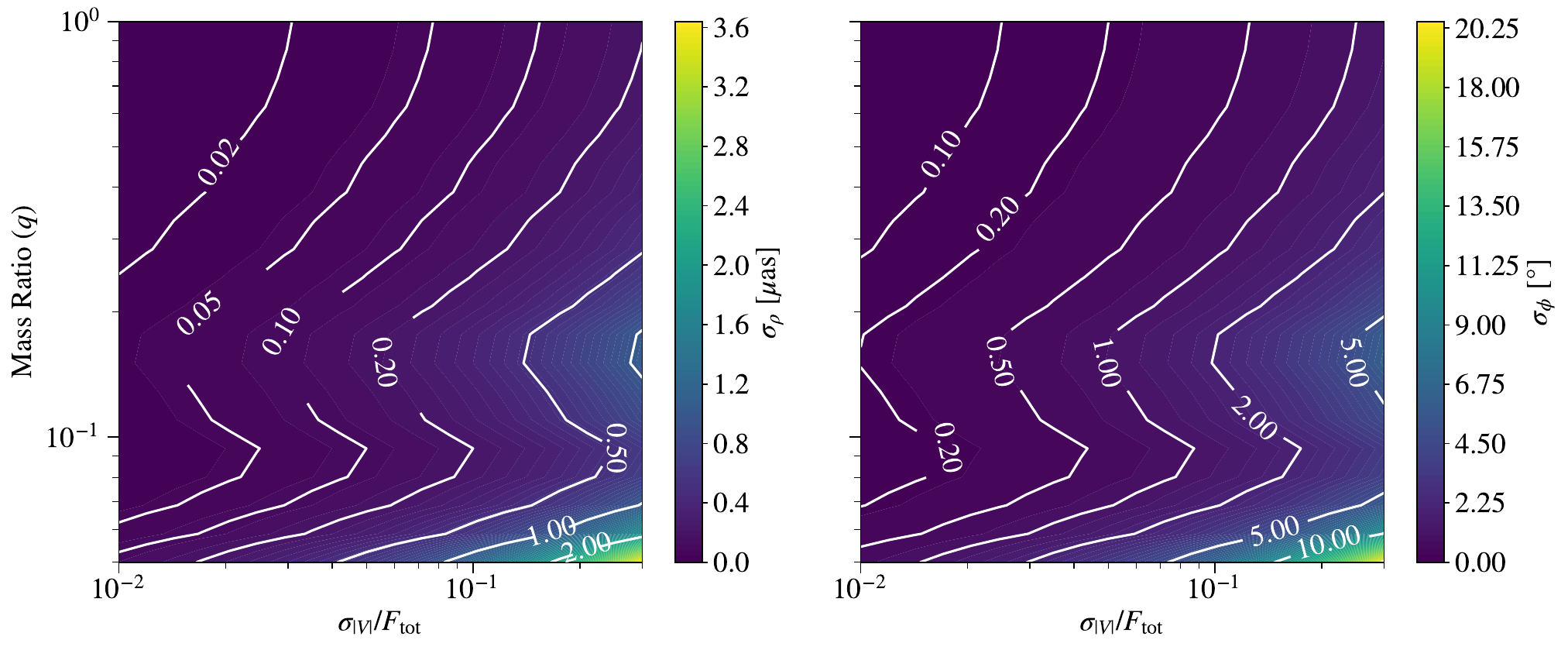}
\caption{Uncertainty (1$\sigma$) from Fisher analysis on angular separation (left) and position angle (right) of an example binary system with $\rho = 6~\mu$as, $\phi = 45\degree$, $M_\text{tot} = 10^9$ \(M_\odot\) and $z = 0.01$. Observations by BHEX and a ground array over four nights with 12 observations-per-night.}
\label{f:fisher}
\end{figure*}

Figure \ref{f:fisher} depicts these uncertainties as a function of mass ratio ($q$) and  Gaussian noise variance ($\sigma_{|V|}$). This is for an example 6~\(\mu\)as separated binary (on the edge of nominal BHEX angular resolution), with a 45\(\degree\) position angle. We assume the same baseline coverage as for Figures \ref{f:bin_resp} and \ref{f:visibility}. A larger ground array consisting of eight locations is considered for this analysis, including BHEX as the spaceborne element. We also make the simplifying assumption that $\sigma_{|V|}$ is common across all baselines, so that we can isolate the effect of improving angular resolution by adding a spaceborne element such as BHEX. The Fisher matrix is a function of flux density ratio ($f$) between the two black hole components. For Figure \ref{f:fisher}, we use the BADAF model and the accretion splitting function presented earlier to calculate $f$ for a system of $M_\text{tot} = 10^9$ \(M_\odot\), at $z = 0.01$, across varying mass ratio ($q$). The non-monotonic behaviour visible in Figure \ref{f:fisher} at $q\sim10^{-1}$ is due to the accretion splitting function which reaches its maximum at this point.

In \citet{hudson_towards_2026}, we demonstrated that semi-major axis and eccentricity could be estimated to within 0.06 dex from three annual observations of a binary with $P_\text{obs} < 10$~years. The average relative astrometric uncertainty in this analysis (from \texttt{ngehtsim} model fitting) was $\sim0.6~\mu$as. Considering the example in Figure \ref{f:fisher}, with an extensive ground array and four nights of observations, the relative position of the secondary with respect to the primary can be characterised to within this uncertainty for $q > 0.05$, depending on the specific noise conditions. This approach allows for easier exploration of the whole binary parameter space, without the need to run expensive synthetic data simulations.

As described in \citet{hudson_towards_2026}, the accuracy with which the relative position between two sources within the same FOV can be constrained is dependent on the number of individual baseline measurements. Figure \ref{f:fisher} is representative of the performance achievable by a BHEX observation with a full ground array. On a single ground baseline (taking the LMT - ALMA baseline assumed earlier), the $\rho$ and $\phi$ uncertainty worsens by approximately a factor of \(\sim\)6 and \(\sim\)10, respectively, to that in Figure \ref{f:fisher}. Adding BHEX to this single baseline, reduces the uncertainties by a factor of \(\sim\)4, in $\rho$ and $\phi$. This shows the benefit of improved angular resolution on the ability of the array to constrain key binary parameters.

We note that the uncertainties derived from the Fisher analysis provide an idealised, lower limit on the ability of a VLBI array to characterise binary systems. We can compare these uncertainties to those from synthetic data simulations to assess how representative the Fisher analysis is of real observations. Table 1 of \citet{hudson_towards_2026} provides $1\sigma$ uncertaines on the separation estimates from synthetic data simulations with \texttt{ngehtsim}. For the 10~$\mu$as separation, 10\% + 5~mJy noise case, \citet{hudson_towards_2026} finds $\sigma_{\rho}=0.13\,\mu$as. Although the treatement of noise is not identical between these two approaches, for an equivalent binary system, we estimate $\sigma_{\rho}=0.115\,\mu$as from the Fisher analysis\footnote{We convert the 10\% + 5~mJy noise case from \citet{hudson_towards_2026}  to $\sigma_{|V|}/F_\mathrm{tot} = 0.24$ for comparison with our Fisher analysis method. This assumes an average sensitivity across the array of 5~mJy for an $F_\mathrm{tot}=50$~mJy source, combined in RMS with the 10\% fractional noise, with 5~mJy (i.e. 10\% of 50~mJy) flat noise added to this.}.

When considering how to include observations of binary candidates in the BHEX schedule, single baseline observations could be used for initial down-selection of binary candidates, with follow-up observations with a more complete ground array to sufficiently constrain separation and position angle to the level needed to perform orbit fitting. 

\section{SMBHB imaging in the multi-messenger context}
\label{s:multi-mess}

\noindent
The subset of SMBHBs studied here, which are amenable to VLBI imaging, has a strong overlap with the population inferred from recent PTA measurements of a GWB \citep{NANO_Astro_Interp+2023}.
Using the population synthesis model of \cite{chen_constraining_2019}, \cite{agarwal_nanograv_2026} decompose the GWB into expected numbers of SMBHBs as a function of $M_\text{tot}$ and $P_\text{obs}$ (see Figure 10 with conversion from GW frequency to $P_\text{obs}$ performed assuming circular orbits). The population model suggests that systems of $9.5 < \text{log}_{10}(M_\text{tot}/M_\odot) < 10$ at $P_\text{obs} \lesssim 6.3$~years are not supported by the data. Therefore, the $P_\text{obs} = 1$~year systems in Figures \ref{f:total_flux} and \ref{f:visibility} at $M_\text{tot} = 10^{10}\,M_\odot$ may not exist within the observable redshift range. However, for $8.5 < \text{log}_{10}(M_\text{tot}/M_\odot) < 9.5$, the model suggests the existence of $\gtrsim10^4$ sources at $P_\text{obs} = 10$~years, and $\sim10^2$ sources at $P_\text{obs} = 1$~year.

The predicted total flux densities of SMBHB sources using the BADAF model do not change greatly with angular separation (see Figure \ref{f:total_flux}). Therefore, the detectability evaluation performed above can be broadly extrapolated to other $P_\text{obs}$, up to a few 10's of years. The wider separations at a given redshift will impact the visibility response (and therefore the dashed lines of Figure \ref{f:visibility}, in general making binary confirmation easier), but Figure \ref{f:ang_sep} can be used to determine the angular separation of a given system and relate it to the parameter combinations depicted in Figure \ref{f:visibility}. For example, a $10^9\,M_\odot$ system with $P_\mathrm{obs}=20$~years at $z=0.1$ has an equivalent angular separation to a $P_\mathrm{obs}=10$~years at $z=0.06$. This system can then be found on Figure \ref{f:visibility} at this redshift to assess likely detectability.

Beyond the GWB, detection of a resolved SMBHB with PTAs (a single system above the noise) is expected in the next few to 10's of years. Population models conditioned on the GWB estimate that there is a 51\% chance of finding such a source in the next 10 years, and a 14\% chance that it is localised to within a few hundred square degrees \citep{Veronsi+2026}. Realisations find such sources in the range $z=$ 0.01--1, with the majority at $z<0.2$ (encompassing our detectable redshift). Masses are predicted to be $9 \lesssim  \text{log}_{10}(M_\mathrm{tot}/M_{\odot}) \lesssim 10.5$, with $P_\mathrm{obs}$ ranging from 5--30 years (2--12 nHz).

With the definition of ``well localised" still relatively poor (a few hundred square degrees, containing 100,000's of galaxies), a method for identifying a small number of candidate systems in electromagnetic observations is still required before VLBI observations. However, with the high chance for false-positives (see discussion on periodic variability signatures in Section \ref{ss:doppler}) in current methods for binary signature identification, a degree of luck, or improved models of binary signatures and their identification methods, will be required in achieving the first indisputable multi-messenger detection of a SMBHB in the near future.

\section{Discussion}
\label{s:discuss}

\noindent Using an ADAF supermassive black hole binary SED model, we have shown that there could exist systems which are directly observable with ground and SVLBI out to $z\sim0.1$. We have used BHEX as an example of future SVLBI, and a ground array representative of the near-future performance of the ngEHT. We find that ground-ground baselines provide access to the largest volume of binary parameter space, due to their enhanced sensitivity. Lower frequency observations (e.g., 86~GHz), also increase the detectable parameter space for the same reason. Conversely, the fine angular resolution provided by spaceborne systems such as BHEX offers significant benefits for source characterisation.

VLBI detectability is heavily dependent on the total mass and mass ratio of the SMBHB. Even on ground baselines, systems with $q=0.05$ are only detectable and confidently distinguishable as a binary structure out to \(\sim\)0.025 redshift. On ground-space baselines, $q=1.0$ systems are detectable out to $z=0.1$ for $M_\text{tot}=10^{10}$ \(M_\odot\), but not confidently distinguishable from a single Gaussian model. This requirement limits the detectable redshift range to $z=0.05$, reducing to $z\sim0.01$ for $M_\text{tot}=10^{9}$ \(M_\odot\). Doppler boosting exhibited by inclined systems may increase the detectable parameter space for some systems (the flux density boost approaches an order of magnitude increase for separations $\sim10^{-4}~$pc and $i>80\degree$), but the same effect will also reduce the brightness of the receding black hole. However, it is possible that for observations over long epochs, each component may be detectable at different times, as a function of the phase of the binary orbit. Of course, the challenge will then be model fitting to time-dependent visibilities, but with the constraint that the relative brightness should follow the orbital phase. We explored the parameter space down to a minimum mass ratio $q=0.05$, beyond which the accretion flow experiences a transition away from our assumed disc model. This low $q$ case may present other detectable signatures but we defer their exploration to future work when models of such systems exist.

As in \citet{dorazio_repeated_2018}, we find the SMBHB population to be flux- rather than angular resolution-limited. Although this reduces the benefit of a spaceborne system such as BHEX (offering reduced sensitivity compared to ground baselines), the finer resolution may still enable observation of some systems unresolvable by the ground. Also, the inclusion of a spaceborne element enables significantly better estimates of binary parameters to be extracted from the visibility data, which is important for orbit reconstruction, a key piece of evidence in confirming that was has been observed is indeed a SMBHB \citep{hudson_towards_2026}. In the example provided, using a Fisher analysis, the accuracy of angular separation and position angle estimates were found to improve by a factor of \(\sim\)4 with a spaceborne element, compared to ground-only observations.

The BADAF SED model does not include any jet components. As discussed briefly in Section \ref{ss:jets}, the inclusion of jets may largen the detectable binary parameter space by simply increasing the flux density of the sources. Alternatively, on larger angular scales, jet components may make binaries observable where the cores would not be detectable on their own, as may be the case for OJ287 \citep{valtonen_identifying_2025}.

A number of assumptions are inherent in the BADAF SED model that should be noted. A specific accretion splitting function is assumed, as are the values for the mass-loss correction factor slope ($s$) and the gas to magnetic pressure ratio ($\beta$). Although \citet{tiede_hot_2025} find that the model is not particularly sensitive to variation in some model parameters (see their Appendix A), the emission characteristics will of course be impacted by the other assumptions. As a preliminary evaluation of parameter sensitivity, in Appendix \ref{A:sens} we have tested the effect of different $s$ and $\beta$ values on the detectable binary parameter space. Beyond SED model parameter variations, increasing the model misspecification error ($f_\mathrm{model}$) to 20\% results in only a slight reduction in the maximum redshift at which the binary is distinguishable from a single Gaussian. It does however reduce the value of $\mathcal{R}$ at $z$ below this maximum, by just under an order of magnitude. See \citet{tiede_hot_2025} for more discussion on the BADAF SED model, and recommendations for next steps. In general, more work is required using full hydrodynamical simulations to assess the likely (sub)millimetre radio emission from SMBHB.

This work deals with a theoretical population of SMBHBs bright in the (sub)millimetre. However, as highlighted by \citet{hudson_towards_2026} and \citet{dorazio_repeated_2018}, the challenge is in identifying candidate sources for future observation. VLBI is not suited to survey-like observations due to the very small field of view, and the computation required to perform correlation for even a small fraction of this. Therefore, VLBI requires specific targets well localised on the sky to perform observations.

Multiple methods have been proposed for the identification of candidate SMBHB systems, including periodic modulation in optical and ultraviolet surveys discussed above \citep{davis_reliable_2024,dorazio_obs_sig_2023}. Monitoring of radio loud AGN variability carried out by the Owens Valley Radio Observatory (OVRO) is also particularly relevant given its observations at GHz frequencies \citep{molina_search_2026,readhead_compelling_2026}. However, intrinsic variability of blazar flares still accounts for the majority of the observed periodicity, with only one candidate (PKS 1309+1154) having a significance of $>2\sigma$. The Event Horizon and Environs (ETHER) database is building a library of SMBHB candidates identified using these methods, with ongoing observations with ground-based VLBI \citep{ramakrishnan_event_2023}. It is envisaged that the approach used in this paper could be applied to the down selection of ETHER's large candidate list to help with identifying promising systems.

A strategy for SMBHB candidate observation could consist of the following:
\begin{enumerate}[itemsep=0pt, topsep=5pt]
    \item Candidate identification: Broad identification of a sky region containing SMBHB from PTA data, with subsequent down-selection using existing methods for identifying binary signatures in electromagnetic observations (e.g., ETHER database). Further down-selection by predicting detectability with the SED model approach used in this work.
    \item Ground radio campaign: Confirmation of detectable flux densities at GHz frequencies of interest (e.g., $\sim$86, $\sim$230 and $\sim$345~GHz for the EHT, ngEHT and BHEX).
    \item Ground VLBI campaign: For those sources with confirmed, detectable (sub)millimetre flux densities, observe with a ground VLBI array.
    \item SVLBI campaign (e.g., BHEX): a) In the best case scenario, if the ground VLBI observations show two compact sources and/or jets, follow-up with BHEX observations to tightly constrain relative positions for subsequent orbit fitting; or b) If the source is still unresolved at the scale of likely binary separation, follow-up with BHEX observations at $>$2x finer resolution than the ground. Due to the competition for space instrument time, it is likely this would only be possible for candidates with very strong additional evidence supporting the binary claim.
    \item Multi-epoch monitoring: In the case of finding a candidate with two distinct sources, observe with ground and/or SVLBI regularly with a cadence of $\sim$6 months/1 year (depending on expected $P_\text{obs}$), to enable orbit fitting as demonstrated by \citet{hudson_towards_2026}.
\end{enumerate}

\section{Conclusions}
\label{s:concl}

\noindent Observation of a sub-parsec supermassive black hole binary (SMBHB) and detection of its orbital motion would be the first direct evidence of such systems existing. Very Long Baseline Interferometry (VLBI) is the only astronomical method capable of meeting this objective. Beyond conclusive proof of their existence, characterisation of a population of sub-parsec SMBHBs would enable evolutionary models across the binary life cycle to be constrained. Combining such measurements with gravitational wave (GW) observations, such as those performed by pulsar timing arrays (PTAs), would broaden our understanding of galactic evolution on a cosmological scale through multi-messenger astronomy.

We have developed an approach for assessing the detectability of SMBHBs with VLBI, and evaluated the prospects of current and near-future arrays. Using an unjetted, binary SED model, we have shown that SMBHB systems could be detectable out to $z\sim0.1$ with ground and SVLBI, under a range of accretion, total mass, mass ratio and orbital conditions. SVLBI, such as that proposed for the Black Hole Explorer (BHEX), is particularly beneficial for characterising the properties of such sources, even though it is unlikely to significantly increase the size of the detectable population. 

SMBHB observation is not a defined science objective of BHEX and as such, the space mission has not been designed with this activity in mind. However, strong binary candidates (e.g., OJ287) are already included in the BHEX target list. Beyond this, SMBHB observation may be an intriguing tertiary activity for the mission, if promising candidate sources can be identified before launch.

Looking to the future, a SVLBI mission specifically designed with SMBHB observation as a primary objective (with sensitivity comparable to that offered by ground arrays) would be a system with unprecedented capabilities, able to tackle some of the most pressing questions in astronomy. 

%% Please use the acknowledgment and contribution environments. This will 
%% be anonomyized when the "anonymous" style option is used. 
\begin{acknowledgments}
\noindent The authors would like to acknowledge the entirety of the BHEX community for their ongoing efforts to realise this exciting mission. BH would like to thank his employer, KISPE Space, for providing him with the flexibility and support to study for a PhD alongside his full-time role. DJD acknowledges support from NSF AAG No. 2511544. The authors are also grateful to the anonymous referee for their constructive comments.
\end{acknowledgments}

%% Following the acknowledgments Section, use the following syntax and the
% \facilities{HST(STIS), Swift(XRT and UVOT), AAVSO, CTIO:1.3m, CTIO:1.5m, CXO}

\software{Astropy \citepcorp{the_astropy_collaboration_astropy_2013,the_astropy_collaboration_astropy_2018,the_astropy_collaboration_astropy_2022}}, BADAF \citep{tiede_hot_2025}, LLAGNSED \citep{pesce_toward_2021}


\begin{appendix}

\section{VLBI Array Properties}
\label{A:arrays}

\noindent Here we define the properties of the observing sites used in the above analysis to evaluate VLBI detectability. SEFD is a key metric in describing the performance of a radio telescope, defined as the strength of a hypothetical incoming signal that would double the receiver's system temperature. The SEFD of a radio antenna is given by

\begin{equation}
\mathrm{SEFD} =
(2 k_{\mathrm{B}} T_{\mathrm{sys}}) / (\eta_{\mathrm{a}} A_{\mathrm{geom}}).
\end{equation}

\noindent Where $T_{\mathrm{sys}}$ is the system temperature, $\eta_{\mathrm{a}}$ is the antenna efficiency, and $A_{\mathrm{geom}}$ is the effective receiving area of the antenna. Whilst the latter are geometric properties or a consequence of the specific antenna design used, $T_{\mathrm{sys}}$ is heavily impacted by weather (for ground sites), environmental noise, and the receiver properties.

At the time of writing, BHEX will have a 3.4~m monolithic antenna, with a surface accuracy of \(\sim\)40~\(\mu\)m, observing across two frequency bands, with a bandwidth of 8~GHz, dual-polarisation. The remaining properties defining the SEFD of the instrument are as specified in \citet{hudson_towards_2026}, and the reader is referred to \citet{johnson_black_2024} for more information.

As in \citet{hudson_towards_2026}, we calculate the SEFDs of BHEX (low-band: 19200 Jy, high-band: 39800 Jy), ALMA (low-band: 65 Jy, high-band: 160 Jy) and LMT (low-band: 170 Jy, high-band: 865 Jy) using \texttt{ngehtsim}. We take the average SEFD across a simulated night of observations, assuming median weather conditions.

\section{Visibility Domain Features}
\label{A:vis_features}
\noindent In this section we provide the derivations of the equations in Table \ref{t:nonim}. This is standard interferometric theory, but rarely presented in the context of binary Gaussian sources, specifically. From Eq. \ref{e:vis_resp}, at zero baseline $\boldsymbol{b}=0$,

\begin{equation}
|V(0,0)| = F_\mathrm{tot}.
\end{equation}

\noindent The phase term in the binary Gaussian visibility is given by $\exp(-2\pi i \mathbf{b}\cdot\boldsymbol{\rho})$. We can write the dot product explicitly as ($\mathbf{b}\cdot\boldsymbol{\rho}
=
b\rho \cos(\gamma-\phi)$), where $\gamma$ is the baseline orientation angle and $\phi$ is the binary position angle. Considering the case where a baseline is aligned with the separation vector ($\gamma=\phi$), constructive interference occurs when $2\pi b\rho = 2\pi n$, giving $b = \frac{n}{\rho}$. Therefore, the spacing between successive maxima is

\begin{equation}
\Delta b = \frac{1}{\rho}.
\end{equation}

\noindent Destructive interference occurs when $2\pi b\rho = (2n+1)\pi$. More generally when, 

\begin{equation}
\mathbf{b}\cdot\boldsymbol{\rho} = n + \tfrac{1}{2}.
\end{equation}

\noindent This links baseline vector to binary separation vector, and enables the angular separation and position angle of the source to be estimated from samples of successive nulls ($n$). Taking the square of the amplitude in the regime where both Gaussian components remain unresolved,

\begin{equation}
|V|^2 = F_1^2 + F_2^2 + 2F_1F_2\cos(2\pi \mathbf{b}\cdot\boldsymbol{\rho}),
\end{equation}

\noindent the maximum occurs when $\cos = +1$: $|V|_{\max} = F_1 + F_2$. The minimum occurs when $\cos = -1$: $|V|_{\min} = |F_1 - F_2|$. Therefore,

\begin{equation}
\frac{|V|_{\min}}{|V|_{\max}} = \frac{|F_1 - F_2|}{F_1 + F_2}.
\end{equation}

\noindent As the two Gaussians become resolved, each contributes a damping of visibility amplitude with baseline length, suppressing the contrast between maxima and minima. In such cases, the separation vector and Gaussian sizes must be fitted simultaneously, with model fitting techniques.

In the short baseline regime, we can substitute the dot product into the phase term and perform a Taylor expansion, under the assumption of small $b$, such that,

\begin{equation} 
\cos(2\pi b\rho\cos(\gamma-\phi)) \approx 1 - 2\pi^2 b^2 \rho^2 \cos^2(\gamma-\phi),
\end{equation} 

\noindent The binary separation introduces an anisotropic quadratic
curvature whose amplitude depends on baseline orientation angle. In
contrast, the short baseline suppression of a circular Gaussian depends only on $b^2$ and FWHM. Therefore,
with sufficient baseline orientation coverage, intrinsic circular
source size and binary separation can remain distinguishable even
without sampling multiple visibility minima and maxima.

\section{Gaussian Best-Fit Approach}
\label{A:bestfit}

\noindent The following single, circular Gaussian model is used in our assessment of binary detectability, defined by the total flux density ($F_{\mathrm G}$) and FWHM ($\theta_{\mathrm G}$),

\begin{equation}
    |V_{\mathrm{G},j}| = F_{\mathrm G}\exp\left[-\frac{\pi^2\theta_{\mathrm G}^2b_j^2}{4\ln 2}\right].
\end{equation}

\noindent The best-fitting
parameters are obtained by minimising

\begin{equation}
    \chi_\mathrm{G}^2 = \sum_{j=1}^{N_{\mathrm{use}}} \, \frac{        \left[|V_{j}|-|V_{\mathrm{G},j}|\left(
    \widehat{F}_{\mathrm G},\widehat{\theta}_{\mathrm G}\right)\right]^2}{\sigma_j^2},
\end{equation}

\noindent where $j$ is the baseline, $N_\mathrm{use}$ is the number of useable baselines' visibility data above the thermal SNR requirement and $\sigma_j$ is the total uncertainty on visibility amplitude.

As the single Gaussian model has two fitted parameters, the number of
degrees of freedom is $\nu=N_\mathrm{use}-2$. The probability of obtaining a value at least as large as the measured
$\chi_\mathrm{G}^2$ under the single Gaussian hypothesis is
\begin{equation}
    p_\mathrm{single}=P\left(\chi_{\nu}^2\geq\chi_{\mathrm G}^2\right)
    =1-F_{\chi_{\nu}^2}\left(\chi_{\mathrm G}^2\right),
\end{equation}

\noindent where $F_{\chi_{\nu}^2}$ is the cumulative distribution function of the
$\chi^2$ distribution, with $\nu$ degrees of freedom.

\section{Fisher Information Analysis}
\label{A:fisher}
\noindent The complex visibility of a two-component circular Gaussian binary can be written as

\begin{equation}
V(b,\gamma)=A_1(b)+A_2(b)\exp[-i\psi(b,\gamma)].
\end{equation}

\noindent $A_1$ and $A_2$ are the baseline-dependent attenuated flux densities, defined as

\begin{equation}
A_1(b)=F_1\exp\!\left(-2\pi^2\sigma_1^2b^2\right),
\end{equation}

\begin{equation}
A_2(b)=F_2\exp\!\left(-2\pi^2\sigma_2^2b^2\right),
\end{equation}

\noindent and the relative binary phase is

\begin{equation}
\psi(b,\gamma)=2\pi b\rho\cos(\gamma-\phi).
\end{equation}

\noindent In an amplitude-only treatment, the observable is the visibility amplitude,

\begin{equation}
|V|=\left(A_1^2+A_2^2+2A_1A_2\cos\psi\right)^{1/2}.
\end{equation}

\noindent We assume independent Gaussian errors on the measured visibility amplitudes,
\begin{equation}
|V_k|^{\mathrm{obs}}=|V_k(\boldsymbol{p})|+n_k,
\end{equation}

\noindent where $\boldsymbol{p}$ is the model parameter vector and $n_k$ is real Gaussian noise with variance $\sigma_{|V|}^2$. The log-likelihood is therefore

\begin{equation}
\ln\mathcal{L}=-\frac{1}{2\sigma_{|V|}^2}\sum_{k=1}^{N}\left[|V_k|^{\mathrm{obs}}-|V_k(\boldsymbol{p})|\right]^2+\mathrm{const}.
\end{equation}

\noindent The corresponding Fisher matrix is

\begin{equation}
F_{ij}=\frac{1}{\sigma_{|V|}^2}\sum_{k=1}^{N}\frac{\partial |V_k|}{\partial p_i}\frac{\partial |V_k|}{\partial p_j}.
\end{equation}

\noindent For the $k$th baseline,

\begin{equation}
|V_k|=\left(A_{1,k}^2+A_{2,k}^2+2A_{1,k}A_{2,k}\cos\psi_k\right)^{1/2}.
\end{equation}

\noindent The derivatives with respect to the separation magnitude and position angle are

\begin{equation}
\frac{\partial |V|}{\partial\rho}=-\frac{A_1A_2}{|V|}\sin\psi\left[2\pi b\cos(\gamma-\phi)\right],
\end{equation}

\begin{equation}
\frac{\partial |V|}{\partial\phi}=-\frac{A_1A_2}{|V|}\sin\psi\left[2\pi b\rho\sin(\gamma-\phi)\right].
\end{equation}

\noindent The amplitude-only Fisher matrix elements are therefore

\begin{equation}
F_{\rho\rho}=\frac{1}{\sigma_{|V|}^2}\sum_{k=1}^{N}\left[\frac{A_{1,k}A_{2,k}}{|V_k|}\right]^2\left[2\pi b_k\cos(\gamma_k-\phi)\right]^2\sin^2\psi_k,
\end{equation}

\begin{equation}
F_{\phi\phi}=\frac{1}{\sigma_{|V|}^2}\sum_{k=1}^{N}\left[\frac{A_{1,k}A_{2,k}}{|V_k|}\right]^2\left[2\pi b_k\rho\sin(\gamma_k-\phi)\right]^2\sin^2\psi_k.
\end{equation}

\noindent The covariance between the separation and position angle is captured by the off-diagonal Fisher matrix element,

\begin{equation}
\begin{aligned}
F_{\rho\phi}=F_{\phi\rho}&=\frac{1}{\sigma_{|V|}^2}\sum_{k=1}^{N}\left[\frac{A_{1,k}A_{2,k}}{|V_k|}\right]^2\left[2\pi b_k\cos(\gamma_k-\phi)\right] \\& \times \left[2\pi b_k\rho\sin(\gamma_k-\phi)\right]\sin^2\psi_k.
\end{aligned}
\end{equation}

\noindent We therefore form the two-parameter Fisher matrix and covariance for the binary geometry,

\begin{equation}
\mathbf{F}_{\rho\phi}=
\begin{pmatrix}
F_{\rho\rho} & F_{\rho\phi}\\
F_{\phi\rho} & F_{\phi\phi}
\end{pmatrix},
\end{equation}

\begin{equation}
\mathbf{C}_{\rho\phi}=\mathbf{F}_{\rho\phi}^{-1}=\frac{1}{F_{\rho\rho}F_{\phi\phi}-F_{\rho\phi}^2}
\begin{pmatrix}
F_{\phi\phi} & -F_{\rho\phi}\\
-F_{\rho\phi} & F_{\rho\rho}
\end{pmatrix}.
\end{equation}

\noindent The marginalised one-sigma uncertainties on $\rho$ and $\phi$ are therefore

\begin{equation}
\sigma_\rho=\sqrt{\frac{F_{\phi\phi}}{F_{\rho\rho}F_{\phi\phi}-F_{\rho\phi}^2}},
\end{equation}
\begin{equation}
\sigma_\phi=\sqrt{\frac{F_{\rho\rho}}{F_{\rho\rho}F_{\phi\phi}-F_{\rho\phi}^2}}.
\end{equation}

\noindent The same Fisher formalism can be used to calculate the diagonal matrix elements for the remaining source parameters. These diagonal terms quantify the conditional information in each parameter when the remaining parameters are held fixed. For the total flux densities ($F_{1/2}$), and angular widths ($\sigma_{1/2}$), these expressions are

\begin{equation}
F_{F_1F_1}=\frac{1}{\sigma_{|V|}^2}\sum_{k=1}^{N}\left\{\frac{\left[A_{1,k}+A_{2,k}\cos\psi_k\right]\exp\!\left(-2\pi^2\sigma_1^2b_k^2\right)}{|V_k|}\right\}^2,
\end{equation}

\begin{equation}
F_{F_2F_2}=\frac{1}{\sigma_{|V|}^2}\sum_{k=1}^{N}\left\{\frac{\left[A_{2,k}+A_{1,k}\cos\psi_k\right]\exp\!\left(-2\pi^2\sigma_2^2b_k^2\right)}{|V_k|}\right\}^2,
\end{equation}

\begin{equation}
F_{\sigma_1\sigma_1}=\frac{1}{\sigma_{|V|}^2}\sum_{k=1}^{N}\left\{\frac{4\pi^2b_k^2\sigma_1A_{1,k}\left[A_{1,k}+A_{2,k}\cos\psi_k\right]}{|V_k|}\right\}^2,
\end{equation}

\begin{equation}
F_{\sigma_2\sigma_2}=\frac{1}{\sigma_{|V|}^2}\sum_{k=1}^{N}\left\{\frac{4\pi^2b_k^2\sigma_2A_{2,k}\left[A_{2,k}+A_{1,k}\cos\psi_k\right]}{|V_k|}\right\}^2.
\end{equation}

\noindent A full multi-parameter characterisation would require the off-diagonal terms involving $F_1$, $F_2$, $\sigma_1$, and $\sigma_2$ to be included and the complete Fisher matrix to be inverted.

\section{Dependence on Model Parameters}
\label{A:sens}

\begin{figure*}
\centering
\includegraphics[width=\textwidth]{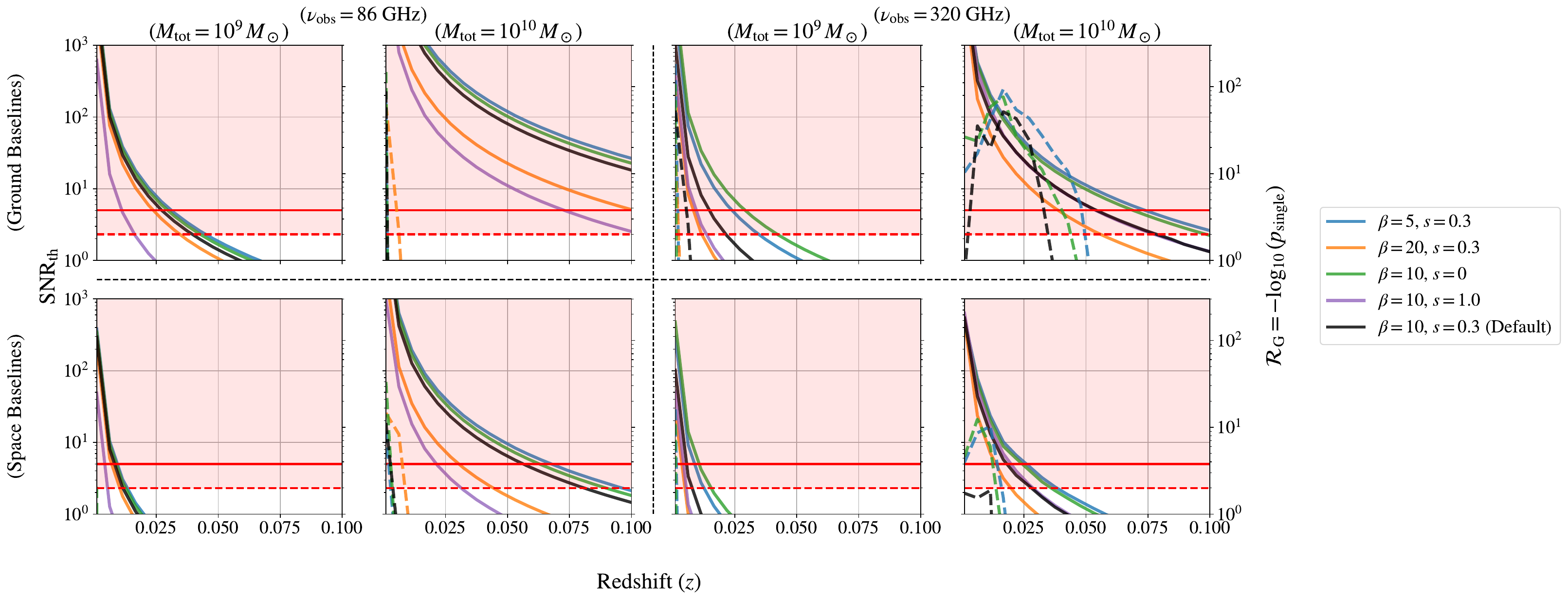}
\caption{Detectability evaluation as shown in Figure \ref{f:visibility} but varying the slope of the mass-loss correction factor ($s$) and the gas pressure to magnetic pressure ratio ($\beta$). Illustrated for a $q=0.1$ system with $P_\mathrm{obs}=10$~years at $\dot{m_0}=10^{-4}$.}
\label{f:parameter_sens}
\end{figure*}

\noindent Increasing $s$ corresponds to higher mass loss and therefore reduces the effective accretion rate used in calculating the SED. Higher values of $\beta$ weaken the magnetic field and consequently the synchrotron emission, the primary contribution to the detectable emission at (sub)millimetre wavelengths. Figure \ref{f:parameter_sens} depicts a subset of the detectable parameter space for variations in $s$ and $\beta$. The default values (as assumed in Figure \ref{f:visibility}) are shown in black. As expected, lower $s$ increases the detectable parameter space. Whereas increased $s$ or $\beta$ significantly reduces the redshift across which the assumed binary can be detected. As stated by \citet{tiede_hot_2025}, we leave investigation of magnetically dominated (i.e. $\beta < 1$) and other parameter variations for future study.

\end{appendix}

%\bibliography{PASPsample701}{}
\bibliography{BHEX-smbhb}{}

@ARTICLE{ClyburnZrake:2026,
       author = {{Clyburn}, Madeline and {Zrake}, Jonathan},
        title = "{Unequal mass binary evolution driven by high-Mach circumbinary discs}",
      journal = {\mnras},
         year = 2026,
        month = may,
       volume = {548},
       number = {1},
          doi = {10.1093/mnras/stag567},
archivePrefix = {arXiv},
       eprint = {2603.27022},
 primaryClass = {astro-ph.HE},
       adsurl = {https://ui.adsabs.harvard.edu/abs/2026MNRAS.548ag567C}
}

@ARTICLE{Liu+2019,
       author = {{Liu}, T. and {Gezari}, S. and {Ayers}, M. and {Burgett}, W. and {Chambers}, K. and {Hodapp}, K. and {Huber}, M.~E. and {Kudritzki}, R.-P. and {Metcalfe}, N. and {Tonry}, J. and {Wainscoat}, R. and {Waters}, C.},
        title = "{Supermassive Black Hole Binary Candidates from the Pan-STARRS1 Medium Deep Survey}",
      journal = {\apj},
         year = 2019,
        month = oct,
       volume = {884},
       number = {1},
          eid = {36},
        pages = {36},
          doi = {10.3847/1538-4357/ab40cb},
archivePrefix = {arXiv},
       eprint = {1906.08315},
 primaryClass = {astro-ph.HE},
       adsurl = {https://ui.adsabs.harvard.edu/abs/2019ApJ...884...36L}
}

@ARTICLE{Graham+2015,
   author = {{Graham}, M.~J. and {Djorgovski}, S.~G. and {Stern}, D. and 
	{Drake}, A.~J. and {Mahabal}, A.~A. and {Donalek}, C. and {Glikman}, E. and 
	{Larson}, S. and {Christensen}, E.},
    title = "{A systematic search for close supermassive black hole binaries in the Catalina Real-time Transient Survey}",
  journal = {\mnras},
archivePrefix = "arXiv",
   eprint = {1507.07603},
     year = 2015,
    month = oct,
   volume = 453,
    pages = {1562-1576},
      doi = {10.1093/mnras/stv1726},
   adsurl = {http://adsabs.harvard.edu/abs/2015MNRAS.453.1562G}
}

@ARTICLE{Charisi:2016,
       author = {{Charisi}, M. and {Bartos}, I. and {Haiman}, Z. and {Price-Whelan}, A.~M. and {Graham}, M.~J. and {Bellm}, E.~C. and {Laher}, R.~R. and {M{\'a}rka}, S.},
        title = "{A population of short-period variable quasars from PTF as supermassive black hole binary candidates}",
      journal = {\mnras},
         year = 2016,
        month = dec,
       volume = {463},
       number = {2},
        pages = {2145-2171},
          doi = {10.1093/mnras/stw1838},
archivePrefix = {arXiv},
       eprint = {1604.01020},
 primaryClass = {astro-ph.GA},
       adsurl = {https://ui.adsabs.harvard.edu/abs/2016MNRAS.463.2145C}
}

@ARTICLE{ChenXin:2020,
       author = {{Chen}, Yu-Ching and {Liu}, Xin and {Liao}, Wei-Ting and {Holgado}, A. Miguel and {Guo}, Hengxiao and {Gruendl}, Robert A. and {Morganson}, Eric and {Shen}, Yue and {Zhang}, Kaiwen and {Abbott}, Tim M.~C. and {Aguena}, Michel and {Allam}, Sahar and {Avila}, Santiago and {Bertin}, Emmanuel and {Bhargava}, Sunayana and {Brooks}, David and {Burke}, David L. and {Carnero Rosell}, Aurelio and {Carollo}, Daniela and {Carrasco Kind}, Matias and {Carretero}, Jorge and {Costanzi}, Matteo and {da Costa}, Luiz N. and {Davis}, Tamara M. and {De Vicente}, Juan and {Desai}, Shantanu and {Diehl}, H. Thomas and {Doel}, Peter and {Everett}, Spencer and {Flaugher}, Brenna and {Friedel}, Douglas and {Frieman}, Joshua and {Garc{\'\i}a-Bellido}, Juan and {Gaztanaga}, Enrique and {Glazebrook}, Karl and {Gruen}, Daniel and {Gutierrez}, Gaston and {Hinton}, Samuel R. and {Hollowood}, Devon L. and {James}, David J. and {Kim}, Alex G. and {Kuehn}, Kyler and {Kuropatkin}, Nikolay and {Lewis}, Geraint F. and {Lidman}, Christopher and {Lima}, Marcos and {Maia}, Marcio A.~G. and {March}, Marisa and {Marshall}, Jennifer L. and {Menanteau}, Felipe and {Miquel}, Ramon and {Palmese}, Antonella and {Paz-Chinch{\'o}n}, Francisco and {Plazas}, Andr{\'e}s A. and {Sanchez}, Eusebio and {Schubnell}, Michael and {Serrano}, Santiago and {Sevilla-Noarbe}, Ignacio and {Smith}, Mathew and {Suchyta}, Eric and {Swanson}, Molly E.~C. and {Tarle}, Gregory and {Tucker}, Brad E. and {Norbert Varga}, Tamas and {Walker}, Alistair R.},
        title = "{Candidate periodically variable quasars from the Dark Energy Survey and the Sloan Digital Sky Survey}",
      journal = {\mnras},
         year = 2020,
        month = dec,
       volume = {499},
       number = {2},
        pages = {2245-2264},
          doi = {10.1093/mnras/staa2957},
archivePrefix = {arXiv},
       eprint = {2008.12329},
 primaryClass = {astro-ph.HE},
       adsurl = {https://ui.adsabs.harvard.edu/abs/2020MNRAS.499.2245C}
}

@ARTICLE{NANO_Astro_Interp+2023,
       author = {{Agazie}, Gabriella and {Anumarlapudi}, Akash and {Archibald}, Anne M. and {Baker}, Paul T. and {B{\'e}csy}, Bence and {Blecha}, Laura and {Bonilla}, Alexander and {Brazier}, Adam and {Brook}, Paul R. and {Burke-Spolaor}, Sarah and {Burnette}, Rand and {Case}, Robin and {Casey-Clyde}, J. Andrew and {Charisi}, Maria and {Chatterjee}, Shami and {Chatziioannou}, Katerina and {Cheeseboro}, Belinda D. and {Chen}, Siyuan and {Cohen}, Tyler and {Cordes}, James M. and {Cornish}, Neil J. and {Crawford}, Fronefield and {Cromartie}, H. Thankful and {Crowter}, Kathryn and {Cutler}, Curt J. and {D'Orazio}, Daniel J. and {Decesar}, Megan E. and {Degan}, Dallas and {Demorest}, Paul B. and {Deng}, Heling and {Dolch}, Timothy and {Drachler}, Brendan and {Ferrara}, Elizabeth C. and {Fiore}, William and {Fonseca}, Emmanuel and {Freedman}, Gabriel E. and {Gardiner}, Emiko and {Garver-Daniels}, Nate and {Gentile}, Peter A. and {Gersbach}, Kyle A. and {Glaser}, Joseph and {Good}, Deborah C. and {G{\"u}ltekin}, Kayhan and {Hazboun}, Jeffrey S. and {Hourihane}, Sophie and {Islo}, Kristina and {Jennings}, Ross J. and {Johnson}, Aaron and {Jones}, Megan L. and {Kaiser}, Andrew R. and {Kaplan}, David L. and {Kelley}, Luke Zoltan and {Kerr}, Matthew and {Key}, Joey S. and {Laal}, Nima and {Lam}, Michael T. and {Lamb}, William G. and {Lazio}, T. Joseph W. and {Lewandowska}, Natalia and {Littenberg}, Tyson B. and {Liu}, Tingting and {Luo}, Jing and {Lynch}, Ryan S. and {Ma}, Chung-Pei and {Madison}, Dustin R. and {McEwen}, Alexander and {McKee}, James W. and {McLaughlin}, Maura A. and {McMann}, Natasha and {Meyers}, Bradley W. and {Meyers}, Patrick M. and {Mingarelli}, Chiara M.~F. and {Mitridate}, Andrea and {Natarajan}, Priyamvada and {Ng}, Cherry and {Nice}, David J. and {Ocker}, Stella Koch and {Olum}, Ken D. and {Pennucci}, Timothy T. and {Perera}, Benetge B.~P. and {Petrov}, Polina and {Pol}, Nihan S. and {Radovan}, Henri A. and {Ransom}, Scott M. and {Ray}, Paul S. and {Romano}, Joseph D. and {Runnoe}, Jessie C. and {Sardesai}, Shashwat C. and {Schmiedekamp}, Ann and {Schmiedekamp}, Carl and {Schmitz}, Kai and {Schult}, Levi and {Shapiro-Albert}, Brent J. and {Siemens}, Xavier and {Simon}, Joseph and {Siwek}, Magdalena S. and {Stairs}, Ingrid H. and {Stinebring}, Daniel R. and {Stovall}, Kevin and {Sun}, Jerry P. and {Susobhanan}, Abhimanyu and {Swiggum}, Joseph K. and {Taylor}, Jacob and {Taylor}, Stephen R. and {Turner}, Jacob E. and {Unal}, Caner and {Vallisneri}, Michele and {Vigeland}, Sarah J. and {Wachter}, Jeremy M. and {Wahl}, Haley M. and {Wang}, Qiaohong and {Witt}, Caitlin A. and {Wright}, David and {Young}, Olivia and {Nanograv Collaboration}},
        title = "{The NANOGrav 15 yr Data Set: Constraints on Supermassive Black Hole Binaries from the Gravitational-wave Background}",
      journal = {\apjl},
         year = 2023,
        month = aug,
       volume = {952},
       number = {2},
          eid = {L37},
        pages = {L37},
          doi = {10.3847/2041-8213/ace18b},
archivePrefix = {arXiv},
       eprint = {2306.16220},
 primaryClass = {astro-ph.HE},
       adsurl = {https://ui.adsabs.harvard.edu/abs/2023ApJ...952L..37A}
}

@ARTICLE{Veronsi+2026,
       author = {{Veronesi}, Niccol{\`o} and {Charisi}, Maria and {Petrov}, Polina and {Taylor}, Stephen R. and {Runnoe}, Jessie and {D'Orazio}, Daniel J. and {Pilawa}, Jacob and {Ma}, Chung-Pei},
        title = "{Prospects of resolving and localising individual supermassive black hole binaries with pulsar timing arrays: the host ranking challenge}",
      journal = {arXiv e-prints},
         year = 2026,
        month = may,
          eid = {arXiv:2606.00218},
        pages = {arXiv:2606.00218},
          doi = {10.48550/arXiv.2606.00218},
archivePrefix = {arXiv},
       eprint = {2606.00218},
 primaryClass = {astro-ph.GA},
       adsurl = {https://ui.adsabs.harvard.edu/abs/2026arXiv260600218V}
}

@article{hudson_towards_2026,
	title = {Towards direct imaging and orbital parameter estimation of supermassive black hole binaries with spaceborne {VLBI}},
	copyright = {https://www.edpsciences.org/en/authors/copyright-and-licensing},
	issn = {0004-6361, 1432-0746},
	url = {https://www.aanda.org/10.1051/0004-6361/202558054},
	doi = {10.1051/0004-6361/202558054},
	urldate = {2026-06-29},
	journal = {A\&A},
	author = {Hudson, B. and Gurvits, L.I. and Mooij, E. and Ricarte, A. and Palumbo, D.},
	month = jun,
	year = {2026},
}

@article{tiede_hot_2025,
	title = {Hot, {Cold}, and {Multicomponent} {Accretion} {Flows} around {Supermassive} {Black} {Hole} {Binaries}},
	volume = {995},
	issn = {0004-637X, 1538-4357},
	url = {https://iopscience.iop.org/article/10.3847/1538-4357/ae17ba},
	doi = {10.3847/1538-4357/ae17ba},
	number = {1},
	urldate = {2026-01-14},
	journal = {ApJ},
	author = {Tiede, Christopher and D’Orazio, Daniel J.},
	month = dec,
	year = {2025},
	pages = {68}
}

@article{dorazio_repeated_2018,
	title = {Repeated {Imaging} of {Massive} {Black} {Hole} {Binary} {Orbits} with {Millimeter} {Interferometry}: {Measuring} {Black} {Hole} {Masses} and the {Hubble} {Constant}},
	volume = {863},
	copyright = {https://iopscience.iop.org/page/copyright},
	issn = {0004-637X, 1538-4357},
	shorttitle = {Repeated {Imaging} of {Massive} {Black} {Hole} {Binary} {Orbits} with {Millimeter} {Interferometry}},
	url = {https://iopscience.iop.org/article/10.3847/1538-4357/aad413},
	doi = {10.3847/1538-4357/aad413},
	number = {2},
	urldate = {2025-05-23},
	journal = {ApJ},
	author = {D’Orazio, Daniel J. and Loeb, Abraham},
	month = aug,
	year = {2018},
	note = {Publisher: American Astronomical Society},
	pages = {185}
}

@article{pesce_toward_2021,
	title = {Toward {Determining} the {Number} of {Observable} {Supermassive} {Black} {Hole} {Shadows}},
	volume = {923},
	copyright = {http://creativecommons.org/licenses/by/4.0/},
	issn = {0004-637X, 1538-4357},
	url = {https://iopscience.iop.org/article/10.3847/1538-4357/ac2eb5},
	doi = {10.3847/1538-4357/ac2eb5},
	number = {2},
	urldate = {2025-05-23},
	journal = {ApJ},
	author = {Pesce, Dominic W. and Palumbo, Daniel C. M. and Narayan, Ramesh and Blackburn, Lindy and Doeleman, Sheperd S. and Johnson, Michael D. and Ma, Chung-Pei and Nagar, Neil M. and Natarajan, Priyamvada and Ricarte, Angelo},
	month = dec,
	year = {2021},
	note = {Publisher: American Astronomical Society},
	pages = {260}
}

@inproceedings{johnson_black_2024,
	address = {Yokohama, Japan},
	title = {The {Black} {Hole} {Explorer}: motivation and vision},
	shorttitle = {The {Black} {Hole} {Explorer}},
	url = {https://www.spiedigitallibrary.org/conference-proceedings-of-spie/13092/3019835/The-Black-Hole-Explorer-motivation-and-vision/10.1117/12.3019835.full},
	doi = {10.1117/12.3019835},
	urldate = {2025-05-23},
	booktitle = {Space {Telescopes} and {Instrumentation} 2024: {Optical}, {Infrared}, and {Millimeter} {Wave}},
	publisher = {SPIE},
	author = {Johnson, Michael and Akiyama, Kazunori and Baturin, Rebecca and Bilyeu, Bryan and Blackburn, Lindy and Boroson, Don and Cárdenas-Avendaño, Alejandro and Chael, Andrew and Chan, Chi-kwan and Chang, Dominic and Cheimets, Peter and Chou, Cathy and Doeleman, Sheperd S. and Farah, Joseph and Galison, Peter and Gamble, Ronald S. and Gammie, Charles F. and Gelles, Zachary and Gómez, José L. and Gralla, Samuel E. and Grimes, Paul K. and Gurvits, Leonid I. and Hadar, Shahar and Haworth, Kari and Hada, Kazuhiro and Hecht, Michael H. and Honma, Mareki and Houston, Janice and Hudson, Ben and Issaoun, Sara and Jia, He and Jorstad, Svetlana and Kauffman, Jens and Kovalev, Yuri and Kurczynski, Peter and Lafon, Robert and Lupsasca, Alexandru and Lehmensiek, Robert and Ma, Chung-Pei and Marrone, Daniel P. and Marscher, Alan P. and Melnick, Gary and Narayan, Ramesh and Niinuma, Kotaro and Noble, Scott C. and Palmer, Eric J. and Palumbo, Daniel C. M. and Paritsky, Lenny and Peretz, Eliad and Pesce, Dominic and Plavin, Alexander and Quataert, Eliot and Rana, Hannah and Ricarte, Angelo and Roelofs, Freek and Shtyrkova, Katia and Sinclair, Laura C. and Small, Jeffrey and Sridharan, Tirupati Kumara and Srinivasan, Ranjani and Strominger, Andrew and Tiede, Paul and Tong, Edward and Wang, Jade and Weintroub, Jonathan and Wielgus, Maciek and Wong, George N.},
	editor = {Coyle, Laura E. and Perrin, Marshall D. and Matsuura, Shuji},
	month = aug,
	year = {2024},
	pages = {90}
}

@article{kelley_massive_2019,
	title = {Massive {BH} binaries as periodically variable {AGN}},
	volume = {485},
	copyright = {https://academic.oup.com/journals/pages/open\_access/funder\_policies/chorus/standard\_publication\_model},
	issn = {0035-8711, 1365-2966},
	url = {https://academic.oup.com/mnras/article/485/2/1579/5304631},
	doi = {10.1093/mnras/stz150},
	language = {en},
	number = {2},
	urldate = {2026-01-23},
	journal = {MNRAS},
	author = {Kelley, Luke Zoltan and Haiman, Zoltán and Sesana, Alberto and Hernquist, Lars},
	month = may,
	year = {2019},
	pages = {1579--1594}
}

@article{gutierrez_non-thermal_2024,
	title = {Non-thermal radiation from dual jet interactions in supermassive black hole binaries},
	volume = {532},
	copyright = {https://creativecommons.org/licenses/by/4.0/},
	issn = {0035-8711, 1365-2966},
	url = {https://academic.oup.com/mnras/article/532/1/506/7692032},
	doi = {10.1093/mnras/stae1473},
	language = {en},
	number = {1},
	urldate = {2025-05-23},
	journal = {MNRAS},
	author = {Gutiérrez, Eduardo M and Combi, Luciano and Romero, Gustavo E and Campanelli, Manuela},
	month = jun,
	year = {2024},
	note = {Publisher: Oxford University Press (OUP)},
	pages = {506--516}
}

@article{chael_interferometric_2018,
	title = {Interferometric {Imaging} {Directly} with {Closure} {Phases} and {Closure} {Amplitudes}},
	volume = {857},
	issn = {0004-637X, 1538-4357},
	url = {https://iopscience.iop.org/article/10.3847/1538-4357/aab6a8},
	doi = {10.3847/1538-4357/aab6a8},
	number = {1},
	urldate = {2024-08-15},
	journal = {ApJ},
	author = {Chael, Andrew A. and Johnson, Michael D. and Bouman, Katherine L. and Blackburn, Lindy L. and Akiyama, Kazunori and Narayan, Ramesh},
	month = apr,
	year = {2018},
	pages = {23}
}

@article{fang_orbit_2022,
	title = {Orbit {Tomography} of {Binary} {Supermassive} {Black} {Holes} with {Very} {Long} {Baseline} {Interferometry}},
	volume = {927},
	issn = {0004-637X, 1538-4357},
	url = {https://iopscience.iop.org/article/10.3847/1538-4357/ac4bd7},
	doi = {10.3847/1538-4357/ac4bd7},
	number = {1},
	urldate = {2025-05-25},
	journal = {ApJ},
	author = {Fang, Yun and Yang, Huan},
	month = mar,
	year = {2022},
	pages = {93}
}

@article{fomalont_sub-milliarcsecond_1999,
	title = {Sub-{Milliarcsecond} {Precision} of {Pulsar} {Motions}: {Using} {In}-{Beam} {Calibrators} with the {VLBA}},
	volume = {117},
	issn = {00046256},
	shorttitle = {Sub-{Milliarcsecond} {Precision} of {Pulsar} {Motions}},
	url = {https://iopscience.iop.org/article/10.1086/300877},
	doi = {10.1086/300877},
	number = {6},
	urldate = {2026-02-16},
	journal = {AJ},
	author = {Fomalont, E. B. and Goss, W. M. and Beasley, A. J. and Chatterjee, S.},
	month = jun,
	year = {1999},
	pages = {3025--3030}
}

@article{gomez_probing_2022,
	title = {Probing the {Innermost} {Regions} of {AGN} {Jets} and {Their} {Magnetic} {Fields} with {RadioAstron}. {V}. {Space} and {Ground} {Millimeter}-{VLBI} {Imaging} of {OJ} 287},
	volume = {924},
	issn = {0004-637X, 1538-4357},
	url = {https://iopscience.iop.org/article/10.3847/1538-4357/ac3bcc},
	doi = {10.3847/1538-4357/ac3bcc},
	number = {2},
	urldate = {2025-04-04},
	journal = {ApJ},
	author = {Gómez, José L. and Traianou, Efthalia and Krichbaum, Thomas P. and Lobanov, Andrei P. and Fuentes, Antonio and Lico, Rocco and Zhao, Guang-Yao and Bruni, Gabriele and Kovalev, Yuri Y. and Lähteenmäki, Anne and Voitsik, Petr A. and Lisakov, Mikhail M. and Angelakis, Emmanouil and Bach, Uwe and Casadio, Carolina and Cho, Ilje and Dey, Lankeswar and Gopakumar, Achamveedu and Gurvits, Leonid I. and Jorstad, Svetlana and Kovalev, Yuri A. and Lister, Matthew L. and Marscher, Alan P. and Myserlis, Ioannis and Pushkarev, Alexander B. and Ros, Eduardo and Savolainen, Tuomas and Tornikoski, Merja and Valtonen, Mauri J. and Zensus, Anton},
	month = jan,
	year = {2022},
	pages = {122}
}

@article{agarwal_nanograv_2026,
	title = {The {NANOGrav} 15 yr {Dataset}: {Targeted} {Searches} for {Supermassive} {Black} {Hole} {Binaries}},
	volume = {998},
	issn = {2041-8205, 2041-8213},
	shorttitle = {The {NANOGrav} 15 yr {Dataset}},
	url = {https://iopscience.iop.org/article/10.3847/2041-8213/ae3719},
	doi = {10.3847/2041-8213/ae3719},
	number = {1},
	urldate = {2026-06-15},
	journal = {ApJL},
	author = {Agarwal, Nikita and Agazie, Gabriella and Anumarlapudi, Akash and Archibald, Anne M. and Arzoumanian, Zaven and Baier, Jeremy G. and Baker, Paul T. and Bécsy, Bence and Blecha, Laura and Brazier, Adam and Brook, Paul R. and Burke-Spolaor, Sarah and Burnette, Rand and Case, Robin and Casey-Clyde, J. Andrew and Chang, Yu-Ting and Charisi, Maria and Chatterjee, Shami and Cohen, Tyler and Coppi, Paolo and Cordes, James M. and Cornish, Neil J. and Crawford, Fronefield and Cromartie, H. Thankful and Crowter, Kathryn and DeCesar, Megan E. and Demorest, Paul B. and Deng, Heling and Dey, Lankeswar and Dolch, Timothy and D’Orazio, Daniel J. and Eisenberg, Ellis and Ferrara, Elizabeth C. and Doskoch, Graham and Fiore, William and Fonseca, Emmanuel and Freedman, Gabriel E. and Gardiner, Emiko C. and Garver-Daniels, Nate and Gentile, Peter A. and Gersbach, Kyle A. and Glaser, Joseph and Graham, Matthew J. and Good, Deborah C. and Gültekin, Kayhan and Harris, C. J. and Hazboun, Jeffrey S. and Hutchison, Forrest and Jennings, Ross J. and Johnson, Aaron D. and Jones, Megan L. and Kaplan, David L. and Kelley, Luke Zoltan and Kerr, Matthew and Key, Joey S. and Laal, Nima and Lam, Michael T. and Lamb, William G. and Larsen, Bjorn and Lazio, T. Joseph W. and Lewandowska, Natalia and Liu, Tingting and Lorimer, Duncan R. and Luo, Jing and Lynch, Ryan S. and Ma, Chung-Pei and Madison, Dustin R. and Matt, Cayenne and McEwen, Alexander and McKee, James W. and McLaughlin, Maura A. and McMann, Natasha and Meyers, Bradley W. and Meyers, Patrick M. and Mingarelli, Chiara M. F. and Mitridate, Andrea and Natarajan, Priyamvada and Ng, Cherry and Nice, David J. and Nichols, Shania and Ocker, Stella Koch and Olum, Ken D. and Pennucci, Timothy T. and Perera, Benetge B. P. and Petrov, Polina and Pol, Nihan S. and Radovan, Henri A. and Ransom, Scott M. and Ray, Paul S. and Romano, Joseph D. and Runnoe, Jessie C. and Saffer, Alexander and Sardesai, Shashwat C. and Schmiedekamp, Ann and Schmiedekamp, Carl and Schmitz, Kai and Semenzato, Federico and Shapiro-Albert, Brent J. and Shivakumar, Rohan and Siemens, Xavier and Simon, Joseph and Sosa Fiscella, Sophia V. and Stairs, Ingrid H. and Stinebring, Daniel R. and Stovall, Kevin and Susobhanan, Abhimanyu and Swiggum, Joseph K. and Taylor, Jacob A. and Taylor, Stephen R. and Thompson, Mercedes S. and Turner, Jacob E. and Vallisneri, Michele and Van Haasteren, Rutger and Vigeland, Sarah J. and Wahl, Haley M. and Willson, London and Wilson, Kevin P. and Witt, Caitlin A. and Wright, David and Young, Olivia and Zheng, Qinyuan and {The NANOGrav Collaboration}},
	month = feb,
	year = {2026},
	pages = {L11},
}

@article{valtonen_identifying_2025,
	title = {Identifying the {Secondary} {Jet} in the {RadioAstron} {Image} of {OJ} 287},
	volume = {992},
	issn = {0004-637X, 1538-4357},
	url = {https://iopscience.iop.org/article/10.3847/1538-4357/ae057e},
	doi = {10.3847/1538-4357/ae057e},
	number = {1},
	urldate = {2026-06-15},
	journal = {ApJ},
	author = {Valtonen, Mauri J. and Dey, Lankeswar and Zola, Staszek and Gupta, Alok C. and Kishore, Shubham and Gopakumar, Achamveedu and Wiita, Paul J. and Gu, Minfeng and Nilsson, Kari and Zhang, Zhongli and Hudec, Rene and Matsumoto, Katsura and Drozdz, Marek and Ogloza, Waldemar and Berdyugin, Andrei V. and Reichart, Daniel E. and Mugrauer, Markus and Pursimo, Tapio and Ciprini, Stefano and Nakaoka, Tatsuya and Uemura, Makoto and Imazawa, Ryo and Zejmo, Michal and Kouprianov, Vladimir V. and Davidson, James W. and Sadun, Alberto and Štrobl, Jan and Jelínek, Martin and Susobhanan, Abhimanyu},
	month = oct,
	year = {2025},
	pages = {110},
}

@article{agazie_nanograv_2023,
	title = {The {NANOGrav} 15 yr {Data} {Set}: {Evidence} for a {Gravitational}-wave {Background}},
	volume = {951},
	copyright = {http://creativecommons.org/licenses/by/4.0/},
	issn = {2041-8205, 2041-8213},
	shorttitle = {The {NANOGrav} 15 yr {Data} {Set}},
	url = {https://iopscience.iop.org/article/10.3847/2041-8213/acdac6},
	doi = {10.3847/2041-8213/acdac6},
	number = {1},
	urldate = {2025-05-23},
	journal = {ApJL},
	author = {Agazie, Gabriella and Anumarlapudi, Akash and Archibald, Anne M. and Arzoumanian, Zaven and Baker, Paul T. and Bécsy, Bence and Blecha, Laura and Brazier, Adam and Brook, Paul R. and Burke-Spolaor, Sarah and Burnette, Rand and Case, Robin and Charisi, Maria and Chatterjee, Shami and Chatziioannou, Katerina and Cheeseboro, Belinda D. and Chen, Siyuan and Cohen, Tyler and Cordes, James M. and Cornish, Neil J. and Crawford, Fronefield and Cromartie, H. Thankful and Crowter, Kathryn and Cutler, Curt J. and DeCesar, Megan E. and DeGan, Dallas and Demorest, Paul B. and Deng, Heling and Dolch, Timothy and Drachler, Brendan and Ellis, Justin A. and Ferrara, Elizabeth C. and Fiore, William and Fonseca, Emmanuel and Freedman, Gabriel E. and Garver-Daniels, Nate and Gentile, Peter A. and Gersbach, Kyle A. and Glaser, Joseph and Good, Deborah C. and Gültekin, Kayhan and Hazboun, Jeffrey S. and Hourihane, Sophie and Islo, Kristina and Jennings, Ross J. and Johnson, Aaron D. and Jones, Megan L. and Kaiser, Andrew R. and Kaplan, David L. and Kelley, Luke Zoltan and Kerr, Matthew and Key, Joey S. and Klein, Tonia C. and Laal, Nima and Lam, Michael T. and Lamb, William G. and W. Lazio, T. Joseph and Lewandowska, Natalia and Littenberg, Tyson B. and Liu, Tingting and Lommen, Andrea and Lorimer, Duncan R. and Luo, Jing and Lynch, Ryan S. and Ma, Chung-Pei and Madison, Dustin R. and Mattson, Margaret A. and McEwen, Alexander and McKee, James W. and McLaughlin, Maura A. and McMann, Natasha and Meyers, Bradley W. and Meyers, Patrick M. and Mingarelli, Chiara M. F. and Mitridate, Andrea and Natarajan, Priyamvada and Ng, Cherry and Nice, David J. and Ocker, Stella Koch and Olum, Ken D. and Pennucci, Timothy T. and Perera, Benetge B. P. and Petrov, Polina and Pol, Nihan S. and Radovan, Henri A. and Ransom, Scott M. and Ray, Paul S. and Romano, Joseph D. and Sardesai, Shashwat C. and Schmiedekamp, Ann and Schmiedekamp, Carl and Schmitz, Kai and Schult, Levi and Shapiro-Albert, Brent J. and Siemens, Xavier and Simon, Joseph and Siwek, Magdalena S. and Stairs, Ingrid H. and Stinebring, Daniel R. and Stovall, Kevin and Sun, Jerry P. and Susobhanan, Abhimanyu and Swiggum, Joseph K. and Taylor, Jacob and Taylor, Stephen R. and Turner, Jacob E. and Unal, Caner and Vallisneri, Michele and Van Haasteren, Rutger and Vigeland, Sarah J. and Wahl, Haley M. and Wang, Qiaohong and Witt, Caitlin A. and Young, Olivia and {The NANOGrav Collaboration}},
	month = jul,
	year = {2023},
	note = {Publisher: American Astronomical Society},
	pages = {L8}
}

@article{begelman_massive_1980,
	title = {Massive black hole binaries in active galactic nuclei},
	volume = {287},
	copyright = {http://www.springer.com/tdm},
	issn = {0028-0836, 1476-4687},
	url = {https://www.nature.com/articles/287307a0},
	doi = {10.1038/287307a0},
	language = {en},
	number = {5780},
	urldate = {2025-05-24},
	journal = {Natur},
	author = {Begelman, M. C. and Blandford, R. D. and Rees, M. J.},
	month = sep,
	year = {1980},
	pages = {307--309},
}

@INCOLLECTION{dorazio_observational_2023,
       author = {{D'Orazio}, Daniel J. and {Charisi}, Maria and {Derdzinski}, Andrea and {Zwick}, Lorenz and {Wevers}, Thomas and {Ryu}, Taheo},
        title = "{Multimessenger astronomy with black holes}",
    booktitle = {Black Holes in the Era of Gravitational-Wave Astronomy},
         year = 2024,
         publisher = {Elsevier},
       editor = {{Arca Sedda}, Manuel and {Bortolas}, Elisa and {Spera}, Mario},
        pages = {379-478},
          doi = {10.1016/B978-0-32-395636-9.00013-X},
       adsurl = {https://ui.adsabs.harvard.edu/abs/2024bheg.book..379D},
      note =  {(updated version:  arXiv:2310.16896)}
}

@article{liao_discovery_2020,
	title = {Discovery of a {Candidate} {Binary} {Supermassive} {Black} {Hole} in a {Periodic} {Quasar} from {Circumbinary} {Accretion} {Variability}},
	copyright = {https://academic.oup.com/journals/pages/open\_access/funder\_policies/chorus/standard\_publication\_model},
	issn = {0035-8711, 1365-2966},
	url = {https://academic.oup.com/mnras/advance-article/doi/10.1093/mnras/staa3055/5920624},
	doi = {10.1093/mnras/staa3055},
	language = {en},
	urldate = {2025-05-23},
	journal = {MNRAS},
	author = {Liao, Wei-Ting and Chen, Yu-Ching and Liu, Xin and Holgado, A Miguel and Guo, Hengxiao and Gruendl, Robert and Morganson, Eric and Shen, Yue and Davis, Tamara and Kessler, Richard and Martini, Paul and McMahon, Richard G and Allam, Sahar and Annis, James and Avila, Santiago and Banerji, Manda and Bechtol, Keith and Bertin, Emmanuel and Brooks, David and Buckley-Geer, Elizabeth and Rosell, Aurelio Carnero and Kind, Matias Carrasco and Carretero, Jorge and Castander, Francisco Javier and Cunha, Carlos and D’Andrea, Chris and Da Costa, Luiz and Davis, Christopher and De Vicente, Juan and Desai, Shantanu and Diehl, H Thomas and Doel, Peter and Eifler, Tim and Evrard, August and Flaugher, Brenna and Fosalba, Pablo and Frieman, Josh and Garcia-Bellido, Juan and Gaztanaga, Enrique and Glazebrook, Karl and Gruen, Daniel and Gschwend, Julia and Gutierrez, Gaston and Hartley, Will and Hollowood, Devon L and Honscheid, Klaus and Hoyle, Ben and James, David and Krause, Elisabeth and Kuehn, Kyler and Lima, Marcos and Maia, Marcio and Marshall, Jennifer and Menanteau, Felipe and Miquel, Ramon and Malagń, Andrés Plazas and Roodman, Aaron and Sanchez, Eusebio and Scarpine, Vic and Schubnell, Michael and Serrano, Santiago and Smith, Mathew and Smith, R Chris and Soares-Santos, Marcelle and Sobreira, Flavia and Suchyta, Eric and Swanson, Molly and Tarle, Gregory and Vikram, Vinu and Walker, Alistair},
	month = oct,
	year = {2020},
	note = {Publisher: Oxford University Press (OUP)}
}

@ARTICLE{gurvits_milliarcsecond_2025,
       author = {{Gurvits}, L.~I. and {Polnarev}, A.~G. and {Frey}, S. and {Titov}, O. and {Osetrova}, A.~A. and {Fan}, X. and {Melnikov}, A.},
        title = "{Milliarcsecond astrometric oscillations in active galactic nuclei as a precursor of multi-messenger gravitational wave events}",
      journal = {\aap},
         year = 2025,
        month = aug,
       volume = {700},
          eid = {A168},
        pages = {A168},
          doi = {10.1051/0004-6361/202450856},
archivePrefix = {arXiv},
       eprint = {2507.02146},
 primaryClass = {astro-ph.CO},
       adsurl = {https://ui.adsabs.harvard.edu/abs/2025A&A...700A.168G}
}

@article{ayzenberg_fundamental_2025,
	title = {Fundamental physics opportunities with future ground-based mm/sub-mm {VLBI} arrays},
	volume = {28},
	copyright = {https://creativecommons.org/licenses/by/4.0},
	issn = {1433-8351},
	url = {https://link.springer.com/10.1007/s41114-025-00057-0},
	doi = {10.1007/s41114-025-00057-0},
	language = {en},
	number = {1},
	urldate = {2025-07-09},
	journal = {LRR},
	author = {Ayzenberg, Dimitry and Blackburn, Lindy and Brito, Richard and Britzen, Silke and Broderick, Avery E. and Carballo-Rubio, Raúl and Cardoso, Vitor and Chael, Andrew and Chatterjee, Koushik and Chen, Yifan and Cunha, Pedro V. P. and Davoudiasl, Hooman and Denton, Peter B. and Doeleman, Sheperd S. and Eichhorn, Astrid and Eubanks, Marshall and Fang, Yun and Foschi, Arianna and Fromm, Christian M. and Galison, Peter and Ghosh, Sushant G. and Gold, Roman and Gurvits, Leonid I. and Hadar, Shahar and Held, Aaron and Houston, Janice and Hu, Yichao and Johnson, Michael D. and Kocherlakota, Prashant and Natarajan, Priyamvada and Olivares, Héctor and Palumbo, Daniel and Pesce, Dominic W. and Rajendran, Surjeet and Roy, Rittick and {Saurabh} and Shao, Lijing and Tahura, Shammi and Tamar, Aditya and Tiede, Paul and Vincent, Frédéric H. and Visinelli, Luca and Wang, Zhiren and Wielgus, Maciek and Xue, Xiao and Yakut, Kadri and Yang, Huan and Younsi, Ziri},
	month = jun,
	year = {2025},
	note = {Publisher: Springer Science and Business Media LLC},
}

@article{ramakrishnan_event_2023,
	title = {Event {Horizon} and {Environs} ({ETHER}): {A} {Curated} {Database} for {EHT} and {ngEHT} {Targets} and {Science}},
	volume = {11},
	copyright = {https://creativecommons.org/licenses/by/4.0/},
	issn = {2075-4434},
	shorttitle = {Event {Horizon} and {Environs} ({ETHER})},
	url = {https://www.mdpi.com/2075-4434/11/1/15},
	doi = {10.3390/galaxies11010015},
	language = {en},
	number = {1},
	urldate = {2025-09-25},
	journal = {Galax},
	author = {Ramakrishnan, Venkatessh and Nagar, Neil and Arratia, Vicente and Hernández-Yévenes, Joaquín and Pesce, Dominic W. and Nair, Dhanya G. and Bandyopadhyay, Bidisha and Medina-Porcile, Catalina and Krichbaum, Thomas P. and Doeleman, Sheperd and Ricarte, Angelo and Fish, Vincent L. and Blackburn, Lindy and Falcke, Heino and Bower, Geoffrey and Natarajan, Priyamvada},
	month = jan,
	year = {2023},
	pages = {15}
}

@article{kardashev_radioastron-telescope_2013,
	title = {“{RadioAstron}”-{A} telescope with a size of 300 000 km: {Main} parameters and first observational results},
	volume = {57},
	issn = {1063-7729, 1562-6881},
	shorttitle = {“{RadioAstron}”-{A} telescope with a size of 300 000 km},
	url = {http://link.springer.com/10.1134/S1063772913030025},
	doi = {10.1134/S1063772913030025},
	language = {en},
	number = {3},
	urldate = {2022-11-19},
	journal = {ARep},
	author = {Kardashev, N. S. and Khartov, V. V. and Abramov, V. V. and Avdeev, V. Yu. and Alakoz, A. V. and Aleksandrov, Yu. A. and Ananthakrishnan, S. and Andreyanov, V. V. and Andrianov, A. S. and Antonov, N. M. and Artyukhov, M. I. and Arkhipov, M. Yu. and Baan, W. and Babakin, N. G. and Babyshkin, V. E. and Bartel’, N. and Belousov, K. G. and Belyaev, A. A. and Berulis, J. J. and Burke, B. F. and Biryukov, A. V. and Bubnov, A. E. and Burgin, M. S. and Busca, G. and Bykadorov, A. A. and Bychkova, V. S. and Vasil’kov, V. I. and Wellington, K. J. and Vinogradov, I. S. and Wietfeldt, R. and Voitsik, P. A. and Gvamichava, A. S. and Girin, I. A. and Gurvits, L. I. and Dagkesamanskii, R. D. and D’Addario, L. and Giovannini, G. and Jauncey, D. L. and Dewdney, P. E. and D’yakov, A. A. and Zharov, V. E. and Zhuravlev, V. I. and Zaslavskii, G. S. and Zakhvatkin, M. V. and Zinov’ev, A. N. and Ilinen, Yu. and Ipatov, A. V. and Kanevskii, B. Z. and Knorin, I. A. and Casse, J. L. and Kellermann, K. I. and Kovalev, Yu. A. and Kovalev, Yu. Yu. and Kovalenko, A. V. and Kogan, B. L. and Komaev, R. V. and Konovalenko, A. A. and Kopelyanskii, G. D. and Korneev, Yu. A. and Kostenko, V. I. and Kotik, A. N. and Kreisman, B. B. and Kukushkin, A. Yu. and Kulishenko, V. F. and Cooper, D. N. and Kut’kin, A. M. and Cannon, W. H. and Larionov, M. G. and Lisakov, M. M. and Litvinenko, L. N. and Likhachev, S. F. and Likhacheva, L. N. and Lobanov, A. P. and Logvinenko, S. V. and Langston, G. and McCracken, K. and Medvedev, S. Yu. and Melekhin, M. V. and Menderov, A. V. and Murphy, D. W. and Mizyakina, T. A. and Mozgovoi, Yu. V. and Nikolaev, N. Ya. and Novikov, B. S. and Novikov, I. D. and Oreshko, V. V. and Pavlenko, Yu. K. and Pashchenko, I. N. and Ponomarev, Yu. N. and Popov, M. V. and Pravin-Kumar, A. and Preston, R. A. and Pyshnov, V. N. and Rakhimov, I. A. and Rozhkov, V. M. and Romney, J. D. and Rocha, P. and Rudakov, V. A. and Räisänen, A. and Sazankov, S. V. and Sakharov, B. A. and Semenov, S. K. and Serebrennikov, V. A. and Schilizzi, R. T. and Skulachev, D. P. and Slysh, V. I. and Smirnov, A. I. and Smith, J. G. and Soglasnov, V. A. and Sokolovskii, K. V. and Sondaar, L. H. and Stepan’yants, V. A. and Turygin, M. S. and Turygin, S. Yu. and Tuchin, A. G. and Urpo, S. and Fedorchuk, S. D. and Finkel’shtein, A. M. and Fomalont, E. B. and Fejes, I. and Fomina, A. N. and Khapin, Yu. B. and Tsarevskii, G. S. and Zensus, J. A. and Chuprikov, A. A. and Shatskaya, M. V. and Shapirovskaya, N. Ya. and Sheikhet, A. I. and Shirshakov, A. E. and Schmidt, A. and Shnyreva, L. A. and Shpilevskii, V. V. and Ekers, R. D. and Yakimov, V. E.},
	month = mar,
	year = {2013},
	pages = {153--194}
}

@article{the_astropy_collaboration_astropy_2013,
	title = {Astropy: {A} community {Python} package for astronomy},
	volume = {558},
	issn = {0004-6361, 1432-0746},
	shorttitle = {Astropy},
	url = {http://www.aanda.org/10.1051/0004-6361/201322068},
	doi = {10.1051/0004-6361/201322068},
	urldate = {2026-03-03},
	journal = {A\&A},
	author = {{The Astropy Collaboration} and Robitaille, Thomas P. and Tollerud, Erik J. and Greenfield, Perry and Droettboom, Michael and Bray, Erik and Aldcroft, Tom and Davis, Matt and Ginsburg, Adam and Price-Whelan, Adrian M. and Kerzendorf, Wolfgang E. and Conley, Alexander and Crighton, Neil and Barbary, Kyle and Muna, Demitri and Ferguson, Henry and Grollier, Frédéric and Parikh, Madhura M. and Nair, Prasanth H. and Günther, Hans M. and Deil, Christoph and Woillez, Julien and Conseil, Simon and Kramer, Roban and Turner, James E. H. and Singer, Leo and Fox, Ryan and Weaver, Benjamin A. and Zabalza, Victor and Edwards, Zachary I. and Azalee Bostroem, K. and Burke, D. J. and Casey, Andrew R. and Crawford, Steven M. and Dencheva, Nadia and Ely, Justin and Jenness, Tim and Labrie, Kathleen and Lim, Pey Lian and Pierfederici, Francesco and Pontzen, Andrew and Ptak, Andy and Refsdal, Brian and Servillat, Mathieu and Streicher, Ole},
	month = oct,
	year = {2013},
	pages = {A33}
}

@article{the_astropy_collaboration_astropy_2018,
	title = {The {Astropy} {Project}: {Building} an {Open}-science {Project} and {Status} of the v2.0 {Core} {Package}$^{\textrm{*}}$},
	volume = {156},
	issn = {0004-6256, 1538-3881},
	shorttitle = {The {Astropy} {Project}},
	url = {https://iopscience.iop.org/article/10.3847/1538-3881/aabc4f},
	doi = {10.3847/1538-3881/aabc4f},
	number = {3},
	urldate = {2026-03-03},
	journal = {AJ},
	author = {{The Astropy Collaboration} and Price-Whelan, A. M. and Sipőcz, B. M. and Günther, H. M. and Lim, P. L. and Crawford, S. M. and Conseil, S. and Shupe, D. L. and Craig, M. W. and Dencheva, N. and Ginsburg, A. and VanderPlas, J. T. and Bradley, L. D. and Pérez-Suárez, D. and De Val-Borro, M. and {(Primary Paper Contributors)} and Aldcroft, T. L. and Cruz, K. L. and Robitaille, T. P. and Tollerud, E. J. and {(Astropy Coordination Committee)} and Ardelean, C. and Babej, T. and Bach, Y. P. and Bachetti, M. and Bakanov, A. V. and Bamford, S. P. and Barentsen, G. and Barmby, P. and Baumbach, A. and Berry, K. L. and Biscani, F. and Boquien, M. and Bostroem, K. A. and Bouma, L. G. and Brammer, G. B. and Bray, E. M. and Breytenbach, H. and Buddelmeijer, H. and Burke, D. J. and Calderone, G. and Rodríguez, J. L. Cano and Cara, M. and Cardoso, J. V. M. and Cheedella, S. and Copin, Y. and Corrales, L. and Crichton, D. and D’Avella, D. and Deil, C. and Depagne, É. and Dietrich, J. P. and Donath, A. and Droettboom, M. and Earl, N. and Erben, T. and Fabbro, S. and Ferreira, L. A. and Finethy, T. and Fox, R. T. and Garrison, L. H. and Gibbons, S. L. J. and Goldstein, D. A. and Gommers, R. and Greco, J. P. and Greenfield, P. and Groener, A. M. and Grollier, F. and Hagen, A. and Hirst, P. and Homeier, D. and Horton, A. J. and Hosseinzadeh, G. and Hu, L. and Hunkeler, J. S. and Ivezić, Ž. and Jain, A. and Jenness, T. and Kanarek, G. and Kendrew, S. and Kern, N. S. and Kerzendorf, W. E. and Khvalko, A. and King, J. and Kirkby, D. and Kulkarni, A. M. and Kumar, A. and Lee, A. and Lenz, D. and Littlefair, S. P. and Ma, Z. and Macleod, D. M. and Mastropietro, M. and McCully, C. and Montagnac, S. and Morris, B. M. and Mueller, M. and Mumford, S. J. and Muna, D. and Murphy, N. A. and Nelson, S. and Nguyen, G. H. and Ninan, J. P. and Nöthe, M. and Ogaz, S. and Oh, S. and Parejko, J. K. and Parley, N. and Pascual, S. and Patil, R. and Patil, A. A. and Plunkett, A. L. and Prochaska, J. X. and Rastogi, T. and Janga, V. Reddy and Sabater, J. and Sakurikar, P. and Seifert, M. and Sherbert, L. E. and Sherwood-Taylor, H. and Shih, A. Y. and Sick, J. and Silbiger, M. T. and Singanamalla, S. and Singer, L. P. and Sladen, P. H. and Sooley, K. A. and Sornarajah, S. and Streicher, O. and Teuben, P. and Thomas, S. W. and Tremblay, G. R. and Turner, J. E. H. and Terrón, V. and Kerkwijk, M. H. Van and De La Vega, A. and Watkins, L. L. and Weaver, B. A. and Whitmore, J. B. and Woillez, J. and Zabalza, V. and {(Astropy Contributors)}},
	month = sep,
	year = {2018},
	pages = {123}
}

@article{the_astropy_collaboration_astropy_2022,
	title = {The {Astropy} {Project}: {Sustaining} and {Growing} a {Community}-oriented {Open}-source {Project} and the {Latest} {Major} {Release} (v5.0) of the {Core} {Package}*},
	volume = {935},
	issn = {0004-637X, 1538-4357},
	shorttitle = {The {Astropy} {Project}},
	url = {https://iopscience.iop.org/article/10.3847/1538-4357/ac7c74},
	doi = {10.3847/1538-4357/ac7c74},
	number = {2},
	urldate = {2026-03-03},
	journal = {ApJ},
	author = {{The Astropy Collaboration} and Price-Whelan, Adrian M. and Lim, Pey Lian and Earl, Nicholas and Starkman, Nathaniel and Bradley, Larry and Shupe, David L. and Patil, Aarya A. and Corrales, Lia and Brasseur, C. E. and Nöthe, Maximilian and Donath, Axel and Tollerud, Erik and Morris, Brett M. and Ginsburg, Adam and Vaher, Eero and Weaver, Benjamin A. and Tocknell, James and Jamieson, William and Van Kerkwijk, Marten H. and Robitaille, Thomas P. and Merry, Bruce and Bachetti, Matteo and Günther, H. Moritz and {Paper Authors} and Aldcroft, Thomas L. and Alvarado-Montes, Jaime A. and Archibald, Anne M. and Bódi, Attila and Bapat, Shreyas and Barentsen, Geert and Bazán, Juanjo and Biswas, Manish and Boquien, Médéric and Burke, D. J. and Cara, Daria and Cara, Mihai and Conroy, Kyle E and Conseil, Simon and Craig, Matthew W. and Cross, Robert M. and Cruz, Kelle L. and D’Eugenio, Francesco and Dencheva, Nadia and Devillepoix, Hadrien A. R. and Dietrich, Jörg P. and Eigenbrot, Arthur Davis and Erben, Thomas and Ferreira, Leonardo and Foreman-Mackey, Daniel and Fox, Ryan and Freij, Nabil and Garg, Suyog and Geda, Robel and Glattly, Lauren and Gondhalekar, Yash and Gordon, Karl D. and Grant, David and Greenfield, Perry and Groener, Austen M. and Guest, Steve and Gurovich, Sebastian and Handberg, Rasmus and Hart, Akeem and Hatfield-Dodds, Zac and Homeier, Derek and Hosseinzadeh, Griffin and Jenness, Tim and Jones, Craig K. and Joseph, Prajwel and Kalmbach, J. Bryce and Karamehmetoglu, Emir and Kałuszyński, Mikołaj and Kelley, Michael S. P. and Kern, Nicholas and Kerzendorf, Wolfgang E. and Koch, Eric W. and Kulumani, Shankar and Lee, Antony and Ly, Chun and Ma, Zhiyuan and MacBride, Conor and Maljaars, Jakob M. and Muna, Demitri and Murphy, N. A. and Norman, Henrik and O’Steen, Richard and Oman, Kyle A. and Pacifici, Camilla and Pascual, Sergio and Pascual-Granado, J. and Patil, Rohit R. and Perren, Gabriel I and Pickering, Timothy E. and Rastogi, Tanuj and Roulston, Benjamin R. and Ryan, Daniel F and Rykoff, Eli S. and Sabater, Jose and Sakurikar, Parikshit and Salgado, Jesús and Sanghi, Aniket and Saunders, Nicholas and Savchenko, Volodymyr and Schwardt, Ludwig and Seifert-Eckert, Michael and Shih, Albert Y. and Jain, Anany Shrey and Shukla, Gyanendra and Sick, Jonathan and Simpson, Chris and Singanamalla, Sudheesh and Singer, Leo P. and Singhal, Jaladh and Sinha, Manodeep and Sipőcz, Brigitta M. and Spitler, Lee R. and Stansby, David and Streicher, Ole and Šumak, Jani and Swinbank, John D. and Taranu, Dan S. and Tewary, Nikita and Tremblay, Grant R. and Val-Borro, Miguel De and Van Kooten, Samuel J. and Vasović, Zlatan and Verma, Shresth and De Miranda Cardoso, José Vinícius and Williams, Peter K. G. and Wilson, Tom J. and Winkel, Benjamin and Wood-Vasey, W. M. and Xue, Rui and Yoachim, Peter and Zhang, Chen and Zonca, Andrea and {Astropy Project Contributors}},
	month = aug,
	year = {2022},
	pages = {167}
}

@article{zhang_accessing_2025,
	title = {Accessing a {New} {Population} of {Supermassive} {Black} {Holes} with {Extensions} to the {Event} {Horizon} {Telescope}},
	volume = {985},
	issn = {0004-637X, 1538-4357},
	url = {https://iopscience.iop.org/article/10.3847/1538-4357/adbd45},
	doi = {10.3847/1538-4357/adbd45},	number = {1},
	urldate = {2026-03-03},
	journal = {ApJ},
	author = {Zhang, Xinyue Alice and Ricarte, Angelo and Pesce, Dominic W. and Johnson, Michael D. and Nagar, Neil and Narayan, Ramesh and Ramakrishnan, Venkatessh and Doeleman, Sheperd and Palumbo, Daniel C. M.},
	month = may,
	year = {2025},
	pages = {41},
}

@article{dorazio_relativistic_2015,
	title = {Relativistic boost as the cause of periodicity in a massive black-hole binary candidate},
	volume = {525},
	issn = {0028-0836, 1476-4687},
	url = {https://www.nature.com/articles/nature15262},
	doi = {10.1038/nature15262},
	language = {en},
	number = {7569},
	urldate = {2025-09-17},
	journal = {Natur},
	author = {D'Orazio, Daniel J. and Haiman, Zoltán and Schiminovich, David},
	month = sep,
	year = {2015},
	pages = {351--353},
}

@article{farris_binary_2014,
	title = {{BINARY} {BLACK} {HOLE} {ACCRETION} {FROM} {A} {CIRCUMBINARY} {DISK}: {GAS} {DYNAMICS} {INSIDE} {THE} {CENTRAL} {CAVITY}},
	volume = {783},
	copyright = {http://iopscience.iop.org/info/page/text-and-data-mining},
	issn = {0004-637X, 1538-4357},
	shorttitle = {{BINARY} {BLACK} {HOLE} {ACCRETION} {FROM} {A} {CIRCUMBINARY} {DISK}},
	url = {https://iopscience.iop.org/article/10.1088/0004-637X/783/2/134},
	doi = {10.1088/0004-637X/783/2/134},
	number = {2},
	urldate = {2025-05-27},
	journal = {ApJ},
	author = {Farris, Brian D. and Duffell, Paul and MacFadyen, Andrew I. and Haiman, Zoltan},
	month = feb,
	year = {2014},
	pages = {134}
}

@article{planck_collaboration_planck_2020,
	title = {\textit{{Planck}} 2018 results: {VI}. {Cosmological} parameters},
	volume = {641},
	copyright = {https://www.edpsciences.org/en/authors/copyright-and-licensing},
	issn = {0004-6361, 1432-0746},
	shorttitle = {\textit{{Planck}} 2018 results},
	url = {https://www.aanda.org/10.1051/0004-6361/201833910},
	doi = {10.1051/0004-6361/201833910},
	urldate = {2026-05-21},
	journal = {A\&A},
	author = {{Planck Collaboration} and Aghanim, N. and Akrami, Y. and Ashdown, M. and Aumont, J. and Baccigalupi, C. and Ballardini, M. and Banday, A. J. and Barreiro, R. B. and Bartolo, N. and Basak, S. and Battye, R. and Benabed, K. and Bernard, J.-P. and Bersanelli, M. and Bielewicz, P. and Bock, J. J. and Bond, J. R. and Borrill, J. and Bouchet, F. R. and Boulanger, F. and Bucher, M. and Burigana, C. and Butler, R. C. and Calabrese, E. and Cardoso, J.-F. and Carron, J. and Challinor, A. and Chiang, H. C. and Chluba, J. and Colombo, L. P. L. and Combet, C. and Contreras, D. and Crill, B. P. and Cuttaia, F. and De Bernardis, P. and De Zotti, G. and Delabrouille, J. and Delouis, J.-M. and Di Valentino, E. and Diego, J. M. and Doré, O. and Douspis, M. and Ducout, A. and Dupac, X. and Dusini, S. and Efstathiou, G. and Elsner, F. and Enßlin, T. A. and Eriksen, H. K. and Fantaye, Y. and Farhang, M. and Fergusson, J. and Fernandez-Cobos, R. and Finelli, F. and Forastieri, F. and Frailis, M. and Fraisse, A. A. and Franceschi, E. and Frolov, A. and Galeotta, S. and Galli, S. and Ganga, K. and Génova-Santos, R. T. and Gerbino, M. and Ghosh, T. and González-Nuevo, J. and Górski, K. M. and Gratton, S. and Gruppuso, A. and Gudmundsson, J. E. and Hamann, J. and Handley, W. and Hansen, F. K. and Herranz, D. and Hildebrandt, S. R. and Hivon, E. and Huang, Z. and Jaffe, A. H. and Jones, W. C. and Karakci, A. and Keihänen, E. and Keskitalo, R. and Kiiveri, K. and Kim, J. and Kisner, T. S. and Knox, L. and Krachmalnicoff, N. and Kunz, M. and Kurki-Suonio, H. and Lagache, G. and Lamarre, J.-M. and Lasenby, A. and Lattanzi, M. and Lawrence, C. R. and Le Jeune, M. and Lemos, P. and Lesgourgues, J. and Levrier, F. and Lewis, A. and Liguori, M. and Lilje, P. B. and Lilley, M. and Lindholm, V. and López-Caniego, M. and Lubin, P. M. and Ma, Y.-Z. and Macías-Pérez, J. F. and Maggio, G. and Maino, D. and Mandolesi, N. and Mangilli, A. and Marcos-Caballero, A. and Maris, M. and Martin, P. G. and Martinelli, M. and Martínez-González, E. and Matarrese, S. and Mauri, N. and McEwen, J. D. and Meinhold, P. R. and Melchiorri, A. and Mennella, A. and Migliaccio, M. and Millea, M. and Mitra, S. and Miville-Deschênes, M.-A. and Molinari, D. and Montier, L. and Morgante, G. and Moss, A. and Natoli, P. and Nørgaard-Nielsen, H. U. and Pagano, L. and Paoletti, D. and Partridge, B. and Patanchon, G. and Peiris, H. V. and Perrotta, F. and Pettorino, V. and Piacentini, F. and Polastri, L. and Polenta, G. and Puget, J.-L. and Rachen, J. P. and Reinecke, M. and Remazeilles, M. and Renzi, A. and Rocha, G. and Rosset, C. and Roudier, G. and Rubiño-Martín, J. A. and Ruiz-Granados, B. and Salvati, L. and Sandri, M. and Savelainen, M. and Scott, D. and Shellard, E. P. S. and Sirignano, C. and Sirri, G. and Spencer, L. D. and Sunyaev, R. and Suur-Uski, A.-S. and Tauber, J. A. and Tavagnacco, D. and Tenti, M. and Toffolatti, L. and Tomasi, M. and Trombetti, T. and Valenziano, L. and Valiviita, J. and Van Tent, B. and Vibert, L. and Vielva, P. and Villa, F. and Vittorio, N. and Wandelt, B. D. and Wehus, I. K. and White, M. and White, S. D. M. and Zacchei, A. and Zonca, A.},
	month = sep,
	year = {2020},
	pages = {A6},
}

@ARTICLE{Traianou+2025AA-OJ287,
       author = {{Traianou}, E. and {G{\'o}mez}, J.~L. and {Cho}, I. and {Chael}, A. and {Fuentes}, A. and {Myserlis}, I. and {Wielgus}, M. and {Zhao}, G.-Y. and {Lico}, R. and {Moriyama}, K. and {Dey}, L. and {Bruni}, G. and {Dahale}, R. and {Toscano}, T. and {Gurvits}, L.~I. and {Lisakov}, M.~M. and {Kovalev}, Y.~Y. and {Lobanov}, A.~P. and {Pushkarev}, A.~B. and {Sokolovsky}, K.~V.},
        title = "{Revealing a ribbon-like jet in OJ 287 with RadioAstron}",
      journal = {\aap},
         year = 2025,
        month = aug,
       volume = {700},
          eid = {A16},
        pages = {A16},
          doi = {10.1051/0004-6361/202554929},
archivePrefix = {arXiv},
       eprint = {2508.01747},
 primaryClass = {astro-ph.HE},
       adsurl = {https://ui.adsabs.harvard.edu/abs/2025A&A...700A..16T}
}

@ARTICLE{Hong+2025SCPMA-LOVEX,
       author = {{Hong}, Xiaoyu and {Wu}, Weiren and {Liu}, Qinghui and {Yu}, Dengyun and {Wang}, Chi and {Shuai}, Tao and {Zhong}, Weiye and {Zhu}, Renjie and {Xie}, Yonghui and {Zhang}, Lihua and {Xiong}, Liang and {Tang}, Yuhua and {Zou}, Yongliao and {Li}, Haitao and {Wang}, Guangli and {Xie}, Jianfeng and {Xue}, Changbin and {Geng}, Hao and {Zhang}, Juan and {Wu}, Xiaojing and {Huang}, Yong and {Zheng}, Weimin and {Liu}, Lei and {Wu}, Fang and {Zhang}, Xiuzhong and {An}, Tao and {Yang}, Xiaolong and {Tong}, Fengxian and {Gurvits}, Leonid I. and {Zheng}, Yong and {Gu}, Minfeng and {Ma}, Xiaofei and {Li}, Liang and {Li}, Peijia and {Zhao}, Shanshan and {Rui}, Ping and {Chen}, Luojing and {Chen}, Guohui and {Li}, Ke and {Zhang}, Chao and {Liu}, Yuanqi and {Jiang}, Yongchen and {Wang}, Jinqing and {Wang}, Wenbin and {Sun}, Yan and {Hao}, Longfei and {Cui}, Lang and {Jiang}, Dongrong and {Qian}, Zhihan and {Ye}, Shuhua},
        title = "{Lunar Orbital VLBI Experiment: Motivation, scientific purposes and status}",
      journal = {SCPMA},
         year = 2025,
        month = nov,
       volume = {69},
       number = {1},
          eid = {219511},
        pages = {219511},
          doi = {10.1007/s11433-025-2751-2},
archivePrefix = {arXiv},
       eprint = {2507.16317},
 primaryClass = {astro-ph.IM},
       adsurl = {https://ui.adsabs.harvard.edu/abs/2025SCPMA..6919511H}
}

@ARTICLE{Levy+1986Sci,
       author = {{Levy}, G.~S. and {Linfield}, R.~P. and {Ulvestad}, J.~S. and {Edwards}, C.~D. and {Jordan}, J.~F. and {di Nardo}, S.~J. and {Christensen}, C.~S. and {Preston}, R.~A. and {Skjerve}, L.~J. and {Stavert}, L.~R. and {Burke}, B.~F. and {Whitney}, A.~R. and {Cappallo}, R.~J. and {Rogers}, A.~E.~E. and {Blaney}, K.~B. and {Maher}, M.~J. and {Ottenhoff}, C.~H. and {Jauncey}, D.~L. and {Peters}, W.~L. and {Nishimura}, T. and {Hayashi}, T. and {Takano}, T. and {Yamada}, T. and {Hirabayashi}, H. and {Morimoto}, M. and {Inoue}, M. and {Shiomi}, T. and {Kawaguchi}, N. and {Kunimori}, H.},
        title = "{Very Long Baseline Interferometric Observations made with an Orbiting Radio Telescope}",
      journal = {Sci},
         year = 1986,
        month = oct,
       volume = {234},
       number = {4773},
        pages = {187-189},
          doi = {10.1126/science.234.4773.187},
       adsurl = {https://ui.adsabs.harvard.edu/abs/1986Sci...234..187L}
}

@ARTICLE{Hirabayashi+1998Sci,
       author = {{Hirabayashi}, H. and {Hirosawa}, H. and {Kobayashi}, H. and {Murata}, Y. and {Edwards}, P.~G. and {Fomalont}, E.~B. and {Fujisawa}, K. and {Ichikawa}, T. and {Kii}, T. and {Lovell}, J.~E.~J. and {Moellenbrock}, G.~A. and {Okayasu}, R. and {Inoue}, M. and {Kawaguchi}, N. and {Kameno}, S. and {Shibata}, K.~M. and {Asaki}, Y. and {Bushimata}, T. and {Enome}, S. and {Horiuchi}, S. and {Miyaji}, T. and {Umemoto}, T. and {Migenes}, V. and {Wajima}, K. and {Nakajima}, J. and {Morimoto}, M. and {Ellis}, J. and {Meier}, D.~L. and {Murphy}, D.~W. and {Preston}, R.~A. and {Smith}, J.~G. and {Tingay}, S.~J. and {Traub}, D.~L. and {Wietfeldt}, R.~D. and {Benson}, J.~M. and {Claussen}, M.~J. and {Flatters}, C. and {Romney}, J.~D. and {Ulvestad}, J.~S. and {D'Addario}, L.~R. and {Langston}, G.~I. and {Minter}, A.~H. and {Carlson}, B.~R. and {Dewdney}, P.~E. and {Jauncey}, D.~L. and {Reynolds}, J.~E. and {Taylor}, A.~R. and {McCulloch}, P.~M. and {Cannon}, W.~H. and {Gurvits}, L.~I. and {Mioduszewski}, A.~J. and {Schilizzi}, R.~T. and {Booth}, R.~S.},
        title = "{Overview and Initial Results of the Very Long Baseline Interferometry Space Observatory Programme}",
      journal = {Sci},
         year = 1998,
        month = sep,
       volume = {281},
        pages = {1825},
          doi = {10.1126/science.281.5384.1825},
       adsurl = {https://ui.adsabs.harvard.edu/abs/1998Sci...281.1825H}
}

@article{the_event_horizon_telescope_collaboration_first_2019,
	title = {First {M87} {Event} {Horizon} {Telescope} {Results}. {I}. {The} {Shadow} of the {Supermassive} {Black} {Hole}},
	volume = {875},
	copyright = {http://creativecommons.org/licenses/by/3.0/},
	issn = {2041-8205, 2041-8213},
	url = {https://iopscience.iop.org/article/10.3847/2041-8213/ab0ec7},
	doi = {10.3847/2041-8213/ab0ec7},
	number = {1},
	urldate = {2025-05-23},
	journal = {ApJL},
	author = {{EHT Collaboration} and Akiyama, Kazunori and Alberdi, Antxon and Alef, Walter and Asada, Keiichi and Azulay, Rebecca and Baczko, Anne-Kathrin and Ball, David and Baloković, Mislav and Barrett, John and Bintley, Dan and Blackburn, Lindy and Boland, Wilfred and Bouman, Katherine L. and Bower, Geoffrey C. and Bremer, Michael and Brinkerink, Christiaan D. and Brissenden, Roger and Britzen, Silke and Broderick, Avery E. and Broguiere, Dominique and Bronzwaer, Thomas and Byun, Do-Young and Carlstrom, John E. and Chael, Andrew and Chan, Chi-kwan and Chatterjee, Shami and Chatterjee, Koushik and Chen, Ming-Tang and Chen 陈, Yongjun 永军 and Cho, Ilje and Christian, Pierre and Conway, John E. and Cordes, James M. and Crew, Geoffrey B. and Cui, Yuzhu and Davelaar, Jordy and Laurentis, Mariafelicia De and Deane, Roger and Dempsey, Jessica and Desvignes, Gregory and Dexter, Jason and Doeleman, Sheperd S. and Eatough, Ralph P. and Falcke, Heino and Fish, Vincent L. and Fomalont, Ed and Fraga-Encinas, Raquel and Freeman, William T. and Friberg, Per and Fromm, Christian M. and Gómez, José L. and Galison, Peter and Gammie, Charles F. and García, Roberto and Gentaz, Olivier and Georgiev, Boris and Goddi, Ciriaco and Gold, Roman and Gu 顾, Minfeng 敏峰 and Gurwell, Mark and Hada, Kazuhiro and Hecht, Michael H. and Hesper, Ronald and Ho 何, Luis C. 子山 and Ho, Paul and Honma, Mareki and Huang, Chih-Wei L. and Huang 黄, Lei 磊 and Hughes, David H. and Ikeda, Shiro and Inoue, Makoto and Issaoun, Sara and James, David J. and Jannuzi, Buell T. and Janssen, Michael and Jeter, Britton and Jiang 江, Wu 悟 and Johnson, Michael D. and Jorstad, Svetlana and Jung, Taehyun and Karami, Mansour and Karuppusamy, Ramesh and Kawashima, Tomohisa and Keating, Garrett K. and Kettenis, Mark and Kim, Jae-Young and Kim, Junhan and Kim, Jongsoo and Kino, Motoki and Koay, Jun Yi and Koch, Patrick M. and Koyama, Shoko and Kramer, Michael and Kramer, Carsten and Krichbaum, Thomas P. and Kuo, Cheng-Yu and Lauer, Tod R. and Lee, Sang-Sung and Li 李, Yan-Rong 彦荣 and Li 李, Zhiyuan 志远 and Lindqvist, Michael and Liu, Kuo and Liuzzo, Elisabetta and Lo, Wen-Ping and Lobanov, Andrei P. and Loinard, Laurent and Lonsdale, Colin and Lu 路, Ru-Sen 如森 and MacDonald, Nicholas R. and Mao 毛, Jirong 基荣 and Markoff, Sera and Marrone, Daniel P. and Marscher, Alan P. and Martí-Vidal, Iván and Matsushita, Satoki and Matthews, Lynn D. and Medeiros, Lia and Menten, Karl M. and Mizuno, Yosuke and Mizuno, Izumi and Moran, James M. and Moriyama, Kotaro and Moscibrodzka, Monika and Müller, Cornelia and Nagai, Hiroshi and Nagar, Neil M. and Nakamura, Masanori and Narayan, Ramesh and Narayanan, Gopal and Natarajan, Iniyan and Neri, Roberto and Ni, Chunchong and Noutsos, Aristeidis and Okino, Hiroki and Olivares, Héctor and Ortiz-León, Gisela N. and Oyama, Tomoaki and Özel, Feryal and Palumbo, Daniel C. M. and Patel, Nimesh and Pen, Ue-Li and Pesce, Dominic W. and Piétu, Vincent and Plambeck, Richard and PopStefanija, Aleksandar and Porth, Oliver and Prather, Ben and Preciado-López, Jorge A. and Psaltis, Dimitrios and Pu, Hung-Yi and Ramakrishnan, Venkatessh and Rao, Ramprasad and Rawlings, Mark G. and Raymond, Alexander W. and Rezzolla, Luciano and Ripperda, Bart and Roelofs, Freek and Rogers, Alan and Ros, Eduardo and Rose, Mel and Roshanineshat, Arash and Rottmann, Helge and Roy, Alan L. and Ruszczyk, Chet and Ryan, Benjamin R. and Rygl, Kazi L. J. and Sánchez, Salvador and Sánchez-Arguelles, David and Sasada, Mahito and Savolainen, Tuomas and Schloerb, F. Peter and Schuster, Karl-Friedrich and Shao, Lijing and Shen 沈, Zhiqiang 志强 and Small, Des and Sohn, Bong Won and SooHoo, Jason and Tazaki, Fumie and Tiede, Paul and Tilanus, Remo P. J. and Titus, Michael and Toma, Kenji and Torne, Pablo and Trent, Tyler and Trippe, Sascha and Tsuda, Shuichiro and Bemmel, Ilse Van and Van Langevelde, Huib Jan and Van Rossum, Daniel R. and Wagner, Jan and Wardle, John and Weintroub, Jonathan and Wex, Norbert and Wharton, Robert and Wielgus, Maciek and Wong, George N. and Wu 吴, Qingwen 庆文 and Young, Ken and Young, André and Younsi, Ziri and Yuan 袁, Feng 峰 and Yuan 袁, Ye-Fei 业飞 and Zensus, J. Anton and Zhao, Guangyao and Zhao, Shan-Shan and Zhu, Ziyan and Algaba, Juan-Carlos and Allardi, Alexander and Amestica, Rodrigo and Anczarski, Jadyn and Bach, Uwe and Baganoff, Frederick K. and Beaudoin, Christopher and Benson, Bradford A. and Berthold, Ryan and Blanchard, Jay M. and Blundell, Ray and Bustamente, Sandra and Cappallo, Roger and Castillo-Domínguez, Edgar and Chang, Chih-Cheng and Chang, Shu-Hao and Chang, Song-Chu and Chen, Chung-Chen and Chilson, Ryan and Chuter, Tim C. and Rosado, Rodrigo Córdova and Coulson, Iain M. and Crawford, Thomas M. and Crowley, Joseph and David, John and Derome, Mark and Dexter, Matthew and Dornbusch, Sven and Dudevoir, Kevin A. and Dzib, Sergio A. and Eckart, Andreas and Eckert, Chris and Erickson, Neal R. and Everett, Wendeline B. and Faber, Aaron and Farah, Joseph R. and Fath, Vernon and Folkers, Thomas W. and Forbes, David C. and Freund, Robert and Gómez-Ruiz, Arturo I. and Gale, David M. and Gao, Feng and Geertsema, Gertie and Graham, David A. and Greer, Christopher H. and Grosslein, Ronald and Gueth, Frédéric and Haggard, Daryl and Halverson, Nils W. and Han, Chih-Chiang and Han, Kuo-Chang and Hao, Jinchi and Hasegawa, Yutaka and Henning, Jason W. and Hernández-Gómez, Antonio and Herrero-Illana, Rubén and Heyminck, Stefan and Hirota, Akihiko and Hoge, James and Huang, Yau-De and Impellizzeri, C. M. Violette and Jiang, Homin and Kamble, Atish and Keisler, Ryan and Kimura, Kimihiro and Kono, Yusuke and Kubo, Derek and Kuroda, John and Lacasse, Richard and Laing, Robert A. and Leitch, Erik M. and Li, Chao-Te and Lin, Lupin C.-C. and Liu, Ching-Tang and Liu, Kuan-Yu and Lu, Li-Ming and Marson, Ralph G. and Martin-Cocher, Pierre L. and Massingill, Kyle D. and Matulonis, Callie and McColl, Martin P. and McWhirter, Stephen R. and Messias, Hugo and Meyer-Zhao, Zheng and Michalik, Daniel and Montaña, Alfredo and Montgomerie, William and Mora-Klein, Matias and Muders, Dirk and Nadolski, Andrew and Navarro, Santiago and Neilsen, Joseph and Nguyen, Chi H. and Nishioka, Hiroaki and Norton, Timothy and Nowak, Michael A. and Nystrom, George and Ogawa, Hideo and Oshiro, Peter and Oyama, Tomoaki and Parsons, Harriet and Paine, Scott N. and Peñalver, Juan and Phillips, Neil M. and Poirier, Michael and Pradel, Nicolas and Primiani, Rurik A. and Raffin, Philippe A. and Rahlin, Alexandra S. and Reiland, George and Risacher, Christopher and Ruiz, Ignacio and Sáez-Madaín, Alejandro F. and Sassella, Remi and Schellart, Pim and Shaw, Paul and Silva, Kevin M. and Shiokawa, Hotaka and Smith, David R. and Snow, William and Souccar, Kamal and Sousa, Don and Sridharan, T. K. and Srinivasan, Ranjani and Stahm, William and Stark, Anthony A. and Story, Kyle and Timmer, Sjoerd T. and Vertatschitsch, Laura and Walther, Craig and Wei, Ta-Shun and Whitehorn, Nathan and Whitney, Alan R. and Woody, David P. and Wouterloot, Jan G. A. and Wright, Melvin and Yamaguchi, Paul and Yu, Chen-Yu and Zeballos, Milagros and Zhang, Shuo and Ziurys, Lucy},
	month = apr,
	year = {2019},
	note = {Publisher: American Astronomical Society},
	pages = {L1},
}

@article{event_horizon_telescope_collaboration_first_2022,
	title = {First {Sagittarius} {A}* {Event} {Horizon} {Telescope} {Results}. {I}. {The} {Shadow} of the {Supermassive} {Black} {Hole} in the {Center} of the {Milky} {Way}},
	volume = {930},
	copyright = {http://creativecommons.org/licenses/by/4.0/},
	issn = {2041-8205, 2041-8213},
	url = {https://iopscience.iop.org/article/10.3847/2041-8213/ac6674},
	doi = {10.3847/2041-8213/ac6674},
	number = {2},
	urldate = {2025-05-23},
	journal = {ApJL},
	author = {{EHT Collaboration} and Akiyama, Kazunori and Alberdi, Antxon and Alef, Walter and Algaba, Juan Carlos and Anantua, Richard and Asada, Keiichi and Azulay, Rebecca and Bach, Uwe and Baczko, Anne-Kathrin and Ball, David and Baloković, Mislav and Barrett, John and Bauböck, Michi and Benson, Bradford A. and Bintley, Dan and Blackburn, Lindy and Blundell, Raymond and Bouman, Katherine L. and Bower, Geoffrey C. and Boyce, Hope and Bremer, Michael and Brinkerink, Christiaan D. and Brissenden, Roger and Britzen, Silke and Broderick, Avery E. and Broguiere, Dominique and Bronzwaer, Thomas and Bustamante, Sandra and Byun, Do-Young and Carlstrom, John E. and Ceccobello, Chiara and Chael, Andrew and Chan, Chi-kwan and Chatterjee, Koushik and Chatterjee, Shami and Chen, Ming-Tang and Chen 陈, Yongjun 永军 and Cheng, Xiaopeng and Cho, Ilje and Christian, Pierre and Conroy, Nicholas S. and Conway, John E. and Cordes, James M. and Crawford, Thomas M. and Crew, Geoffrey B. and Cruz-Osorio, Alejandro and Cui 崔, Yuzhu 玉竹 and Davelaar, Jordy and Laurentis, Mariafelicia De and Deane, Roger and Dempsey, Jessica and Desvignes, Gregory and Dexter, Jason and Dhruv, Vedant and Doeleman, Sheperd S. and Dougal, Sean and Dzib, Sergio A. and Eatough, Ralph P. and Emami, Razieh and Falcke, Heino and Farah, Joseph and Fish, Vincent L. and Fomalont, Ed and Ford, H. Alyson and Fraga-Encinas, Raquel and Freeman, William T. and Friberg, Per and Fromm, Christian M. and Fuentes, Antonio and Galison, Peter and Gammie, Charles F. and García, Roberto and Gentaz, Olivier and Georgiev, Boris and Goddi, Ciriaco and Gold, Roman and Gómez-Ruiz, Arturo I. and Gómez, José L. and Gu 顾, Minfeng 敏峰 and Gurwell, Mark and Hada, Kazuhiro and Haggard, Daryl and Haworth, Kari and Hecht, Michael H. and Hesper, Ronald and Heumann, Dirk and Ho 何, Luis C. 子山 and Ho, Paul and Honma, Mareki and Huang, Chih-Wei L. and Huang 黄, Lei 磊 and Hughes, David H. and Ikeda, Shiro and Impellizzeri, C. M. Violette and Inoue, Makoto and Issaoun, Sara and James, David J. and Jannuzi, Buell T. and Janssen, Michael and Jeter, Britton and Jiang 江, Wu 悟 and Jiménez-Rosales, Alejandra and Johnson, Michael D. and Jorstad, Svetlana and Joshi, Abhishek V. and Jung, Taehyun and Karami, Mansour and Karuppusamy, Ramesh and Kawashima, Tomohisa and Keating, Garrett K. and Kettenis, Mark and Kim, Dong-Jin and Kim, Jae-Young and Kim, Jongsoo and Kim, Junhan and Kino, Motoki and Koay, Jun Yi and Kocherlakota, Prashant and Kofuji, Yutaro and Koch, Patrick M. and Koyama, Shoko and Kramer, Carsten and Kramer, Michael and Krichbaum, Thomas P. and Kuo, Cheng-Yu and Bella, Noemi La and Lauer, Tod R. and Lee, Daeyoung and Lee, Sang-Sung and Leung, Po Kin and Levis, Aviad and Li 李, Zhiyuan 志远 and Lico, Rocco and Lindahl, Greg and Lindqvist, Michael and Lisakov, Mikhail and Liu 刘, Jun 俊 and Liu, Kuo and Liuzzo, Elisabetta and Lo, Wen-Ping and Lobanov, Andrei P. and Loinard, Laurent and Lonsdale, Colin J. and Lu 路, Ru-Sen 如森 and Mao 毛, Jirong 基荣 and Marchili, Nicola and Markoff, Sera and Marrone, Daniel P. and Marscher, Alan P. and Martí-Vidal, Iván and Matsushita, Satoki and Matthews, Lynn D. and Medeiros, Lia and Menten, Karl M. and Michalik, Daniel and Mizuno, Izumi and Mizuno, Yosuke and Moran, James M. and Moriyama, Kotaro and Moscibrodzka, Monika and Müller, Cornelia and Mus, Alejandro and Musoke, Gibwa and Myserlis, Ioannis and Nadolski, Andrew and Nagai, Hiroshi and Nagar, Neil M. and Nakamura, Masanori and Narayan, Ramesh and Narayanan, Gopal and Natarajan, Iniyan and Nathanail, Antonios and Fuentes, Santiago Navarro and Neilsen, Joey and Neri, Roberto and Ni, Chunchong and Noutsos, Aristeidis and Nowak, Michael A. and Oh, Junghwan and Okino, Hiroki and Olivares, Héctor and Ortiz-León, Gisela N. and Oyama, Tomoaki and Özel, Feryal and Palumbo, Daniel C. M. and Paraschos, Georgios Filippos and Park, Jongho and Parsons, Harriet and Patel, Nimesh and Pen, Ue-Li and Pesce, Dominic W. and Piétu, Vincent and Plambeck, Richard and PopStefanija, Aleksandar and Porth, Oliver and Pötzl, Felix M. and Prather, Ben and Preciado-López, Jorge A. and Psaltis, Dimitrios and Pu, Hung-Yi and Ramakrishnan, Venkatessh and Rao, Ramprasad and Rawlings, Mark G. and Raymond, Alexander W. and Rezzolla, Luciano and Ricarte, Angelo and Ripperda, Bart and Roelofs, Freek and Rogers, Alan and Ros, Eduardo and Romero-Cañizales, Cristina and Roshanineshat, Arash and Rottmann, Helge and Roy, Alan L. and Ruiz, Ignacio and Ruszczyk, Chet and Rygl, Kazi L. J. and Sánchez, Salvador and Sánchez-Argüelles, David and Sánchez-Portal, Miguel and Sasada, Mahito and Satapathy, Kaushik and Savolainen, Tuomas and Schloerb, F. Peter and Schonfeld, Jonathan and Schuster, Karl-Friedrich and Shao, Lijing and Shen 沈, Zhiqiang 志强 and Small, Des and Sohn, Bong Won and SooHoo, Jason and Souccar, Kamal and Sun 孙, He 赫 and Tazaki, Fumie and Tetarenko, Alexandra J. and Tiede, Paul and Tilanus, Remo P. J. and Titus, Michael and Torne, Pablo and Traianou, Efthalia and Trent, Tyler and Trippe, Sascha and Turk, Matthew and Van Bemmel, Ilse and Van Langevelde, Huib Jan and Van Rossum, Daniel R. and Vos, Jesse and Wagner, Jan and Ward-Thompson, Derek and Wardle, John and Weintroub, Jonathan and Wex, Norbert and Wharton, Robert and Wielgus, Maciek and Wiik, Kaj and Witzel, Gunther and Wondrak, Michael F. and Wong, George N. and Wu 吴, Qingwen 庆文 and Yamaguchi, Paul and Yoon, Doosoo and Young, André and Young, Ken and Younsi, Ziri and Yuan 袁, Feng 峰 and Yuan 袁, Ye-Fei 业飞 and Zensus, J. Anton and Zhang, Shuo and Zhao, Guang-Yao and Zhao 赵, Shan-Shan 杉杉 and Agurto, Claudio and Allardi, Alexander and Amestica, Rodrigo and Araneda, Juan Pablo and Arriagada, Oriel and Berghuis, Jennie L. and Bertarini, Alessandra and Berthold, Ryan and Blanchard, Jay and Brown, Ken and Cárdenas, Mauricio and Cantzler, Michael and Caro, Patricio and Castillo-Domínguez, Edgar and Chan, Tin Lok and Chang, Chih-Cheng and Chang, Dominic O. and Chang, Shu-Hao and Chang, Song-Chu and Chen, Chung-Chen and Chilson, Ryan and Chuter, Tim C. and Ciechanowicz, Miroslaw and Colin-Beltran, Edgar and Coulson, Iain M. and Crowley, Joseph and Degenaar, Nathalie and Dornbusch, Sven and Durán, Carlos A. and Everett, Wendeline B. and Faber, Aaron and Forster, Karl and Fuchs, Miriam M. and Gale, David M. and Geertsema, Gertie and González, Edouard and Graham, Dave and Gueth, Frédéric and Halverson, Nils W. and Han, Chih-Chiang and Han, Kuo-Chang and Hasegawa, Yutaka and Hernández-Rebollar, José Luis and Herrera, Cristian and Herrero-Illana, Ruben and Heyminck, Stefan and Hirota, Akihiko and Hoge, James and Hostler Schimpf, Shelbi R. and Howie, Ryan E. and Huang, Yau-De and Jiang, Homin and Jinchi, Hao and John, David and Kimura, Kimihiro and Klein, Thomas and Kubo, Derek and Kuroda, John and Kwon, Caleb and Lacasse, Richard and Laing, Robert and Leitch, Erik M. and Li, Chao-Te and Liu, Ching-Tang and Liu, Kuan-Yu and Lin, Lupin C.-C. and Lu, Li-Ming and Mac-Auliffe, Felipe and Martin-Cocher, Pierre and Matulonis, Callie and Maute, John K. and Messias, Hugo and Meyer-Zhao, Zheng and Montaña, Alfredo and Montenegro-Montes, Francisco and Montgomerie, William and Moreno Nolasco, Marcos Emir and Muders, Dirk and Nishioka, Hiroaki and Norton, Timothy J. and Nystrom, George and Ogawa, Hideo and Olivares, Rodrigo and Oshiro, Peter and Pérez-Beaupuits, Juan Pablo and Parra, Rodrigo and Phillips, Neil M. and Poirier, Michael and Pradel, Nicolas and Qiu, Richard and Raffin, Philippe A. and Rahlin, Alexandra S. and Ramírez, Jorge and Ressler, Sean and Reynolds, Mark and Rodríguez-Montoya, Iván and Saez-Madain, Alejandro F. and Santana, Jorge and Shaw, Paul and Shirkey, Leslie E. and Silva, Kevin M. and Snow, William and Sousa, Don and Sridharan, T. K. and Stahm, William and Stark, Anthony A. and Test, John and Torstensson, Karl and Venegas, Paulina and Walther, Craig and Wei, Ta-Shun and White, Chris and Wieching, Gundolf and Wijnands, Rudy and Wouterloot, Jan G. A. and Yu, Chen-Yu and Yu (于威), Wei and Zeballos, Milagros},
	month = may,
	year = {2022},
	note = {Publisher: American Astronomical Society},
	pages = {L12},
}

@article{rioja_transformational_2023,
	title = {The {Transformational} {Power} of {Frequency} {Phase} {Transfer} {Methods} for {ngEHT}},
	volume = {11},
	copyright = {https://creativecommons.org/licenses/by/4.0/},
	issn = {2075-4434},
	url = {https://www.mdpi.com/2075-4434/11/1/16},
	doi = {10.3390/galaxies11010016},
	language = {en},
	number = {1},
	urldate = {2025-05-23},
	journal = {Galax},
	author = {Rioja, María J. and Dodson, Richard and Asaki, Yoshiharu},
	month = jan,
	year = {2023},
	note = {Publisher: MDPI AG},
	pages = {16},
}

@article{event_horizon_telescope_collaboration_first_2022_II,
	title = {First {Sagittarius} {A}* {Event} {Horizon} {Telescope} {Results}. {II}. {EHT} and {Multiwavelength} {Observations}, {Data} {Processing}, and {Calibration}},
	volume = {930},
	issn = {2041-8205, 2041-8213},
	url = {https://iopscience.iop.org/article/10.3847/2041-8213/ac6675},
	doi = {10.3847/2041-8213/ac6675},
	number = {2},
	urldate = {2026-01-30},
	journal = {ApJL},
	author = {{EHT Collaboration} and Akiyama, Kazunori and Alberdi, Antxon and Alef, Walter and Algaba, Juan Carlos and Anantua, Richard and Asada, Keiichi and Azulay, Rebecca and Bach, Uwe and Baczko, Anne-Kathrin and Ball, David and Baloković, Mislav and Barrett, John and Bauböck, Michi and Benson, Bradford A. and Bintley, Dan and Blackburn, Lindy and Blundell, Raymond and Bouman, Katherine L. and Bower, Geoffrey C. and Boyce, Hope and Bremer, Michael and Brinkerink, Christiaan D. and Brissenden, Roger and Britzen, Silke and Broderick, Avery E. and Broguiere, Dominique and Bronzwaer, Thomas and Bustamante, Sandra and Byun, Do-Young and Carlstrom, John E. and Ceccobello, Chiara and Chael, Andrew and Chan, Chi-kwan and Chatterjee, Koushik and Chatterjee, Shami and Chen, Ming-Tang and Chen 陈, Yongjun 永军 and Cheng, Xiaopeng and Cho, Ilje and Christian, Pierre and Conroy, Nicholas S. and Conway, John E. and Cordes, James M. and Crawford, Thomas M. and Crew, Geoffrey B. and Cruz-Osorio, Alejandro and Cui 崔, Yuzhu 玉竹 and Davelaar, Jordy and De Laurentis, Mariafelicia and Deane, Roger and Dempsey, Jessica and Desvignes, Gregory and Dexter, Jason and Dhruv, Vedant and Doeleman, Sheperd S. and Dougal, Sean and Dzib, Sergio A. and Eatough, Ralph P. and Emami, Razieh and Falcke, Heino and Farah, Joseph and Fish, Vincent L. and Fomalont, Ed and Ford, H. Alyson and Fraga-Encinas, Raquel and Freeman, William T. and Friberg, Per and Fromm, Christian M. and Fuentes, Antonio and Galison, Peter and Gammie, Charles F. and García, Roberto and Gentaz, Olivier and Georgiev, Boris and Goddi, Ciriaco and Gold, Roman and Gómez-Ruiz, Arturo I. and Gómez, José L. and Gu 顾, Minfeng 敏峰 and Gurwell, Mark and Hada, Kazuhiro and Haggard, Daryl and Haworth, Kari and Hecht, Michael H. and Hesper, Ronald and Heumann, Dirk and Ho 何, Luis C. 子山 and Ho, Paul and Honma, Mareki and Huang, Chih-Wei L. and Huang 黄, Lei 磊 and Hughes, David H. and Ikeda, Shiro and Impellizzeri, C. M. Violette and Inoue, Makoto and Issaoun, Sara and James, David J. and Jannuzi, Buell T. and Janssen, Michael and Jeter, Britton and Jiang 江, Wu 悟 and Jiménez-Rosales, Alejandra and Johnson, Michael D. and Jorstad, Svetlana and Joshi, Abhishek V. and Jung, Taehyun and Karami, Mansour and Karuppusamy, Ramesh and Kawashima, Tomohisa and Keating, Garrett K. and Kettenis, Mark and Kim, Dong-Jin and Kim, Jae-Young and Kim, Jongsoo and Kim, Junhan and Kino, Motoki and Koay, Jun Yi and Kocherlakota, Prashant and Kofuji, Yutaro and Koch, Patrick M. and Koyama, Shoko and Kramer, Carsten and Kramer, Michael and Krichbaum, Thomas P. and Kuo, Cheng-Yu and Bella, Noemi La and Lauer, Tod R. and Lee, Daeyoung and Lee, Sang-Sung and Leung, Po Kin and Levis, Aviad and Li 李, Zhiyuan 志远 and Lico, Rocco and Lindahl, Greg and Lindqvist, Michael and Lisakov, Mikhail and Liu 刘, Jun 俊 and Liu, Kuo and Liuzzo, Elisabetta and Lo, Wen-Ping and Lobanov, Andrei P. and Loinard, Laurent and Lonsdale, Colin J. and Lu 路, Ru-Sen 如森 and Mao 毛, Jirong 基荣 and Marchili, Nicola and Markoff, Sera and Marrone, Daniel P. and Marscher, Alan P. and Martí-Vidal, Iván and Matsushita, Satoki and Matthews, Lynn D. and Medeiros, Lia and Menten, Karl M. and Michalik, Daniel and Mizuno, Izumi and Mizuno, Yosuke and Moran, James M. and Moriyama, Kotaro and Moscibrodzka, Monika and Müller, Cornelia and Mus, Alejandro and Musoke, Gibwa and Myserlis, Ioannis and Nadolski, Andrew and Nagai, Hiroshi and Nagar, Neil M. and Nakamura, Masanori and Narayan, Ramesh and Narayanan, Gopal and Natarajan, Iniyan and Nathanail, Antonios and Fuentes, Santiago Navarro and Neilsen, Joey and Neri, Roberto and Ni, Chunchong and Noutsos, Aristeidis and Nowak, Michael A. and Oh, Junghwan and Okino, Hiroki and Olivares, Héctor and Ortiz-León, Gisela N. and Oyama, Tomoaki and Özel, Feryal and Palumbo, Daniel C. M. and Paraschos, Georgios Filippos and Park, Jongho and Parsons, Harriet and Patel, Nimesh and Pen, Ue-Li and Pesce, Dominic W. and Piétu, Vincent and Plambeck, Richard and PopStefanija, Aleksandar and Porth, Oliver and Pötzl, Felix M. and Prather, Ben and Preciado-López, Jorge A. and Psaltis, Dimitrios and Pu, Hung-Yi and Ramakrishnan, Venkatessh and Rao, Ramprasad and Rawlings, Mark G. and Raymond, Alexander W. and Rezzolla, Luciano and Ricarte, Angelo and Ripperda, Bart and Roelofs, Freek and Rogers, Alan and Ros, Eduardo and Romero-Cañizales, Cristina and Roshanineshat, Arash and Rottmann, Helge and Roy, Alan L. and Ruiz, Ignacio and Ruszczyk, Chet and Rygl, Kazi L. J. and Sánchez, Salvador and Sánchez-Argüelles, David and Sánchez-Portal, Miguel and Sasada, Mahito and Satapathy, Kaushik and Savolainen, Tuomas and Schloerb, F. Peter and Schonfeld, Jonathan and Schuster, Karl-Friedrich and Shao, Lijing and Shen 沈, Zhiqiang 志强 and Small, Des and Sohn, Bong Won and SooHoo, Jason and Souccar, Kamal and Sun 孙, He 赫 and Tazaki, Fumie and Tetarenko, Alexandra J. and Tiede, Paul and Tilanus, Remo P. J. and Titus, Michael and Torne, Pablo and Traianou, Efthalia and Trent, Tyler and Trippe, Sascha and Turk, Matthew and Van Bemmel, Ilse and Van Langevelde, Huib Jan and Van Rossum, Daniel R. and Vos, Jesse and Wagner, Jan and Ward-Thompson, Derek and Wardle, John and Weintroub, Jonathan and Wex, Norbert and Wharton, Robert and Wielgus, Maciek and Wiik, Kaj and Witzel, Gunther and Wondrak, Michael F. and Wong, George N. and Wu 吴, Qingwen 庆文 and Yamaguchi, Paul and Yoon, Doosoo and Young, André and Young, Ken and Younsi, Ziri and Yuan 袁, Feng 峰 and Yuan 袁, Ye-Fei 业飞 and Zensus, J. Anton and Zhang, Shuo and Zhao, Guang-Yao and Zhao 赵, Shan-Shan 杉杉 and Agurto, Claudio and Araneda, Juan Pablo and Arriagada, Oriel and Bertarini, Alessandra and Berthold, Ryan and Blanchard, Jay and Brown, Ken and Cárdenas, Mauricio and Cantzler, Michael and Caro, Patricio and Chuter, Tim C. and Ciechanowicz, Miroslaw and Coulson, Iain M. and Crowley, Joseph and Degenaar, Nathalie and Dornbusch, Sven and Durán, Carlos A. and Forster, Karl and Geertsema, Gertie and González, Edouard and Graham, Dave and Gueth, Frédéric and Han, Chih-Chiang and Herrera, Cristian and Herrero-Illana, Ruben and Heyminck, Stefan and Hoge, James and Huang, Yau-De and Jiang, Homin and John, David and Klein, Thomas and Kubo, Derek and Kuroda, John and Kwon, Caleb and Laing, Robert and Liu, Ching-Tang and Liu, Kuan-Yu and Mac-Auliffe, Felipe and Martin-Cocher, Pierre and Matulonis, Callie and Messias, Hugo and Meyer-Zhao, Zheng and Montenegro-Montes, Francisco and Montgomerie, William and Muders, Dirk and Nishioka, Hiroaki and Norton, Timothy J. and Olivares, Rodrigo and Pérez-Beaupuits, Juan Pablo and Parra, Rodrigo and Poirier, Michael and Pradel, Nicolas and Raffin, Philippe A. and Ramírez, Jorge and Reynolds, Mark and Saez-Madain, Alejandro F. and Santana, Jorge and Silva, Kevin M. and Sousa, Don and Stahm, William and Torstensson, Karl and Venegas, Paulina and Walther, Craig and Wieching, Gundolf and Wijnands, Rudy and Wouterloot, Jan G. A.},
	month = may,
	year = {2022},
	pages = {L13},
}

@article{kim_limb-brightened_2018,
	title = {The limb-brightened jet of {M87} down to the 7 {Schwarzschild} radii scale},
	volume = {616},
	copyright = {https://www.edpsciences.org/en/authors/copyright-and-licensing},
	issn = {0004-6361, 1432-0746},
	url = {https://www.aanda.org/10.1051/0004-6361/201832921},
	doi = {10.1051/0004-6361/201832921},
	urldate = {2025-05-27},
	journal = {A\&A},
	author = {Kim, J.-Y. and Krichbaum, T. P. and Lu, R.-S. and Ros, E. and Bach, U. and Bremer, M. and De Vicente, P. and Lindqvist, M. and Zensus, J. A.},
	month = aug,
	year = {2018},
	pages = {A188},
}

@book{thompson_interferometry_2017,
	address = {Cham},
	series = {Astronomy and {Astrophysics} {Library}},
	title = {Interferometry and {Synthesis} in {Radio} {Astronomy}},
	copyright = {https://creativecommons.org/licenses/by-nc/4.0},
	isbn = {978-3-319-44429-1 978-3-319-44431-4},
	url = {http://link.springer.com/10.1007/978-3-319-44431-4},
	language = {en},
	urldate = {2025-11-14},
	publisher = {Springer International Publishing},
	author = {Thompson, A. Richard and Moran, James M. and Swenson, George W.},
	year = {2017},
	doi = {10.1007/978-3-319-44431-4},
}

@article{antoniadis_international_2022,
	title = {The {International} {Pulsar} {Timing} {Array} second data release: {Search} for an isotropic gravitational wave background},
	volume = {510},
	copyright = {https://academic.oup.com/journals/pages/open\_access/funder\_policies/chorus/standard\_publication\_model},
	issn = {0035-8711, 1365-2966},
	shorttitle = {The {International} {Pulsar} {Timing} {Array} second data release},
	url = {https://academic.oup.com/mnras/article/510/4/4873/6503453},
	doi = {10.1093/mnras/stab3418},
	language = {en},
	number = {4},
	urldate = {2026-07-02},
	journal = {MNRAS},
	author = {Antoniadis, J and Arzoumanian, Z and Babak, S and Bailes, M and Bak Nielsen, A-S and Baker, P T and Bassa, C G and Bécsy, B and Berthereau, A and Bonetti, M and Brazier, A and Brook, P R and Burgay, M and Burke-Spolaor, S and Caballero, R N and Casey-Clyde, J A and Chalumeau, A and Champion, D J and Charisi, M and Chatterjee, S and Chen, S and Cognard, I and Cordes, J M and Cornish, N J and Crawford, F and Cromartie, H T and Crowter, K and Dai, S and DeCesar, M E and Demorest, P B and Desvignes, G and Dolch, T and Drachler, B and Falxa, M and Ferrara, E C and Fiore, W and Fonseca, E and Gair, J R and Garver-Daniels, N and Goncharov, B and Good, D C and Graikou, E and Guillemot, L and Guo, Y J and Hazboun, J S and Hobbs, G and Hu, H and Islo, K and Janssen, G H and Jennings, R J and Johnson, A D and Jones, M L and Kaiser, A R and Kaplan, D L and Karuppusamy, R and Keith, M J and Kelley, L Z and Kerr, M and Key, J S and Kramer, M and Lam, M T and Lamb, W G and Lazio, T J W and Lee, K J and Lentati, L and Liu, K and Luo, J and Lynch, R S and Lyne, A G and Madison, D R and Main, R A and Manchester, R N and McEwen, A and McKee, J W and McLaughlin, M A and Mickaliger, M B and Mingarelli, C M F and Ng, C and Nice, D J and Osłowski, S and Parthasarathy, A and Pennucci, T T and Perera, B B P and Perrodin, D and Petiteau, A and Pol, N S and Porayko, N K and Possenti, A and Ransom, S M and Ray, P S and Reardon, D J and Russell, C J and Samajdar, A and Sampson, L M and Sanidas, S and Sarkissian, J M and Schmitz, K and Schult, L and Sesana, A and Shaifullah, G and Shannon, R M and Shapiro-Albert, B J and Siemens, X and Simon, J and Smith, T L and Speri, L and Spiewak, R and Stairs, I H and Stappers, B W and Stinebring, D R and Swiggum, J K and Taylor, S R and Theureau, G and Tiburzi, C and Vallisneri, M and van der Wateren, E and Vecchio, A and Verbiest, J P W and Vigeland, S J and Wahl, H and Wang, J B and Wang, J and Wang, L and Witt, C A and Zhang, S and Zhu, X J},
	month = jan,
	year = {2022},
	pages = {4873--4887},
}

@article{epta_collaboration_and_inpta_collaboration_second_2024,
	title = {The second data release from the {European} {Pulsar} {Timing} {Array}: {V}. {Search} for continuous gravitational wave signals},
	volume = {690},
	copyright = {https://creativecommons.org/licenses/by/4.0},
	issn = {0004-6361, 1432-0746},
	shorttitle = {The second data release from the {European} {Pulsar} {Timing} {Array}},
	url = {https://www.aanda.org/10.1051/0004-6361/202348568},
	doi = {10.1051/0004-6361/202348568},
	urldate = {2026-07-02},
	journal = {A\&A},
	author = {{EPTA Collaboration and InPTA Collaboration} and Antoniadis, J. and Arumugam, P. and Arumugam, S. and Babak, S. and Bagchi, M. and Bak Nielsen, A.-S. and Bassa, C. G. and Bathula, A. and Berthereau, A. and Bonetti, M. and Bortolas, E. and Brook, P. R. and Burgay, M. and Caballero, R. N. and Chalumeau, A. and Champion, D. J. and Chanlaridis, S. and Chen, S. and Cognard, I. and Dandapat, S. and Deb, D. and Desai, S. and Desvignes, G. and Dhanda-Batra, N. and Dwivedi, C. and Falxa, M. and Ferranti, I. and Ferdman, R. D. and Franchini, A. and Gair, J. R. and Goncharov, B. and Gopakumar, A. and Graikou, E. and Grießmeier, J.-M. and Guillemot, L. and Guo, Y. J. and Gupta, Y. and Hisano, S. and Hu, H. and Iraci, F. and Izquierdo-Villalba, D. and Jang, J. and Jawor, J. and Janssen, G. H. and Jessner, A. and Joshi, B. C. and Kareem, F. and Karuppusamy, R. and Keane, E. F. and Keith, M. J. and Kharbanda, D. and Kikunaga, T. and Kolhe, N. and Kramer, M. and Krishnakumar, M. A. and Lackeos, K. and Lee, K. J. and Liu, K. and Liu, Y. and Lyne, A. G. and McKee, J. W. and Maan, Y. and Main, R. A. and Manzini, S. and Mickaliger, M. B. and Niţu, I. C. and Nobleson, K. and Paladi, A. K. and Parthasarathy, A. and Perera, B. B. P. and Perrodin, D. and Petiteau, A. and Porayko, N. K. and Possenti, A. and Prabu, T. and Quelquejay Leclere, H. and Rana, P. and Samajdar, A. and Sanidas, S. A. and Sesana, A. and Shaifullah, G. and Singha, J. and Speri, L. and Spiewak, R. and Srivastava, A. and Stappers, B. W. and Surnis, M. and Susarla, S. C. and Susobhanan, A. and Takahashi, K. and Tarafdar, P. and Theureau, G. and Tiburzi, C. and Van Der Wateren, E. and Vecchio, A. and Venkatraman Krishnan, V. and Verbiest, J. P. W. and Wang, J. and Wang, L. and Wu, Z.},
	month = oct,
	year = {2024},
	pages = {A118},
}

@article{xu_searching_2023,
	title = {Searching for the {Nano}-{Hertz} {Stochastic} {Gravitational} {Wave} {Background} with the {Chinese} {Pulsar} {Timing} {Array} {Data} {Release} {I}},
	volume = {23},
	issn = {1674-4527, 2397-6209},
	url = {https://iopscience.iop.org/article/10.1088/1674-4527/acdfa5},
	doi = {10.1088/1674-4527/acdfa5},
	number = {7},
	urldate = {2026-07-02},
	journal = {RAA},
	author = {Xu, Heng and Chen, Siyuan and Guo, Yanjun and Jiang, Jinchen and Wang, Bojun and Xu, Jiangwei and Xue, Zihan and Nicolas Caballero, R. and Yuan, Jianping and Xu, Yonghua and Wang, Jingbo and Hao, Longfei and Luo, Jingtao and Lee, Kejia and Han, Jinlin and Jiang, Peng and Shen, Zhiqiang and Wang, Min and Wang, Na and Xu, Renxin and Wu, Xiangping and Manchester, Richard and Qian, Lei and Guan, Xin and Huang, Menglin and Sun, Chun and Zhu, Yan},
	month = jul,
	year = {2023},
	pages = {075024},
}

@article{chen_constraining_2019,
	title = {Constraining astrophysical observables of galaxy and supermassive black hole binary mergers using pulsar timing arrays},
	volume = {488},
	copyright = {https://academic.oup.com/journals/pages/open\_access/funder\_policies/chorus/standard\_publication\_model},
	issn = {0035-8711, 1365-2966},
	url = {https://academic.oup.com/mnras/article/488/1/401/5521897},
	doi = {10.1093/mnras/stz1722},
	language = {en},
	number = {1},
	urldate = {2026-07-08},
	journal = {MNRAS},
	author = {Chen, Siyuan and Sesana, Alberto and Conselice, Christopher J},
	month = sep,
	year = {2019},
	pages = {401--418},
}

@ARTICLE{Britzen+2018MNRAS,
       author = {{Britzen}, S. and {Fendt}, C. and {Witzel}, G. and {Qian}, S.-J. and {Pashchenko}, I.~N. and {Kurtanidze}, O. and {Zajacek}, M. and {Martinez}, G. and {Karas}, V. and {Aller}, M. and {Aller}, H. and {Eckart}, A. and {Nilsson}, K. and {Ar{\'e}valo}, P. and {Cuadra}, J. and {Subroweit}, M. and {Witzel}, A.},
        title = "{OJ287: deciphering the `Rosetta stone of blazars}",
      journal = {\mnras},
         year = 2018,
        month = aug,
       volume = {478},
       number = {3},
        pages = {3199-3219},
          doi = {10.1093/mnras/sty1026},
       adsurl = {https://ui.adsabs.harvard.edu/abs/2018MNRAS.478.3199B}
}

@misc{dorazio_obs_sig_2023,
	title = {Observational {Signatures} of {Supermassive} {Black} {Hole} {Binaries}},
	copyright = {Creative Commons Attribution 4.0 International},
	url = {https://arxiv.org/abs/2310.16896},
	doi = {10.48550/ARXIV.2310.16896},
	urldate = {2025-05-23},
	publisher = {arXiv},
	author = {D'Orazio, Daniel J. and Charisi, Maria},
	year = {2023},
	note = {Version Number: 2},
}

@article{davis_reliable_2024,
	title = {Reliable {Identification} of {Binary} {Supermassive} {Black} {Holes} from {Rubin} {Observatory} {Time}-domain {Monitoring}},
	volume = {965},
	issn = {0004-637X, 1538-4357},
	url = {https://iopscience.iop.org/article/10.3847/1538-4357/ad276e},
	doi = {10.3847/1538-4357/ad276e},
	number = {1},
	urldate = {2026-07-28},
	journal = {ApJ},
	author = {Davis, Megan C. and Grace, Kaylee E. and Trump, Jonathan R. and Runnoe, Jessie C. and Henkel, Amelia and Blecha, Laura and Brandt, W. N. and Casey-Clyde, J. Andrew and Charisi, Maria and Witt, Caitlin A.},
	month = apr,
	year = {2024},
	pages = {34},
}

@article{dorazio_transition_2016,
	title = {A transition in circumbinary accretion discs at a binary mass ratio of 1:25},
	volume = {459},
	issn = {0035-8711, 1365-2966},
	shorttitle = {A transition in circumbinary accretion discs at a binary mass ratio of 1},
	url = {https://academic.oup.com/mnras/article-lookup/doi/10.1093/mnras/stw792},
	doi = {10.1093/mnras/stw792},
	language = {en},
	number = {3},
	urldate = {2025-05-27},
	journal = {MNRAS},
	author = {D'Orazio, Daniel J. and Haiman, Zoltán and Duffell, Paul and MacFadyen, Andrew and Farris, Brian},
	month = jul,
	year = {2016},
	pages = {2379--2393},
}

@article{molina_search_2026,
	title = {A {Search} for {Supermassive} {Black} {Hole} {Binary} {Candidates} in 46 yr {Radio} {Light} {Curves} of 83 {Blazars}},
	volume = {1000},
	issn = {0004-637X, 1538-4357},
	url = {https://iopscience.iop.org/article/10.3847/1538-4357/ae3141},
	doi = {10.3847/1538-4357/ae3141},
	number = {1},
	urldate = {2026-08-03},
	journal = {ApJ},
	author = {Molina, B. and Mróz, P. and De La Parra, P. V. and Readhead, A. C. S. and Surti, T. and Aller, M. F. and Scargle, J. D. and Reeves, R. A. and Aller, H. and Begelman, M. C. and Blandford, R. D. and Ding, Y. and Graham, M. J. and Harrison, F. and Hovatta, T. and Liodakis, I. and Lister, M. L. and Max-Moerbeck, W. and Pavlidou, V. and Pearson, T. J. and Ravi, V. and Sullivan, A. G. and Synani, A. and Tassis, K. and Tremblay, S. E. and Zensus, J. A.},
	month = mar,
	year = {2026},
	pages = {72},
}

@article{readhead_compelling_2026,
	title = {Compelling {Evidence} for a {Harmonic} in the {Light} {Curve} of the {Supermassive} {Black} {Hole} {Binary} {Candidate} {PKS} {J1309}+1154},
	volume = {996},
	issn = {2041-8205, 2041-8213},
	url = {https://iopscience.iop.org/article/10.3847/2041-8213/ae2656},
	doi = {10.3847/2041-8213/ae2656},
	number = {2},
	urldate = {2026-08-03},
	journal = {ApJL},
	author = {Readhead, A. C. S. and Aller, M. F. and Sullivan, A. G. and Blandford, R. D. and Mróz, P. and De La Parra, P. V. and Molina, B. and Most, E. R. and Lister, M. L. and Synani, A. and Aller, H. and Begelman, M. C. and Ding, Y. and Graham, M. J. and Harrison, F. and Hovatta, T. and Liodakis, I. and Max-Moerbeck, W. and Pavlidou, V. and Pearson, T. J. and Ravi, V. and Reeves, R. A. and Surti, T. and Tassis, K. and Tremblay, S. E. and Zensus, J. A.},
	month = jan,
	year = {2026},
	pages = {L39},
}

@article{kelley_gravitational_2021,
	title = {Gravitational self-lensing in populations of massive black hole binaries},
	volume = {508},
	copyright = {https://academic.oup.com/journals/pages/open\_access/funder\_policies/chorus/standard\_publication\_model},
	issn = {0035-8711, 1365-2966},
	url = {https://academic.oup.com/mnras/article/508/2/2524/6380509},
	doi = {10.1093/mnras/stab2776},
	language = {en},
	number = {2},
	urldate = {2026-08-03},
	journal = {MNRAS},
	author = {Kelley, Luke Zoltan and D’Orazio, Daniel J and Di Stefano, Rosanne},
	month = oct,
	year = {2021},
	pages = {2524--2536},
}

@article{roedig_observational_2014,
	title = {{OBSERVATIONAL} {SIGNATURES} {OF} {BINARY} {SUPERMASSIVE} {BLACK} {HOLES}},
	volume = {785},
	copyright = {http://iopscience.iop.org/info/page/text-and-data-mining},
	issn = {0004-637X, 1538-4357},
	url = {https://iopscience.iop.org/article/10.1088/0004-637X/785/2/115},
	doi = {10.1088/0004-637X/785/2/115},
	number = {2},
	urldate = {2026-08-03},
	journal = {ApJ},
	author = {Roedig, Constanze and Krolik, Julian H. and Miller, M. Coleman},
	month = apr,
	year = {2014},
	pages = {115},
}

@article{paczynski_model_1977,
	title = {A model of accretion disks in close binaries},
	volume = {216},
	issn = {0004-637X, 1538-4357},
	url = {http://adsabs.harvard.edu/doi/10.1086/155526},
	doi = {10.1086/155526},
	language = {en},
	urldate = {2026-08-03},
	journal = {ApJ},
	author = {Paczynski, B.},
	month = sep,
	year = {1977},
	pages = {822},
}

@article{oneill_unanticipated_2022,
	title = {The {Unanticipated} {Phenomenology} of the {Blazar} {PKS} 2131–021: {A} {Unique} {Supermassive} {Black} {Hole} {Binary} {Candidate}},
	volume = {926},
	issn = {2041-8205, 2041-8213},
	shorttitle = {The {Unanticipated} {Phenomenology} of the {Blazar} {PKS} 2131–021},
	url = {https://iopscience.iop.org/article/10.3847/2041-8213/ac504b},
	doi = {10.3847/2041-8213/ac504b},
	number = {2},
	urldate = {2026-08-06},
	journal = {ApJL},
	author = {O’Neill, S. and Kiehlmann, S. and Readhead, A. C. S. and Aller, M. F. and Blandford, R. D. and Liodakis, I. and Lister, M. L. and Mróz, P. and O’Dea, C. P. and Pearson, T. J. and Ravi, V. and Vallisneri, M. and Cleary, K. A. and Graham, M. J. and Grainge, K. J. B. and Hodges, M. W. and Hovatta, T. and Lähteenmäki, A. and Lamb, J. W. and Lazio, T. J. W. and Max-Moerbeck, W. and Pavlidou, V. and Prince, T. A. and Reeves, R. A. and Tornikoski, M. and Vergara De La Parra, P. and Zensus, J. A.},
	month = feb,
	year = {2022},
	pages = {L35},
}

@article{issaoun_first_2025,
	title = {First {Frequency} {Phase} {Transfer} from the 3 mm to the 1 mm {Band} on an {Earth}-sized {Baseline}},
	volume = {169},
	issn = {0004-6256, 1538-3881},
	url = {https://iopscience.iop.org/article/10.3847/1538-3881/adbb55},
	doi = {10.3847/1538-3881/adbb55},
	number = {4},
	urldate = {2026-07-15},
	journal = {The Astronomical Journal},
	author = {Issaoun, Sara and Pesce, Dominic W. and Rioja, María J. and Dodson, Richard and Blackburn, Lindy and Keating, Garrett K. and Doeleman, Sheperd S. and Sohn, Bong Won and Jiang 江, Wu 悟 and Hoak, Dan and Yu 于, Wei 威 and Torne, Pablo and Rao, Ramprasad and Tilanus, Remo P. J. and Martí-Vidal, Iván and Jung, Taehyun and Fitzpatrick, Garret and Sánchez-Portal, Miguel and Sánchez, Salvador and Weintroub, Jonathan and Gurwell, Mark and Kramer, Carsten and Durán, Carlos and John, David and Santaren, Juan L. and Kubo, Derek and Han, Chih-Chiang and Rottmann, Helge and SooHoo, Jason and Fish, Vincent L. and Zhao, Guang-Yao and Algaba, Juan Carlos and Lu 路, Ru-Sen 如森 and Cho, Ilje and Matsushita, Satoki and Schuster, Karl-Friedrich},
	month = apr,
	year = {2025},
	pages = {229},
}
\bibliographystyle{aasjournalv7}

\end{document}